\documentclass[preprint,prd,floats,aps,amsmath,amssymb,superscriptaddress,11pt,nofootinbib,longbibliography]{revtex4}
\pdfoutput=1
\usepackage[normalem]{ulem}
\usepackage{circledsteps}
\usepackage{tikzsymbols}
\usepackage{natbib}
\usepackage{xcolor}
\usepackage[hidelinks]{hyperref}
\hypersetup{
    colorlinks,
    linkcolor={blue!99!black},
    citecolor={blue!99!black},
    urlcolor={blue!99!black}
}
\usepackage{float}
\usepackage{comment}
\usepackage[caption=false]{subfig}
\usepackage[utf8]{inputenc}
\definecolor{lime}{HTML}{A6CE39}
\DeclareRobustCommand{\orcidicon}{
	\begin{tikzpicture}
	\draw[lime, fill=lime] (0,0) 
	circle [radius=0.2] 
	node[white] {{\fontfamily{qag}\selectfont \tiny ID}};
	\draw[white, fill=white] (-0.0625,0.095) 
	circle [radius=0.007];
	\end{tikzpicture}
	\hspace{-2mm}
}

\foreach \x in {A, ..., Z}{\expandafter\xdef\csname orcid\x\endcsname{\noexpand\href{https://orcid.org/\csname orcidauthor\x\endcsname}
			{\noexpand\orcidicon}}
}

\newcommand{\be}{\begin{equation}}
\newcommand{\ee}{\end{equation}}
\newcommand{\bea}{\begin{eqnarray}}
\newcommand{\eea}{\end{eqnarray}}

\def\x{\xi}

\allowdisplaybreaks

\begin{document}

\title{\Large Primordial non-Gaussianity as a probe of seesaw and leptogenesis
}

\author{Chee Sheng Fong}
\email{sheng.fong@ufabc.edu.br}
\affiliation{\footnotesize{Centro de Ciências Naturais e Humanas,
Universidade Federal do ABC, 09.210-170, Santo André, SP, Brazil}}

\author{Anish Ghoshal}
\email{anish.ghoshal@fuw.edu.pl}
%\affiliation{INFN Rome, Italy}
\affiliation{\footnotesize{Institute of Theoretical Physics, Faculty of Physics, University of Warsaw, \\ ul. Pasteura 5, 02-093 Warsaw, Poland}}

\author{Abhishek Naskar}
\email{abhiatrkmrc@gmail.com}
%\affiliation{INFN Rome, Italy}
\affiliation{\footnotesize{Indian Institute of Technology - Mumbai, India}}

\author{\\ Moinul Hossain Rahat}
\email{m.h.rahat@soton.ac.uk}
\affiliation{\footnotesize{School of Physics \& Astronomy, University of Southampton, Southampton SO17 1BJ, UK }}

\author{Shaikh Saad}
\email{shaikh.saad@unibas.ch}
\affiliation{\footnotesize{Department of Physics, University of Basel, Klingelbergstrasse\ 82, CH-4056 Basel, Switzerland\\ \vspace{1cm}}}

\begin{abstract}
We present the possibility that the seesaw mechanism and nonthermal leptogenesis can be  {investigated} via primordial non-Gaussianities in the context of a majoron curvaton model. Originating as a massless Nambu-Goldstone boson from the spontaneous breaking of the global baryon ($B$) minus lepton ($L$) number symmetry at a scale $v_{B-L}$, majoron becomes massive when it couples to a new confining 
sector through anomaly. Acting as a curvaton, majoron produces the observed red-tilted curvature power spectrum without relying on any inflaton contribution, and its decay in the post-inflationary era gives rise to a nonthermal population of right-handed neutrinos that participate in leptogenesis. A distinctive feature of the mechanism is the generation of observable non-Gaussianity, 
{in the parameter space where the red-tilted power spectrum and sufficient baryon asymmetry are produced.}   We {find} that the non-Gaussianity parameter $f_{\rm NL} \gtrsim \mathcal{O} (0.1)$ is produced for high-scale seesaw ($v_{B-L}$ at $\mathcal{O}(10^{14-17})$ GeV) and leptogenesis ($M_1 \gtrsim \mathcal{O}(10^6)$ GeV) where the latter represents the lightest right-handed neutrino mass. While the current bounds on local non-Gaussianity excludes some part of parameter space, the rest can be fully probed by future experiments like CMB-S4, LSST, and 21 cm tomography.

\end{abstract}

\maketitle
\tableofcontents

%%%%%%%%%%%%%%%%%%%%%%%%%%%%%%%%%%%%%%%%%%%%%%%%%%%%%%%%%%%%%%%%%
\section{Introduction}
\label{sec:intro}
%%%%%%%%%%%%%%%%%%%%%%%%%%%%%%%%%%%%%%%%%%%%%%%%%%%%%%%%%%%%%%%%%
Two clear evidence of physics beyond the Standard Model (SM) are the presence of nonzero neutrino mass, as demonstrated by the oscillation experiments \cite{Super-Kamiokande:1998kpq, SNO:2001kpb, SNO:2002tuh} (see the current global fit~\cite{Esteban:2020cvm,NuFIT}), and the  observation of cosmic baryon asymmetry, which is quite precisely inferred from the abundances of light elements during Big Bang Nucleosynthesis (BBN) \cite{Fields:2019pfx} and the measurement of the Cosmic Microwave Background (CMB) by Planck 2018 \cite{Planck:2018vyg}.

These two observations can be elegantly explained from a minimal extension of the SM with two or more Majorana right-handed neutrinos (RHNs). Tiny neutrino masses are generated through the type-I seesaw mechanism \cite{Minkowski:1977sc,Yanagida:1979as,GellMann1979,Glashow:1979nm,Mohapatra:1979ia}, and baryon asymmetry can be explained from baryogenesis via leptogenesis \cite{Fukugita:1986hr}. In both cases, the Majorona nature of the RHN mass plays a crucial role. This mass could arise from an explicit or a spontaneous violation of the gauged/global lepton number (or equivalently, baryon minus lepton number) symmetry. 

Due to the intimate connection between the neutrino mass scale and the CP violation for leptogenesis, the lightest RHN mass is required to exceed the Davidson-Ibarra bound, $M_1 \gtrsim 10^9$ GeV, if one assumes that the RHN mass spectrum is hierarchical, and lepton flavor effects~\cite{Nardi:2006fx,Abada:2006fw,Abada:2006ea} are completely absent~\cite{Davidson:2002qv,Giudice:2003jh}. By considering mildly hierarchical RHN mass with lepton flavor effects, one could reduce the bound down to $\sim 10^6$ GeV with certain tuning, such that the tree-level and one-loop contributions to neutrino mass are equally important~\cite{Moffat:2018wke}. If we were to consider quasi-degenerate mass spectrum such that CP violation can be resonantly enhanced~\cite{Covi:1996wh,Pilaftsis:1997jf,Pilaftsis:2003gt}, the Davidson-Ibarra bound is completely evaded. In that case, the lower bound on RHN mass scale is set by the temperature when the electroweak sphaleron processes decouple at $T = 132$ GeV~\cite{DOnofrio:2014rug}. Only the lepton asymmetry generated above $T > 132$ GeV can induce a nonzero baryon asymmetry, and hence, the RHN mass scale cannot be much lower than this temperature such that a sufficient amount of lepton asymmetry can be generated from the RHN decay. For such a low scale, one should also consider another source of lepton asymmetry generation from the RHN oscillations~\cite{Akhmedov:1998qx}. Combining both the contributions from decays and oscillations, it is shown in refs.~\cite{Klaric:2020phc,Drewes:2021nqr} that the RHN mass scale can be as low as 50 MeV. Low-scale RHNs can be searched in colliders, or in high-intensity experiments (see refs. \cite{Abdullahi:2022jlv, Beacham:2019nyx} for the recent status).

Unfortunately, any direct detection of the theoretically motivated scenario of high-scale seesaw and leptogenesis lies beyond the energy reach of the current or foreseeable future experiments. The two indirect probes that immediately come to mind are lepton number violation through neutrinoless double beta decay~\cite{Cirigliano:2022oqy} and CP violation in neutrino oscillation~\cite{Endoh:2002wm}. From the theoretical side, one can rely on the structure of couplings that are consistent with SO (10) Grand Unified Theories~\cite{DiBari:2008mp,Bertuzzo:2010et,Buccella:2012kc,Altarelli:2013aqa,Fong:2014gea,Mummidi:2021anm,Patel:2022xxu} or from the requirement of Higgs vacuum (meta)stability in the early Universe~\cite{Ipek:2018sai,Croon:2019dfw}. Another front is through the cosmological probes; CMB spectral indices ~~\cite{Ghoshal:2022fud} or gravitational waves from local cosmic strings~\cite{Dror:2019syi, Saad:2022mzu, DiBari:2023mwu}, global cosmic strings \cite{Fu:2023nrn}, domain walls~\cite{Barman:2022yos, King:2023cgv}, nucleating and colliding vacuum bubbles~\cite{Dasgupta:2022isg,Borah:2022cdx}, other topological defects~\cite{Dunsky:2021tih}, graviton bremmstrahlung~\cite{Ghoshal:2022kqp}, inflationary tensor perturbations~\cite{Berbig:2023yyy} and primordial blackholes~\cite{Perez-Gonzalez:2020vnz,Datta:2020bht,JyotiDas:2021shi,Barman:2021ost,Bernal:2022pue,Bhaumik:2022pil}, and investigating the oscillatory features of the curvature trispectrum \cite{Cui:2021iie} have been proposed to verify or constrain high-scale seesaw and leptogenesis. Under the circumstances, it is highly necessary, albeit challenging, to find new and complementary ways to probe the scale of the seesaw mechanism and leptogenesis.

In this work, we propose a novel approach to explore the scale of seesaw and leptogenesis by investigating their imprint on primordial non-Gaussianities. In our setup, the RHN masses are generated from the spontaneous breaking of a global $U(1)_{B-L}$ symmetry at a high scale $v_{B-L}$ around the GUT scale. A massless Nambu-Goldstone boson, dubbed as the ``majoron'', is generated in the process~\cite{Chikashige:1980ui,Schechter:1981cv}. We assume that the $U(1)_{B-L}$ is anomalous under a new force that is confined at a scale $\Lambda$ below $v_{B-L}$. As a result of this anomaly, the majoron experiences a periodic potential and becomes massive. We show that the majoron can act as a curvaton \cite{Linde:1996gt, Enqvist:2001zp, Lyth:2001nq, Moroi:2001ct}, which produces the entirety of the observed red-tilted scalar density perturbation without any contribution from the inflaton, following closely the scenario of an axionlike curvaton in ref.~\cite{Kobayashi:2020xhm}.\footnote{See refs. \cite{Sasaki:2006kq,  Enqvist:2008gk,   Chingangbam:2009xi, Enqvist:2009zf,  Enqvist:2009ww, Mazumdar:2010sa, Kawasaki:2011pd,  Kawasaki:2012gg, Byrnes:2014xua,    Takahashi:2022bqc, Ghoshal:2023lly} for some other curvaton scenarios.} 
In this setup, inflation \cite{Guth:1980zm,Starobinsky:1980te,Linde:1981mu, Albrecht:1982wi, Linde:1983gd} is driven by the inflaton and majoron remains a subdominant field during inflation, \textcolor{black}{with its quantum fluctuations converted to classical perturbations at horizon exit with a flat spectrum. The perturbations in the majoron field are converted into curvature perturbations at late times in the post-inflationary era, when the inflaton decay products have redshifted away and majoron dominates the energy density of the Universe.} This process generates observable local non-Gaussianity within the reach of future CMB and LSS experiments \cite{Munchmeyer:2018eey} and 21 cm tomography \cite{Munoz:2015eqa}. Intriguingly, we find that the decay of the majoron at the end of the curvaton dynamics can produce a nonthermal population of RHNs which participate in successful leptogenesis. The scale of leptogenesis, primarily determined by the lightest RHN mass, and the seesaw scale, identified as the $B-L$ breaking scale, leave nontrivial imprint on the non-Gaussianity parameter $f_{\rm NL}$. {Although such observable $f_{\rm NL}$ may arise from other scenarios, nonobservation of $f_{\rm NL}$ predicted in our scenario would certainly falsify the proposed majoron-as-curvaton mechanism.} 

The paper is organized as follows. In section \ref{sec:2} we provide a concise overview of the majoron model, briefly examining its key aspects. In section \ref{sec:3} we discuss how the majoron can act as a curvaton and generate the curvature power spectrum and spectral index, and how this leads to observable non-Gaussianity. Section \ref{sec:4} investigates the scenario of nonthermal leptogenesis after the majoron decays. We present our key results in section \ref{sec:5} and conclude in section \ref{sec:6}.

%%%%%%%%%%%%%%%%%%%%%%%%%%%%%%%%%%%%%%%%%%%%%%%%%%%%%%%%%%%%%%%%%
\section{Majoron model for neutrino mass genesis}
\label{sec:2}
We will consider the simplest majoron model \cite{Chikashige:1980ui,Schechter:1981cv}
with a global baryon
minus lepton number $U(1)_{B-L}$ symmetry under which three SM singlet
fermions $N_{i}$ carry charge $-1$ and a complex scalar field $\sigma$
carries charge $2$. The relevant new interactions for us is
\begin{eqnarray}
-{\cal L} & \supset & \frac{1}{2}\xi_{i}\sigma\overline{N_{i}^{c}}N_{i}+\lambda_{\alpha i}\overline{\ell_{\alpha i}}\epsilon H^{\dagger}N_{i}+y_{\alpha}\overline{\ell_{\alpha}}He_{\alpha}+\textrm{H.c.},
\label{eq:Majoron_model}
\end{eqnarray}
where $\ell_{\alpha}$ and $H$ are respectively the SM lepton and
Higgs doublet ($\epsilon$ is the $SU(2)$ antisymmetry tensor). Without loss of generality, we work in the basis where dimensionless couplings $\xi_{i}$
and $y_{\alpha}$ are real %and diagonal 
while $\lambda$ remains a
complex 3$\times3$ matrix. After $\sigma$ acquires a vacuum expectation value $v_{B-L}$ along the radial direction, 
\begin{eqnarray}
\sigma & = & \left(v_{B-L}+\rho\right)e^{i\chi/v_{B-L}},
\end{eqnarray}
we have
\begin{eqnarray}
-{\cal L} & \supset & \frac{1}{2}M_{i}\overline{N_{i}^{c}}N_{i}e^{i\chi/v_{B-L}}+\frac{1}{2}\xi_{i}\rho e^{i\chi/v_{B-L}}\overline{N_{i}^{c}}N_{i}+\lambda_{\alpha i}\overline{\ell_{\alpha i}}\epsilon H^{\dagger}N_{i}+y_{\alpha}\overline{\ell_{\alpha}}He_{\alpha}+\textrm{H.c.},
\end{eqnarray}
where $M_{i}=\xi_{i}v_{B-L}$. We will assume that the radial field has mass $m_{\rho}>T_{\textrm{max}}$,
where $T_{\textrm{max}}$ is the maximum temperature from reheating
after inflation and ignore the contribution
from $\rho$. At this point, the majoron $\chi$ is a Nambu-Goldstone (NG) boson which remains massless. It only couples to $N_{i}$
and an elegant way to deal with this nonlinear parametrization is
to carry out field-dependent redefinition of fermionic fields, $f\to fe^{iq_{f}^{B-L}\chi/\left(2v_{B-L}\right)}$,
where $q_{f}^{B-L}$ is the $B-L$ charge of the fermion $f$. From the kinetic terms, we obtain
\begin{eqnarray}
{\cal L} & \supset & -\frac{\partial_{\mu}\chi}{2v_{B-L}}J_{B-L}^{\mu},
\end{eqnarray}
where $J_{B-L}^{\mu}=\sum_{f}q_{f}^{B-L}\overline{f}\gamma^{\mu}f$
is the $B-L$ current. Carrying out integration by parts in the action
and discarding the surface term, we have
\begin{eqnarray}
{\cal L} & \supset & \frac{\chi}{2v_{B-L}}\partial_{\mu}J_{B-L}^{\mu}=\frac{i\chi}{v_{B-L}}M_{i}\overline{N_{i}^{c}}N_{i},
\end{eqnarray}
where the nonconservation of $B-L$ current comes from the Majorana
mass term and $U(1)_{B-L}$ has no anomaly with respect to the SM gauge interactions. 

Now, we will give mass $m_\chi$ to $\chi$ by assuming that $U(1)_{B-L}$ is anomalous under a new gauge interaction which becomes strong at $\Lambda < v_{B-L}$. $\chi$ will couple to the new gauge field as follows
\begin{eqnarray}
{\cal L} & \supset & \frac{\chi}{v_{B-L}}\frac{g_{X}^{2}n}{16\pi^{2}}G\widetilde{G},\label{eq:anomaly}
\end{eqnarray}
where $g_{X}$ is the gauge coupling of the new gauge interaction,
$\widetilde{G}$ is its dual field strength of $G$ (with both Lorentz
and gauge indices suppressed) while $n$ is the anomaly coefficient. In the explicit majoron model in this work, the decay of $\chi$ to the new sector through eq.~\eqref{eq:anomaly} is loop-suppressed or forbidden in the absence of states with masses lighter than $m_{\chi}/2$. On the other hand, if $m_{\chi}>2M_{i}$ for some $i$, the decay width of $\chi\to N_{i}N_{i}$ is given by\footnote{If $m_{\chi}<2M_{i}$ for all $i$, we can still have the decay $\chi\to\ell\ell HH$ through off-shell $N_{i}$. 
Since we are interested in the production of $N_{i}$ through decay of $\chi$, we will not consider this possibility further.
}
\begin{eqnarray}
\Gamma_{\chi} & = & \frac{m_{\chi}M_{i}^{2}}{16\pi v_{B-L}^{2}}\sqrt{1-\frac{4M_{i}^{2}}{m_{\chi}^{2}}}.
\label{eq:decaywidth_to_N}
\end{eqnarray}
For definiteness, we will assume that $m_{\chi}>2M_{1}$ while $m_{\chi}\ll M_{2}, M_{3}$ such that the dominant channel is $\chi\to N_{1}N_{1}$. 
{Relaxing this assumption will give only order of one changes to the parameter space of our analysis.}
Hence we will consider only the decay $\chi\to N_{1}N_{1}$ with decay width given by eq.~\eqref{eq:decaywidth_to_N}.

{
To complete the story, after the electroweak symmetry breaking, light neutrino mass is generated through the type-I seesaw mechanism \cite{Minkowski:1977sc,Yanagida:1979as,GellMann1979,Glashow:1979nm,Mohapatra:1979ia} with light neutrino mass matrix given by 
\begin{equation}
m_\nu = - v^2 \lambda M^{-1} \lambda^T,
\end{equation}
with $v=174$ GeV the Higgs vacuum expectation value.
}

%--------------------------------
\section{Majoron as a curvaton}\label{sec:3}
%\subsection{Review of axion as curvaton}

In this section we discuss how majoron can play the role of curvaton to generate the power spectrum, spectral index and non-Gaussianity. 

\subsection{Cosmological evolution}
As we discussed in the previous section, from the spontaneous breaking of global $U(1)_{B-L}$ symmetry at a scale $v_{B-L}$,
one obtains a massless majoron $\chi$.
If $U(1)_{B-L}$ is anomalous under a new gauge interaction which
becomes strong at a scale $\Lambda<v_{B-L}$, $\chi$ will acquire
a periodic potential \cite{DiLuzio:2020wdo,Kobayashi:2020xhm}
\begin{eqnarray}
V\left(\chi\right) & = & m_{\chi}^{2}v_{B-L}^{2}\left(1-\cos\frac{\chi}{v_{B-L}}\right).
\end{eqnarray}
Here, we model the majoron mass, following ref.~\cite{Kobayashi:2020xhm}, 
\begin{align}
    m_{\chi} = m_{\chi 0 } \left(\frac{\Lambda}{T}\right)^p, \qquad {\rm and} \qquad p = p_0\ \Theta (T-\Lambda),
\end{align}
where $p_0 \neq 0$  and $\Theta$ is the Heaviside step function $\Theta (x) = 1$ for $x > 0$ and $0$ otherwise. If the new gauge sector is identified as QCD, $p_0 = 4$ \cite{DiLuzio:2020wdo}. $\chi$ is now a pseudo-NG boson
with zero temperature mass 
\begin{eqnarray}
m_{\chi0} & = & \frac{\Lambda^{2}}{v_{B-L}}=10^{12}\,\textrm{GeV}\left(\frac{\Lambda}{10^{14}\,\textrm{GeV}}\right)^{2}\left(\frac{10^{16}\,\textrm{GeV}}{v_{B-L}}\right).
\end{eqnarray}

We assume that the de Sitter temperature during inflation $T_{\textrm{inf}}=H_{\textrm{inf}}/({2\pi})$
lies below both $v_{B-L}$ and $\Lambda$
\begin{eqnarray}
T_{\textrm{inf}} & < & \Lambda<v_{B-L}.
\end{eqnarray}
This implies that during inflation, the mass of the majoron is $m_{\chi 0}$. The equation of motion of the majoron can be written as
\begin{align}
    \ddot{\chi} + 3H_{\rm inf} \dot{\chi} + \frac{dV}{d\chi} = 0, \label{EOMcurv0}
\end{align}
where the first term can be neglected assuming slow roll during inflation, and the Hubble rate during inflation, $H_{\rm inf}$, can be considered constant. Solving this equation, we can determine the majoron field value at the end of inflation $\chi_{\rm end}$.

As soon as inflation ends, the inflaton instantaneously decays into radiation and reheats the Universe. The maximum radiation temperature is obtained
by setting the inflaton energy density equal that of the radiation
$3H_{\textrm{inf}}^{2}M_{\textrm{Pl}}^{2}=\frac{\pi^{2}}{30}g_{\star}T^{4}$,
and solving for the temperature 
\begin{eqnarray}
T_{\textrm{max}} & = & \left(\frac{90}{\pi^{2}g_{\star}}\right)^{1/4}\sqrt{H_{\textrm{inf}}M_{\textrm{Pl}}}=8.4\times10^{15}\,\textrm{GeV}\left(\frac{H_{\textrm{inf}}}{10^{14}\,\textrm{GeV}}\right)^{1/2}\left(\frac{106.75}{g_{\star}}\right)^{1/4},\label{eq:Tmax}
\end{eqnarray}
where the reduced Planck mass is $M_{\textrm{Pl}}=2.43\times10^{18}$
GeV and $g_{\star}=106.75$ assuming the SM relativistic degrees of
freedom. We make a further assumption  
\begin{eqnarray}
\Lambda & < & T_{\textrm{max}}<v_{B-L},
\end{eqnarray}
so that the potential of $\chi$ diminishes after inflation and its field value is approximately frozen at $\chi_{\textrm{end}}$.  In the post-inflationary period, the evolution of the Hubble rate can be tracked from the Friedmann equation
\begin{align}
    H^2 = \frac{1}{3M_{\rm Pl}^2} (\rho_r+\rho_\chi), \label{Hubble}
\end{align}
where the energy densities of radiation and majoron are given by
\begin{align}
    \dot{\rho}_r + 4H\rho_r = 0, \qquad {\rm and} \qquad \rho_{\chi} = \frac{1}{2} \dot{\chi}^2 + V(\chi). \label{EDrchi}
\end{align}
The majoron field evolves according to 
\begin{align}
    \ddot{\chi} + 3H \dot{\chi} + \frac{dV}{d\chi} = 0, \label{EOMcurv1}
\end{align}
where the first term can no longer be neglected, and $H$ evolves with time. 
{In the equations above, we have not included the decay terms of $\chi$ since we will be using the sudden decay approximation when $H = \Gamma_\chi $. This has negligible effect on the calculation of $f_{\rm NL}$, as shown in ref.~\cite{Kitajima:2014xna}.}

As the Universe cools down sufficiently, majoron begins to oscillate about the potential minimum when $H\simeq m_{\chi}$. 
Assuming that this happens at $T_{\textrm{osc}}$ when the Universe is still radiation-dominated by setting $3m_{\chi}^{2}M_{\textrm{Pl}}^{2}=\frac{\pi^{2}}{30}g_{\star}T_{\textrm{osc}}^{4}$, we have
\begin{align}
    T_{\rm osc} &= \Lambda \left[ \sqrt{\frac{90}{g_\star}} \frac{M_{\rm Pl}}{\pi v_{B-L}}\right]^{\frac{1}{p+2}}.
\end{align}
As long as $v_{B-L} < \sqrt{\frac{90}{\pi^2 g_\star}} M_{\rm Pl} \approx 0.3M_{\rm Pl}$, $T_{\rm osc} > \Lambda$, and the majoron potential at the onset of oscillation would depend on $p \neq 0$. This dependence continues until $T = \Lambda = \sqrt{m_{\chi 0} v_{B-L}}$.

At the onset of oscillation, the number
density of $\chi$ can be approximated as
\begin{eqnarray}
n_{\chi}\left(T_{\textrm{osc}}\right) & = & \left.\frac{V\left(\chi_{\textrm{end}}\right)}{m_{\chi}}\right|_{T=T_{\textrm{osc}}}=m_{\chi0}\left(\frac{\Lambda}{T_{\textrm{osc}}}\right)^{p}v_{B-L}^{2}\left(1-\cos\frac{\chi_{\rm end}}{v_{B-L}}\right).
\end{eqnarray}
From this point onwards, the number of $\chi$ per comoving volume
is conserved 
\begin{eqnarray}
Y_{\chi} & = & \left.\frac{n_{\chi}}{s}\right|_{T=T_{\textrm{osc}}}=\frac{45}{2\pi^{2}g_\star}\frac{m_{\chi0}v_{B-L}^{2}}{T_{\textrm{osc}}^{3}}\left(\frac{\Lambda}{T_{\textrm{osc}}}\right)^{p}\left(1-\cos\frac{\chi_{\textrm{end}}}{v_{B-L}}\right),\label{eq:Ychi}
\end{eqnarray}
where $s=\frac{2\pi^{2}}{45}g_{\star}T^{3}$ is the cosmic entropy
density. In fact, $Y_{\chi}$ is not so sensitive to $p$ and for
$p\to\infty$, the value saturates to
\begin{eqnarray}
Y_{\chi} & \to & \frac{45}{2\pi^{2}g_\star}\sqrt{\frac{\pi^{2}g_{\star}}{90}}\frac{v_{B-L}^{2}}{M_{\textrm{Pl}}\Lambda}\left(1-\cos\frac{\chi_{\textrm{end}}}{v_{B-L}}\right).\label{eq:Ychi_sat}
\end{eqnarray}
For our calculation, we will fix $p=4$.

Assuming sudden decay approximation, $\chi$ decays at $T_{\textrm{dec}}$
when $H\simeq\Gamma_{\chi}$ and from energy conservation, we have
\begin{eqnarray}
3\Gamma_{\chi}^{2}M_{\textrm{Pl}}^{2} & = & m_{\chi}Y_{\chi}s\left(T_{\textrm{dec}}\right)+\frac{\pi^{2}}{30}g_{\star}T_{\textrm{dec}}^{4}=\frac{\pi^{2}}{30}g_{\star}\tilde{T}^{4}, \label{Tdecay}
\end{eqnarray}
where $\tilde{T}$ is the temperature after all $\chi$ particles
have decayed.

The decay of the majoron injects entropy and dilutes the number density of other species. The dilution factor from entropy injection is
\begin{eqnarray}
d & \equiv & \left(\frac{\tilde{T}}{T_{\textrm{dec}}}\right)^{3}=\left(1+\frac{4}{3}\frac{m_{\chi}Y_{\chi}}{T_{\textrm{dec}}}\right)^{3/4}\equiv\left(1+R\right)^{3/4},\label{eq:dilution}
\end{eqnarray}
where $R$ is the energy density of $\chi$ in comparison to the radiation
density at $T_{\textrm{dec}}$. 
One can solve for $T_\textrm{dec}$ and hence $d$ for a given $\Gamma_{\chi}$ and $Y_{\chi}$.

\subsection{Power spectrum, spectral index and non-Gaussianity}
Majoron remains a sub-dominant source of energy density during inflation and in the early reheating period, at least until it starts to oscillate. The energy density of the oscillating majoron eventually supersedes the radiation energy density. The super-horizon field fluctuations of the majoron created during inflation can be converted into curvature perturbations as the majoron dominates the energy density of the Universe. The generation of the curvature perturbation continues until the majoron finally decays at $H \simeq \Gamma_\chi$. If the majoron is to act as a curvaton, it is crucial for it to be sufficiently long-lived so that it can dominate the energy density of the Universe before its decay.

The curvature perturbation induced by the majoron can be computed using the $\delta N$ formalism \cite{Starobinsky:1982ee, Salopek:1990jq, Sasaki:1995aw, Lyth:2004gb, Lyth:2005fi, Sugiyama:2012tj}. The quantity $\delta N$ is defined as the  perturbation in the number of e-folds between a spatially flat slice and an  uniform density slice and the curvature perturbation depends linearly on $\delta N$. 
Expanding in terms of the field fluctuations, the curvature perturbation can be expressed as
\begin{align}
    \zeta(\mathbf{x}) = \delta N = \frac{\partial N}{\partial \chi_{\rm in}} \delta \chi_{\rm in} + \frac{1}{2} \frac{\partial^2N}{\partial\chi_{\rm in}^2} \delta \chi_{\rm in}^2  + \cdots. \label{zetaf}
\end{align}
Here $\chi_{\rm in}$ is the initial majoron field value at time $t_{\rm in}$ when the CMB modes exit the horizon, and $N$ is the total number of e-folds between $t_{\rm in}$ and the time when the majoron decays. Defining the Fourier transform of the curvature perturbation as
\begin{align}
    \zeta_\mathbf{k} = \int d^3 \mathbf{x}\ e^{-i \mathbf{k}\cdot \mathbf{x}} \zeta (\mathbf{x}), \label{zetak}
\end{align}
the two- and three-point correlation functions are given by \cite{Maldacena:2002vr}
\begin{align}
    \langle \zeta_\mathbf{k_1} \zeta_\mathbf{k_2} \rangle' &= \frac{2\pi^2}{k_1^3} P_\zeta (k_1), \label{2pcf}  \\
    \langle \zeta_\mathbf{k_1} \zeta_\mathbf{k_2} \zeta_\mathbf{k_3}\rangle' &= \frac{6}{5} f_{\rm NL} (k_1, k_2, k_3) \left[\frac{(2\pi^2)^2}{k_1^3 k_2^3}P_\zeta(k_1)P_\zeta(k_2) + \text{cyclic}\right], \label{3pcf}
\end{align}
where prime denotes the delta function $(2\pi)^3 \delta^{(3)}(\sum_i \mathbf{k_i})$ stripped off the correlation functions. Plugging eqs.~\eqref{zetaf} and \eqref{zetak} into eqs.~\eqref{2pcf} and \eqref{3pcf}, one can derive the following expressions for the power spectrum and the non-Gaussianity parameter \cite{Chingangbam:2009xi,Kawasaki:2011pd,Lyth:2005qk},
\begin{align}
    P_\zeta &\simeq P_{\delta \chi_{\rm in}} \left(\frac{\partial N}{\partial \chi_{\rm in}}\right)^2, \qquad
    f_{\rm NL} \simeq \frac{5}{6} \frac{\partial^2 N}{\partial\chi_{\rm in}^2} \left(\frac{\partial N}{\partial \chi_{\rm in}}\right)^{-2}, \label{PzetafNL} 
\end{align}
up to leading order in $\delta \chi_{\rm in}$. Here, power spectrum of the majoron field fluctuations on spatially-flat hypersurfaces at horizon exit is assumed to be Gaussian, and is given by $P_{\delta \chi_{\rm in}} = H_{\rm inf}^2/(2\pi)^2$ \cite{Byrnes:2006fr}. Here we have assumed that the Hubble rate during inflation remains nearly constant.

Typically it takes $50-60$ e-folds of inflation to resolve the horizon problem \cite{Remmen:2014mia}. For definiteness, we will take $N = -50$ at the time when the CMB modes leave the horizon. Then, identifying $N=0$ as the e-folding number at the end of inflation, we denote $N_f$ as the `final' e-folding number that marks the end of the curvaton dynamics when the majoron decays.

To calculate the power spectrum and the non-Gaussianity parameter, we need to know the first and second order derivatives of the `total' e-folding number $N$ with respect to the initial field value $\chi_{\rm in}$. Since the number of e-folds before inflation is independent of the majoron field value $\chi_{\rm in}$, we can instead take the derivatives of the final e-folding number $N_f$ at the time of majoron decay. It is done by carefully tracking the evolution of the majoron field value, Hubble rate, and the energy density of the radiation bath and of the majoron field with the coupled system of equations \eqref{EOMcurv0} \eqref{Hubble}, \eqref{EDrchi} and \eqref{EOMcurv1}, taking the non-trivial temperature dependence of the majoron potential at different stages of evolution into account. 

For convenience, we define the dimensionless field $\theta \equiv \chi/v_{B-L}$. During inflation, the dynamics of the majoron can be described by eq.~\eqref{EOMcurv0}, dropping the first term assuming slow roll. In terms of the e-folding number $N = \log{a(t)}$, it can be written as 
\begin{align}
    3H_{\rm inf}^2 \frac{d\theta}{d N} + m_{\chi 0}^2 \sin{\theta} = 0, \label{EOMinf}
\end{align}
with the initial condition $\theta (N=-50) = \theta_{\rm in} = \chi_{\rm in}/v_{B-L}$. Here we have taken the majoron mass to be $m_{\chi 0}$ assuming $T_{\rm inf} < \Lambda$.

We denote the end of inflation with $N=0$. To analyze the post-inflationary dynamics, we
express the Hubble rate $H$ in terms of the e-folding number $N$, and introduce the dimensionless variable $x \equiv m_{\chi 0}t$,. Then, eqs.~\eqref{Hubble}, \eqref{EDrchi} and \eqref{EOMcurv1} can be condensed into the following two coupled equations,
\begin{align}
    &\theta''(x) + N'(x)\ \theta'(x) + \left( \frac{\Lambda}{T} \right)^{2p} \sin{\theta(x)} = 0, \label{EOM1}\\
    &N'(x) = \left[ \tilde{\rho}_{r, \rm ini}\ e^{-4 N(x)} + \frac{v_{B-L}^2}{3M_{\rm Pl}^2} \left\{ \frac{1}{2} \theta'(x)^2 + \left( \frac{\Lambda}{T} \right)^{2p} \left(1- \cos{\theta(x)}\right) \right\} \right]^{1/2}. \label{EOM2} 
\end{align}
Here prime denotes derivative w.r.t. $x$, and $\tilde{\rho}_{r, \rm ini} = {\rho_{r, \rm ini}}/{(3 m_{\chi 0}^2 M_{\rm Pl}^2)}$,
where $\rho_{r, \rm ini} = 3H_{\rm inf}^2 M_{\rm Pl}^2$ is the radiation energy density after instantaneous reheating at the end of inflation. 

For any particular initial condition $\pi/2 \leq \theta_{\rm in} \leq \pi$ at $N=-50$, the solution of eq.~\eqref{EOMinf} yields $\theta_{\rm end}$, the majoron field value at the end of inflation. This, together with its first derivative calculated from eq.~\eqref{EOMinf}, are fed into eqs.~\eqref{EOM1} and \eqref{EOM2} as initial conditions at $N=0$ at $x=m_{\chi 0}/(2H_{\rm inf})$. In order to solve eqs.~\eqref{EOM1} and \eqref{EOM2}, the post-inflationary period can be divided into two regimes, (i) $T > \Lambda$ where $p\neq 0$, and (ii) $T < \Lambda$ where $p=0$. The first regime occurs from the end of inflation at $x = m_{\chi 0}/(2H_{\rm inf})$ to $x=x_p$ when $T = \Lambda$, where 
\begin{align}
    x_p = \frac{1}{2\pi} \sqrt{\frac{90}{g_\star}} \left( \frac{M_{\rm Pl}}{v_{B-L}} \right). 
\end{align}
The Universe is radiation dominated during this regime, and the relation between $x$ and temperature $T$ can be determined  from
\begin{align}
    \rho_r = \frac{\pi^2}{30} g_\star T^4 = 3H^2 M_{\rm Pl}^2 = \frac{3}{4} \frac{M_{\rm Pl}^2 m_{\chi 0}^2}{x^2}, 
\end{align}
where we have used $t= 1/(2H)$ in writing the last equality. 
The second regime then starts and continues until the majoron decays at $H = \Gamma_{\chi}$, corresponding to $x = m_{\chi 0}/\Gamma_{\chi}$, at which point we evaluate the `final' e-folding number $N_f$.

Since the majoron field value oscillates until it decays, solving the system of equations \eqref{EOM1} and \eqref{EOM2} is numerically very challenging. We use an approximate method, elaborated and justified in appendix \ref{app:A}, to circumvent this problem. 

Scanning over $0 \leq \theta_{\rm in} \leq \pi$ and determining $N_f$ for each $\theta_{\rm in}$, we can approximate $dN_f/d\theta_{\rm in}$ and $d^2N_f/d\theta_{\rm in}^2$ to calculate the power spectrum and $f_{\rm NL}$ {from eq.~\eqref{PzetafNL}},
\begin{align}
    P_\zeta \simeq \left(\frac{H_{\rm inf}}{2\pi v_{B-L}}\right)^2 \left(\frac{\partial N_f}{\partial \theta_{\rm in}}\right)^2, \qquad f_{\rm NL} \simeq \frac{5}{6} \frac{\partial^2 N_f}{\partial\theta_{\rm in}^2} \left(\frac{\partial N_f}{\partial\theta_{\rm in}}\right)^{-2}. \label{Pfnl}
\end{align}
The spectral index of the power spectrum is defined as $n_s - 1 = d\log{P_\zeta}/d\log{k}$. Using $d\log{k} \simeq H_{\rm inf}/m_{\chi 0} dx$, and assuming slow-roll during inflation, it can be expressed as
\begin{align}
    n_s -1 \simeq \frac{2}{3} \left(\frac{m_{\chi 0}}{H_{\rm inf}}\right)^2 \cos{\theta_{\rm in}} + 2\frac{\dot{H}_{\rm inf}}{H_{\rm inf}^2}, \label{nseq}
\end{align}
where the first (second) term comes from the curvaton (inflaton) dynamics. 

\begin{figure}
    \centering        \includegraphics[width=0.99\textwidth]{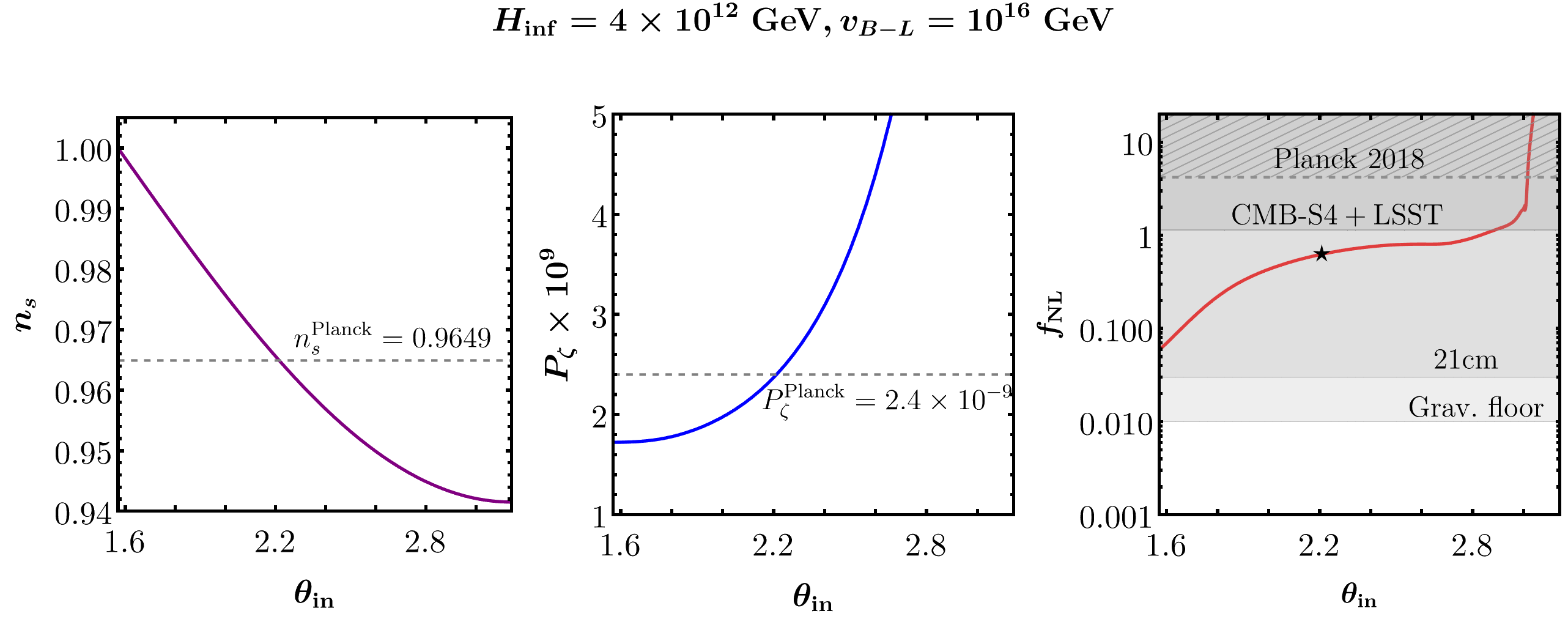}
        \caption{ An example of determining the initial dimensionless field value $\theta_{\rm in} \simeq 2.21$ and majoron mass $m_\chi \simeq 1.18 \times 10^{12}$ GeV, for which the spectral index and the scalar power spectrum matches with Planck 2018 values $n_s = 0.9649$ and $P_\zeta^{\rm CMB} = 2.4 \times 10^{-9}$. Local non-Gaussianity at this point is $0.63$. These results are insensitive to the decay width as long as the ratio of majoron's energy density before decay to radiation energy density is sufficiently large, $R \gg 1$. Shaded hatched region in the third plot shows Planck 2018 bound on positive local non-Gaussianity $f_{\rm NL}^{\rm local} < 4.2$ \cite{Planck:2019kim}, and other shaded regions represent the sensitivity of future experiments CMB-S4+LSST, 21 cm tomography, and ``gravitational floor''.} %\csf{from Planck?}. }. \csf{Is it standard to italicize the caption?}
    \label{fig:MatchingTheta}
\end{figure}

There are five free parameters in our analysis, the Hubble rate during inflation $H_{\rm inf}$, the $B-L$ breaking scale $v_{B-L}$, the majoron mass $m_{\chi 0}$, the decay width $\Gamma_\chi$, and the intial field value $\chi_{\rm in}$, or its dimensionless version $\theta_{\rm in} \equiv \chi_{\rm in}/v_{B-L}$. As long as $\Gamma_\chi$ is sufficiently small, so that the majoron's lifetime is sufficiently large and it can dominate the energy density of the Universe, the results do not depend on $\Gamma_\chi$.  We assume that the majoron as a curvaton is responsible for generating the entirety of the spectral index $n_s$ and the scalar power spectrum $P_\zeta$ observed at CMB, namely $n_s = 0.9649$ and $P_\zeta^{\rm CMB} = 2.4 \times 10^{-9}$ \cite{Planck:2018jri}.\footnote{This is a strong requirement, as it is quite possible phenomenologically to get only a fraction of these quantities from the curvaton dynamics, while the rest is contributed by the inflaton and/or spectator
fields \cite{Kinney:2012ik, Fonseca:2012cj, Enqvist:2013paa, Ellis:2013iea, Lodman:2023yrc}.} 
For $n_s$, this implies that $\cos{\theta_{\rm in}} < 0$, hence $\pi/2 \leq \theta_{\rm in} \leq \pi$.  
In fig.~\ref{fig:MatchingTheta}, we show how $n_s$, $P_\zeta$ and $f_{\rm NL}$ varies with $\theta_{\rm in}$ for given $H_{\rm inf}$ and $v_{B-L}$. We see that $n_s$ and $P_\zeta$ generated entirely from the majoron acting as a curvaton matches the Planck 2018 observed values when $\theta_{\rm in} \simeq 2.2$ and $m_{\chi 0} \simeq 0.29 H_{\rm inf}$. This results in a prediction for the non-Gaussianity parameter $f_{\rm NL} \simeq 0.56$, which is consistent with the current estimate on local-type non-Gaussianity from Planck 2018 is $f_{\rm NL}^{\rm local} = -0.9 \pm 5.1$ \cite{Planck:2019kim}, and can be probed at future $21$ cm tomography sensitive up to $f_{\rm NL}^{\rm local} \simeq 0.03$ \cite{Munoz:2015eqa}, nearing the ``gravitational floor'', $f_{\rm NL}^{\rm local} \simeq 0.01$ \cite{Acquaviva:2002ud, Maldacena:2002vr, Cabass:2016cgp}, which is the non-Gaussianity generated by purely gravitational interactions among inflaton fluctuations in single-field inflation.

\section{The dominance of nonthermal leptogenesis}\label{sec:4}
RHNs can be produced during two stages of the cosmological evolution in our setup. One from the thermal bath during the reheating period, and the other from the decay of the majoron at the end of its dyanamics.

After inflation, the reheating temperature \eqref{eq:Tmax} is in general higher than the mass of $N_i$ (in particularly $N_1$). 
Unavoidably, $N_i$ will be produced through the Yukawa interactions which depend on $\lambda_{\alpha i}$ in eq.~\eqref{eq:Majoron_model}.
As a result, thermal leptogenesis will take place and taking into account the dilution from entropy injection in eq.~\eqref{eq:dilution}, the $B-L$ asymmetry generated from $N_1$ can written as
\begin{eqnarray}
	Y_{B-L}^{\textrm{thermal}} & = & \frac{1}{d} \epsilon_{1}\eta_{1}Y_{N_{1}}^{\textrm{eq}}\left(T\gg M_{1}\right),
	\label{eq:thermal_B-L}
\end{eqnarray}
where $\epsilon_1$ parametrizes the CP violation from $N_1$ decay,  $Y_{N_{1}}^{\textrm{eq}}\left(T\gg M_{1}\right)=\frac{45}{2\pi^{4} g_\star}$ and $\eta_{1}\leq1$ is the efficiency factor taking into account the washout processes. 

On the other hand, nonthermal leptogenesis will occur much later at $T \ll M_1$ from $N_1$ particles which are produced from the decays of $\chi$. The temperature of the bath after majoron decay can be determined by combining eq.~\eqref{eq:decaywidth_to_N} with eq.~\eqref{Tdecay}, 
\begin{align}
    \left(\frac{\tilde{T}}{M_1}\right)^4 = \frac{45}{128 \pi^4 g_\star} \frac{m_\chi^2 M_{\rm Pl}^2}{v_{B-L}^4} \left( 1 - \frac{4M_1^2}{m_{\chi}^2}\right). \label{Ttilde1}
\end{align}
We have verified that for the parameter space interesting for the curvaton dynamics yields $\tilde{T} < M_1$, for which only nonthermal production of the RHN is relevant.

Another constraint on $\tilde{T}$ is that it should be greater than the temperature where the Big Bang Nucleosynthesis (BBN) commences $\tilde T  \gtrsim T_\textrm{BBN} \sim $ MeV. However, since
we are interested in leptogenesis from $N_1$ decays, we require $\tilde T > 132$ GeV~\cite{DOnofrio:2014rug}
when the electroweak sphalerons are still active, so that the generated $B-L$ asymmetry can be converted into baryon asymmetry.

In general, since the decay width of $N_1$ is much larger than that of $\chi$, $\Gamma_{N_1} \gg \Gamma_{\chi}$, $N_1$ particles decay almost instantaneously after being produced. With $T \ll M_1$, the washout process is completely suppressed and the $B-L$ asymmetry produced by these new population of $N_1$ is 
\begin{eqnarray}
	Y_{B-L}^{\textrm{nonthermal}} & = & \frac{2}{d}\epsilon_{1}Y_{\chi},
	\label{eq:nonthermal_B-L}
\end{eqnarray}
where the factor of 2 arises since each $\chi$ decays to two $N_{1}$.
From eq.~\eqref{eq:Ychi}, we verify that nonthermal contribution above will dominate over thermal contribution \eqref{eq:thermal_B-L} as long as $\eta_1 \lesssim 0.08$. This is generally realized due to either strong washout (large $\lambda_{\alpha 1}$) or inefficiency in $N_1$ production (small $\lambda_{\alpha 1}$) when the efficiency is suppressed $\eta_1 \ll 0.1$. 
To highlight the role played by $\chi$ in realizing nonthermal leptogenesis, we will assume the thermal contribution is subdominant in the rest of the work.

Finally, at $T \sim 132$ GeV when the electroweak sphaleron freezes out after the electroweak phase transition at $160$ GeV~\cite{DOnofrio:2014rug}, the baryon asymmetry is given by
\begin{equation}
    Y_B = 0.315\ Y_{B-L}^\textrm{nonthermal}, 
\end{equation}
where we have taken into account the finite top mass.
This value is to be matched with the observed value $Y_{B}^{\textrm{obs}}\simeq 8.7\times10^{-11}$~\cite{Planck:2018vyg}.

Assuming hierarchical masses of $N_{i}$, the CP asymmetry parameter
from the decays of $N_{1}$ is bounded from above by the Davidson-Ibarra
bound~\cite{Davidson:2002qv}
\begin{eqnarray}
\left|\epsilon_{1}\right| & \leq & \frac{3M_{1}}{16\pi v^{2}}\frac{\left|\Delta m_{\textrm{atm}}^{2}\right|}{m_{h}+m_{l}},
\label{eq:DI_bound}
\end{eqnarray}
where the atmospheric mass squared splitting is $\left|\Delta m_{\textrm{atm}}^{2}\right|\simeq2.5\times10^{-3}\,\textrm{eV}^{2}$ 
and $m_{h}+m_{l}$
is the sum of the masses of the lightest and the heaviest light neutrinos.
We will consider the most optimistic scenario by saturating the bound in eq.~\eqref{eq:DI_bound} by
setting $m_{h}+m_{l}=0.05$ eV which gives
\begin{align}
    |\epsilon_1| \leq 4.9 \times 10^{-6}\ \left( \frac{M_1}{10^{10}\ \text{GeV}} \right). \label{DIbound2}
\end{align}

%------------
\section{Imprints of seesaw  and leptogenesis on non-Gaussianity}
\label{sec:5}
The massive majoron acting as a curvaton to give rise to the primordial curvature power spectrum  also generates observable non-Gaussianity. On the other hand, at the end of the curvaton dynamics, the decay of the majoron yields a nonthermal population of RHNs, which participate in leptogenesis to generate the observed baryon asymmetry of the Universe. Hence, this setup provides a way to explain three observables, the magnitude and spectral index of the scalar power spectrum, and the baryon asymmetry observed at CMB. A testable prediction of this mechanism is the non-Gaussianity parameter $f_{\rm NL}$ that can be probed in upcoming CMB experiments and 21 cm tomography. 
{Our main result is to show that in the viable parameter space to explain the aforementioned three observables, non-Gaussianity at an observable level is produced.}

As mentioned earlier, there are five undetermined parameters in our setup, namely, the inflation scale $H_{\rm inf}$, $B-L$ breaking scale $v_{B-L}$, the majoron mass $m_{\chi}$, initial field value $\chi_{\rm in}$ (or equivalently, its dimensionless counterpart $\theta_{\rm in}$) during inflation when the CMB modes leave the horizon, and the RHN mass $M_1$. We are specifically interested in the scenario where both the observed value of the spectral index $n_s$ and the power spectrum $P_\zeta$ at the CMB scale are entirely generated by the majoron. Our strategy for the parameter scan is as follows. For a given $v_{B-L}$, $H_{\rm inf}$ and $M_1$, we determine the combination $(m_\chi, \theta_{\rm in})$ for which $n_s \simeq 0.9649$ and $P_\zeta \simeq 2.4 \times 10^{-9}$. We then scan over $M_1$ from 1 TeV to a large scale bounded by $M_1 < m_\chi/2 < H_{\rm inf}/2$, and show how $f_{\rm NL}$ and $Y_{B}$ varies.

\begin{figure}
    \centering        \includegraphics[width=0.99\textwidth]{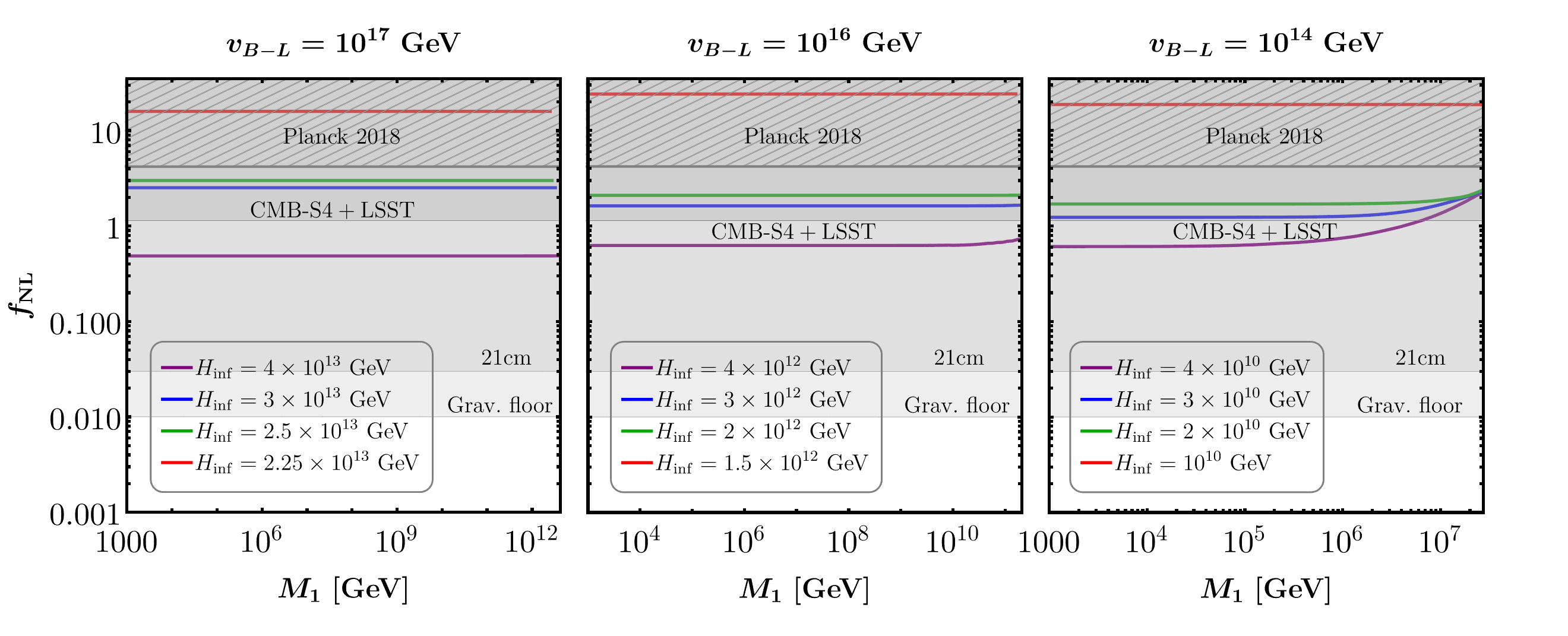} 
    \caption{$f_{\rm NL}$ as a function of the lightest RHN mass $M_1$ for three choices of the $B-L$ breaking scale $v_{B-L} = 10^{17}$ GeV (left), $10^{16}$ GeV (middle) and $10^{14}$ GeV (right). In each case we show results for four choices of $H_{\rm inf}$, shown in the inset, with an appropriate choice of $m_{\chi 0}$ that is required to generate both the scalar power spectrum, $P_\zeta$, and the spectral index, $n_s$, observed at CMB scales, entirely from the curvaton dynamics. For comparison we show the upper bound on local non-Gaussianity from Planck 2018 \cite{Planck:2019kim} (hatched region), sensitivities of future experiments CMB-S4+LSST  and 21 cm tomography, and the ``gravitational floor''.%\ag{gravitational floor ?}
    }
    \label{fig:fNLvsM}
\end{figure}

In fig.~\ref{fig:fNLvsM} we show $f_{\rm NL}$ as a function of $M_1$ for the three cases $v_{B-L} = 10^{17}$ GeV (left), $10^{16}$ GeV (middle), and $10^{14}$ GeV (right). In each case, we show four examples of different $H_{\rm inf}$.\footnote{The corresponding $m_{\chi 0}$ and $\theta_{\rm in}$ are shown in appendix \ref{app:C}.} For comparison, we show the region excluded by Planck 2018 data \cite{Planck:2019kim} for non-observation of non-Gaussianity (hatched region), and sensitivity of future experiments CMB-S4+LSST \cite{Munchmeyer:2018eey} and 21 cm tomography \cite{Munoz:2015eqa}, and the ``gravitational floor'' \cite{Acquaviva:2002ud, Maldacena:2002vr, Cabass:2016cgp} representing the $f_{\rm NL}$ generated by purely gravitational interactions of the inflaton (gray regions). 

For each case of $v_{B-L}$, Planck 2018 data puts an upper bound on $H_{\rm inf}$.\footnote{Since the curvaton dynamics does not affect the tensor power spectrum, it is given by the usual relation $P_T = {2H_{\rm inf}^2}/{(\pi^2 M_{\rm Pl}^2)}$. Using the observed value of the scalar power spectrum at CMB scales, the tensor-to-scalar ratio, $r \equiv P_T/P_\zeta$, which is bounded from Planck 2018 \cite{Planck:2018vyg} and BICEP2/Keck Array BK18 \cite{BICEP:2021xfz} at $r < 0.032$ \cite{Tristram:2021tvh}, yields an upper bound on the scale of inflation, $H_{\rm inf} < 4.67 \times 10^{13}$ GeV. All benchmark points we have considered in this work are within this bound.} On the other hand, smaller $H_{\rm inf}$ allows for observable non-Gaussianity of $f_{\rm NL} \sim \mathcal{O}(0.1 \sim 1)$ to be probed by CMB-S4+LSST and 21 cm tomography. %We note that for a given $v_{B-L}$, $f_{\rm NL}$ is very sensitive to $H_{\rm inf}$. 

Keeping everything else fixed, $f_{\rm NL}$ does not depend on $M_1$ unless it is sufficiently large. This is consistent with our expectation that smaller $M_1$ results in a smaller decay width, therefore, longer lifetime of the majoron, facilitating that it can dominate the energy density of the Universe before it decays. In this case the ratio of the energy density of the majoron to that of the radiation bath, $R$, is so large that one can effectively take the limit $R \rightarrow \infty$. On the other hand, if $M_1$ is sufficiently large, energy density of the majoron dominates for a short period. In this case, the ratio of energy densities, $R$ is not very large and one gets a non-negligible contribution to $f_{\rm NL}$ that increases with larger $M_1$. This effect is more prominent when $v_{B-L} \ll M_{\rm Pl}$.

\begin{figure}
    \centering        \includegraphics[width=0.99\textwidth]{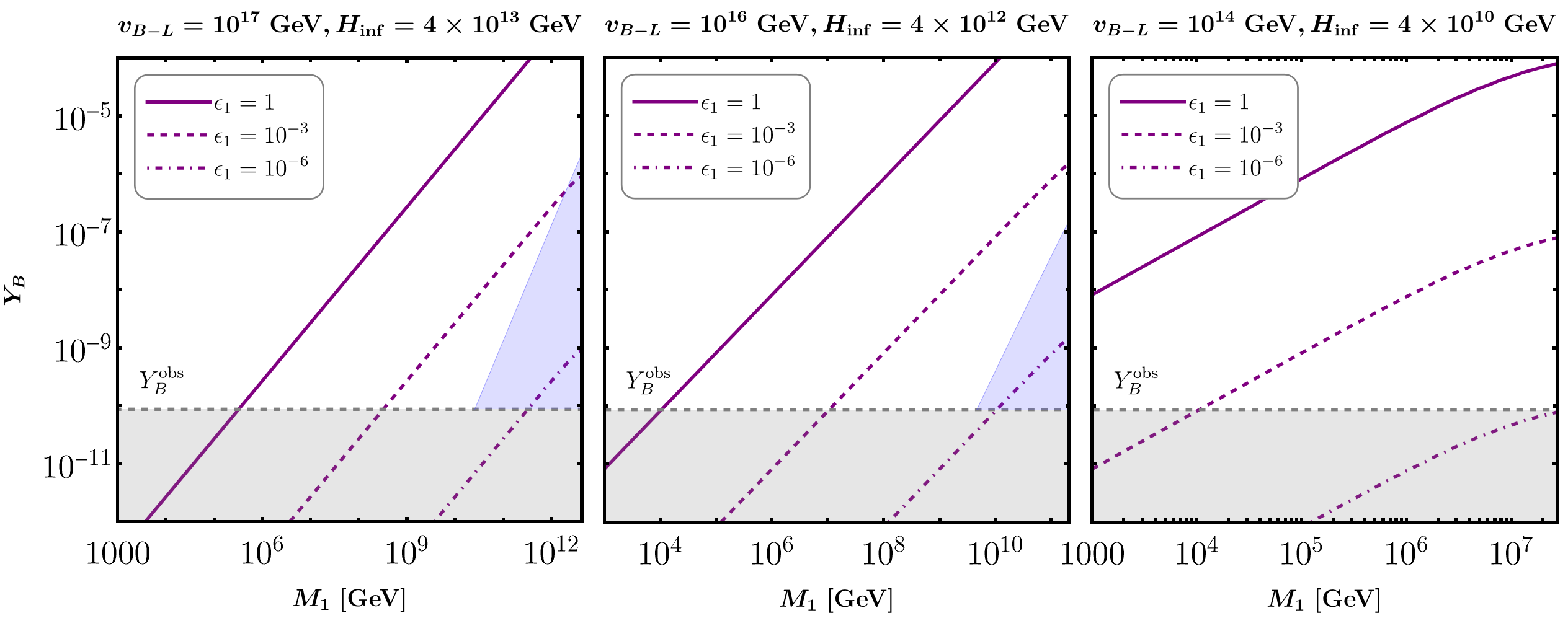} %\includegraphics[width=0.48\textwidth]%{YvsMzoom.png}
    \caption{Baryon asymmetry from nonthermal leptogenesis. Purple lines show the generated asymmetry for different CP asymmetry parameters. The gray shaded region shows the parameter space where generated asymmetry is smaller than the observed asymmetry. Blue shaded region corresponds to successful leptogenesis where the RHN mass is above the Davidson-Ibarra bound. %\st{This region shrinks with decreasing $v_{B-L}$ and vanishes for $v_{B-L} = 10^{14}$ GeV.} \saad{is this line necessary?}
    }
    \label{fig:YBvsM}
\end{figure}
{Next, let us discuss the baryon asymmetry generated in this model.}
In fig.~\ref{fig:YBvsM}, we show 
the baryon asymmetry $Y_B$ as a function of the RHN mass $M_1$ for the same set of benchmark  values of $v_{B-L}$. 
Baryon asymmetry depends weakly on $H_{\rm inf}$, only through the field value at the end of inflation, $\chi_{\rm end}$ in eq.~\eqref{eq:Ychi}. The resulting $Y_B$ for different $H_{\rm inf}$ considered in fig.~\eqref{fig:fNLvsM} for each $v_{B-L}$ are very close. Hence, in fig.~\ref{fig:YBvsM}, we only show results for one representative $H_{\rm inf}$, but for three choices of the CP asymmetry parameter $\epsilon_1$. The gray region represents baryon asymmetry below the observed value $Y_B^{\rm obs}$. Here the case $\epsilon_1 = 1$ represents the theoretical maximum of the baryon asymmetry that can be produced from nonthermal leptogenesis in this model. This sets a lower bound on $M_1$ in fig.~\ref{fig:YBvsM}. For example, for the benchmark scenario in the left panel, $M_1 \lesssim 3 \times 10^{5}$ GeV could never produce the observed $Y_B$. Similarly, for a given $\epsilon_1$, one can determine a lower bound on $M_1$ for successful leptogenesis. 
The shaded blue region represents the Davidson-Ibarra bound assuming a hierarchical mass spectrum of $N_i$ and negligible lepton flavor effects, given by eq.~\eqref{DIbound2}, which is satisfied only for smaller $\epsilon_1$ and larger $M_1$. 
{Beyond this assumption, we notice that there is a large parameter space for successful leptogenesis, which by itself is a nontrivial result.}

Figs.~\ref{fig:fNLvsM} and \ref{fig:YBvsM} show that the same parameter space that generates the entirety of the curvature power spectrum and the spectral index observed at CMB scales yields testable non-Gaussianity, and results in successful leptogenesis. Furthermore, these results depend crucially on the seesaw scale $v_{B-L}$. Hence, the massive majoron model provides us an interesting way to  {investigate} the high scale of seesaw and leptogenesis through primordial non-Gaussianities observable at future CMB, LSS experiments and 21 cm tomography.
{While nonobservation of local non-Gaussianities will rule out our {proposed} scenario, a positive signature will only be a suggestive evidence  since there are other mechanisms where the non-Gaussianity can be produced.}

\section{Discussion and Conclusion}\label{sec:6}
We have presented an interesting way to {investigate} the high scale of seesaw and nonthermal leptogenesis through their imprints on the primordial non-Gaussianity. In our setup, the right-handed neutrino Majorana masses are generated from the spontaneous breaking of a global $U(1)_{B-L}$ symmetry at a high scale $v_{B-L}$ close to the GUT scale, that yields the small masses of the Standard Model neutrinos through the type I seesaw mechanism. \textcolor{black}{The corresponding massless Nambu-Goldstone boson, called majoron, gains a periodic potential and becomes massive due to nonperturbative dynamics from its coupling to a confining new sector through anomaly.} We have shown that the massive majoron can perform the role of a curvaton, a heavy particle \textcolor{black}{whose field fluctuations are subdominant during inflation
but are amplified to yield the observed {red-tilted} curvature power spectrum in the post-inflationary era when majoron dominates the energy budget of the Universe.} Interestingly, at the end of the curvaton dynamics, majoron decays into right-handed neutrinos, facilitating a production of baryon asymmetry from nonthermal leptogenesis. We therefore bring three scenarios under one umbrella, generation of neutrino masses through the seesaw mechanism, creation of the observed curvature power spectrum at CMB, and the production of baryon asymmetry in the same model. 

A consequence of the majoron acting as a curvaton is the generation of non-Gaussianity, which nontrivially depends on the scale of seesaw (related to $B-L$ breaking scale) and leptogenesis (related to the lightest RHN mass), thus promoting the observation of non-Gaussianity as a unique probe of the scale of seesaw and leptogenesis. We have shown that the majoron acting as a curvaton allows for an ample parameter space for seesaw and leptogenesis to happen, with $v_{B-L} \sim \mathcal{O}(10^{14} - 10^{17})$ GeV and $M_1 \gtrsim \mathcal{O}(10^{3-6})$ GeV depending on $v_{B-L}$, see figs.~\ref{fig:fNLvsM} and \ref{fig:YBvsM}.  
{While a negative detection of the predicted non-Gaussianity $f_{\rm NL} \gtrsim \mathcal{O}(0.1)$, at future CMB and LSS experiments and 21 cm tomography will falsify our scenario, a positive signature will give a strong support of the mechanism which will then deserve further investigation.}

\begin{acknowledgments}
MHR acknowledges financial support from the STFC Consolidated Grant ST/T000775/1,
and from the European Union's Horizon 2020 Research and Innovation
Programme under Marie Sk\l odowska-Curie grant agreement HIDDeN European ITN project (H2020-MSCA-ITN-2019//860881-HIDDeN).
CSF acknowledges the support by Fundação de Amparo à Pesquisa do Estado de São Paulo (FAPESP) Contracts No. 2019/11197-6 and 2022/00404-3, Conselho Nacional de Desenvolvimento Científico e Tecnológico (CNPq) under Contract No. 407149/2021-0, and the support from the ICTP through the Associates Programme (2023-2028) while this work was being completed. 
CSF and AG are grateful to the Mainz Institute for Theoretical Physics (MITP) of the Cluster of Excellence PRISMA$^+$ (Project ID 39083149), for its hospitality and support during ``New Proposals for Baryogenesis'' workshop where part of the work was carried out. AN acknowledges financial support from the ISRO Respond Grant.
We are grateful to Takeshi Kobayashi for discussions.
\end{acknowledgments}

\appendix
\section{Numerical solution of the curvaton EOM} \label{app:A}

In this appendix we outline our strategy for numerically solving the equations of motion of the majoron in eqs.~\eqref{EOM1} and \eqref{EOM2}. 
 
Because of the temperature-suppressed potential, $\theta$ evolves slowly in the period $x\leq x_p$, and the coupled equations \eqref{EOM1} and \eqref{EOM2} can be numerically solved. However, for $x\gg x_p$, $\theta(x)$ becomes very oscillatory and it becomes numerically challenging to solve the system of equations exactly. 

The first and second terms in the r.h.s. of eq.~\eqref{EOM2} are proportional to radiation and curvaton energy densities, respectively. During oscillation, the curvaton energy density can be treated as matter and its average energy density falls off as $a^{-3}$, if the potential is quadratic in $\theta$. While we consider a cosine potential here, close to the minima of the potential, it can be treated as a quadratic potential. As the curvaton oscillates around the minima, the amplitude of oscillation decreases and frequency increases. We approximate that when the percent difference between the quadratic potential and the cosine potential is below $0.01\%$, we can evolve the curvaton energy density in eq.~\eqref{EOM2} with a $e^{-3N(x)}$ factor. Suppose the onset of this occurs at $x_{\rm quad}$, eq.~\eqref{EOM2} can be approximated as
\begin{align}
    N'(x) = \left[ \tilde{\rho}_{r, \rm ini}\ e^{-4 N(x)} + \frac{v_{B-L}^2}{3M_{\rm Pl}^2} \left\{ \frac{1}{2} \theta'(x)^2|_{x_{\rm quad}} +  \left(1- \cos{\theta(x_{\rm quad})}\right) \right\} e^{3(N(x_{\rm quad}) - N(x))}\right]^{1/2}.
    \label{EOM2a} 
\end{align}
Here we have dropped the temperature dependent factor as typically $x_{\rm quad} > x_{p}$. The exponential factor in the second term ensures that $N'(x)$ remains same for $x = x_{\rm quad}$ in both eqs.~\eqref{EOM2} and \eqref{EOM2a}. 

\begin{figure}
    \centering        \includegraphics[width=0.6\textwidth]{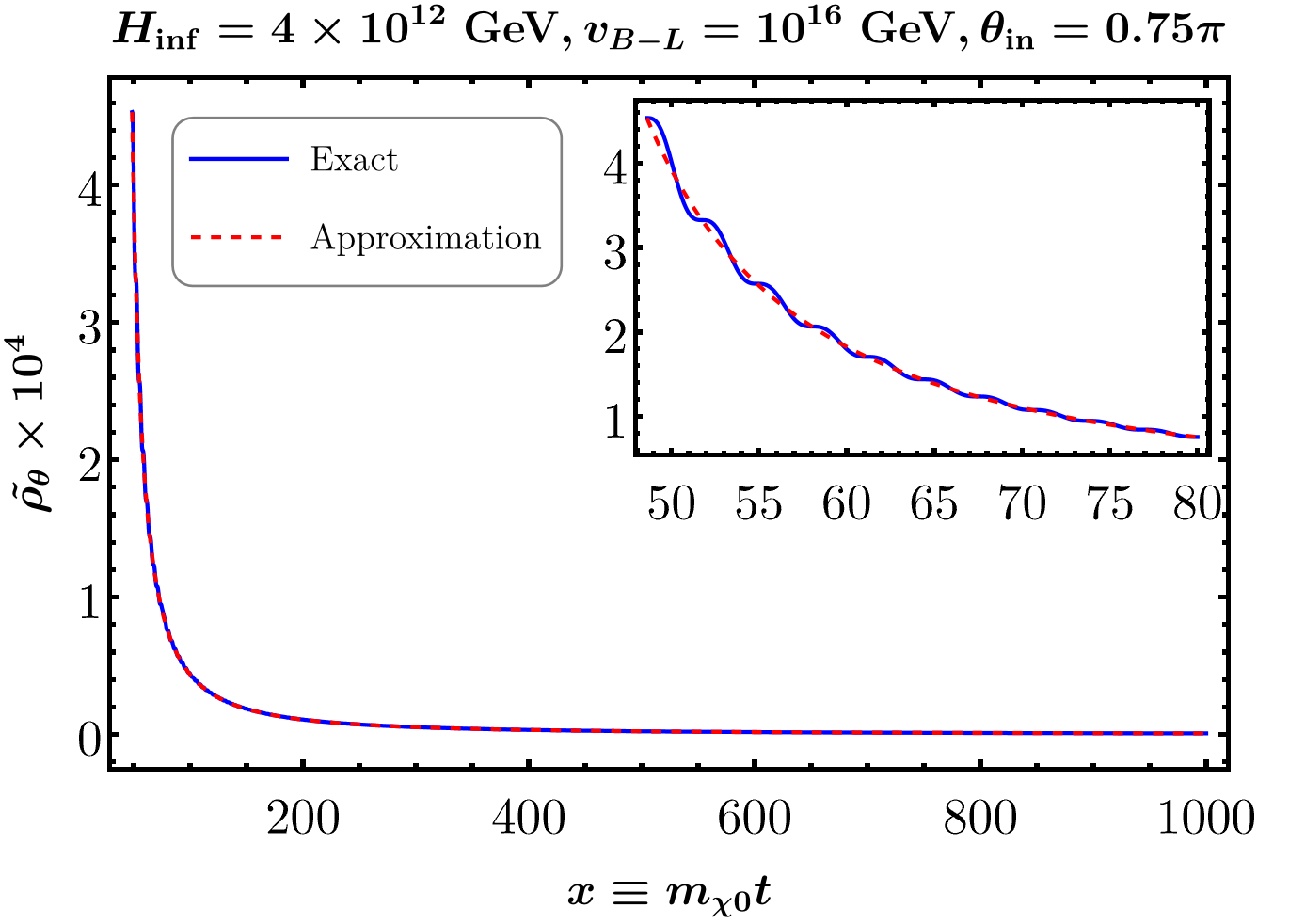}
    \caption{Comparison of the exact numerical solution vs. the quadratic approximation of the equation of motion.}
    \label{fig:compare}
\end{figure}

This approximation can be verified by evolving the system of equations exactly up to the point $x_{\rm max}$, beyond which numerical solution becomes difficult because of the oscillatory nature of the solution $\theta(x)$, and then comparing the result to the approximate solution obtained by using the strategy above for $x_{\rm quad} < x < x_{\rm max}$. In fig.~\ref{fig:compare}, we show the normalized curvaton energy density
\begin{align}
    \tilde{\rho}_\theta(x) = \frac{1}{2} \theta'(x)^2 + 1-\cos{\theta(x)}
\end{align}
for both cases.
The energy density of curvaton obtained from exact numerical solution shown tiny oscillations with a decreasing average value, which is captured remarkably well by the approximate result.

\section{Constraints on the parameter space for simultaneously explaining $n_s$ and $P_\zeta$} \label{app:B}

\begin{figure}
    \centering        \includegraphics[width=0.49\textwidth]{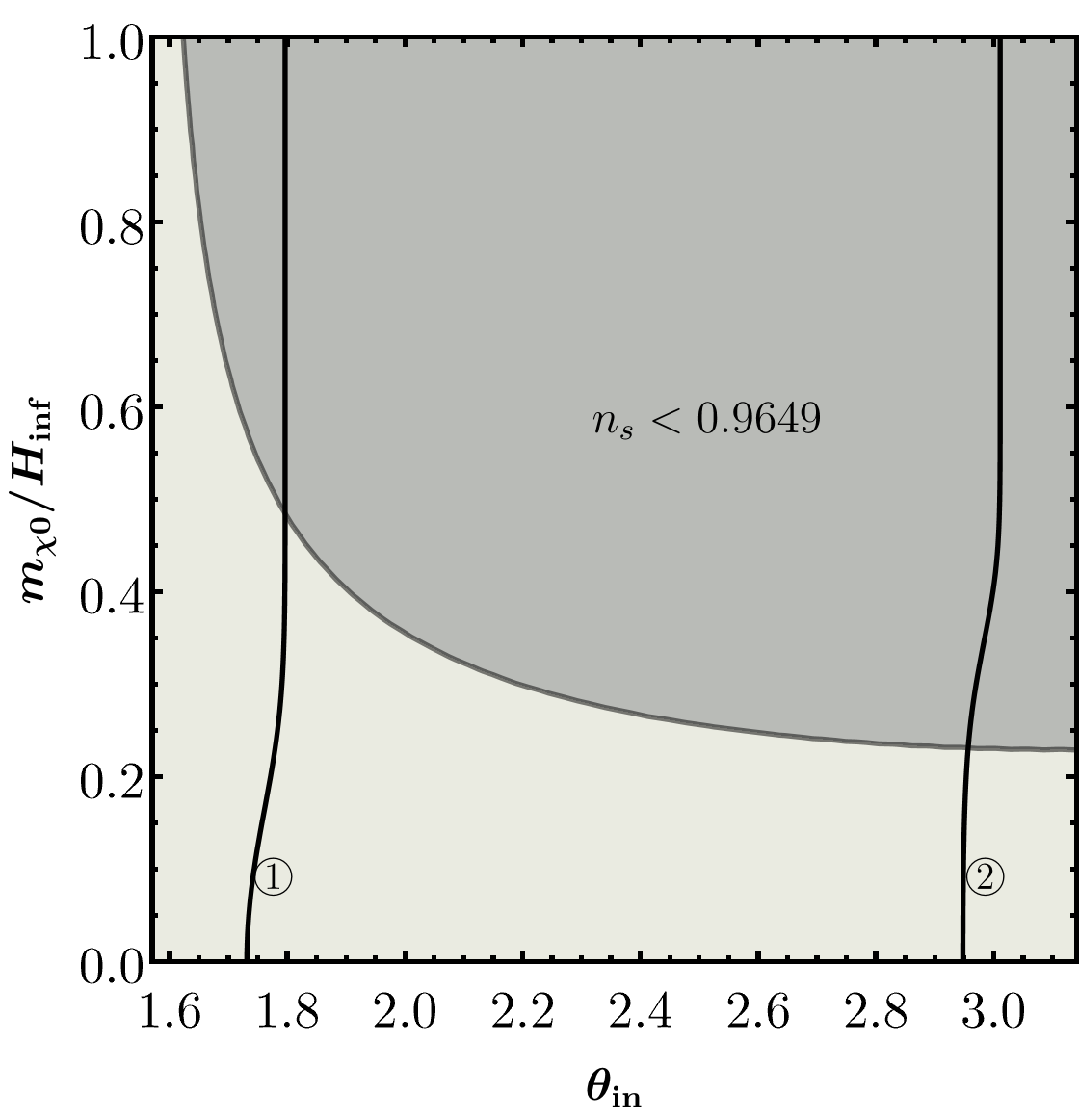}
    \caption{Constraints from matching spectral index $n_s = 0.9649$ and power spectrum $P_\zeta = 2.4 \times 10^{-9}$ \cite{Planck:2018jri} entirely from the majoron. Since the inflaton's contribution to $n_s$ is always negative, the darker shaded region corresponding to $n_s < 0.9649$ contributed by the curvaton is ruled out. While we do not consider it here, it is possible to supplement the majoron contribution from the inflaton in the light shaded region to reproduce the observed $n_s$. The almost-vertical lines are contours of $P_\zeta = 2.4\times 10^{-9}$ corresponding to the benchmark points \Circled{1} $H_{\rm inf}=4.5 \times 10^{12}$ GeV, $v_{B-L} = 10^{16}$ GeV, and \Circled{2} $H_{\rm inf}=2\times 10^{12}$ GeV.}    \label{fig:nsPconstraints}
\end{figure}

Requiring that the entirety of the observed $n_s$ and $P_\zeta$ are produced by majoron oscillations constrains the parameter space of $H_{\rm inf}, v_{B-L}, m_{\chi 0}$ and $\theta_{\rm in}$. In fig.~\ref{fig:nsPconstraints} we show the contours of Planck 2018 best fit values, $n_s = 0.9649$ (solid line) and $P_\zeta = 2.4 \times 10^{-9}$ (dashed line) \cite{Planck:2018jri}, in the plane of $m_{\chi 0}/H_{\rm inf}$ and $\theta_{\rm in}$. 

We restrict $m_{\chi 0} < H_{\rm inf}$, which is necessary to ensure that the majoron does not dominate the energy density during inflation, {such that it does not affect the inflation itself.} In that case, eq.~\eqref{nseq} implies that $\pi/2 \leq \theta_{\rm in} \leq \pi$ if the dominant contribution to $n_s$ is to come from the majoron. Furthermore, in eq.~\eqref{nseq}, since the inflaton contribution from the second term to $n_s$ is always negative as $\dot{H}_{\rm inf} < 0$, the majoron contribution cannot be below the observed value $0.9649$. This rules out the dark shaded region in fig.~\ref{fig:nsPconstraints}. On the other hand, in the light shaded region, the majoron contribution exceeds the observed value, which can be tamed with an appropriate subleading contribution from the inflaton dynamics. In this paper we do not consider this possibility, and expect $n_s$ to come from completely from the majoron contribution. In that case, we find that $m_{\chi 0} \gtrsim 0.23 H_{\rm inf}$ if $n_s$ is to be entirely generated from the majoron.  

The dashed line \Circled{1} corresponds to the contour of $P_\zeta = 2.4 \times 10^{-9}$ generated by the majoron for $H_{\rm inf} = 4.5\times 10^{12}$ GeV and $v_{B-L} =  10^{16}$ GeV. For this benchmark point, we see that $\theta_{\rm in} \simeq 2.05$ and $m_\chi = 0.35 H_{\rm inf}$ yields the observed $n_s$ and $P_\zeta$ entirely from the majoron. 

Reducing $H_{\rm inf}$ while keeping $v_{B-L}$ same shifts the $P_\zeta$ contour to the right, as seen for benchmark point \Circled{2}, where $H_{\rm inf} = 2\times 10^{12}$ GeV. This results in a $\theta_{\rm in}$ close to $\pi$ and $m_\chi/H_{\rm inf}$ close to the lower bound $0.23$, implying that there is a lower bound on $H_{\rm inf}$ for a given $v_{B-L}$. 
Furthermore, we find that  $P_\zeta$ dominantly depends on the ratio $H_{\rm inf}/v_{B-L}$, with a subleading dependence on $m_\chi$, as suggested by eq.~\eqref{Pfnl}.

\section{Majoron mass and initial field value from parameter scan} \label{app:C}
\begin{figure}[!ht]
    \centering        \includegraphics[width=0.99\textwidth]{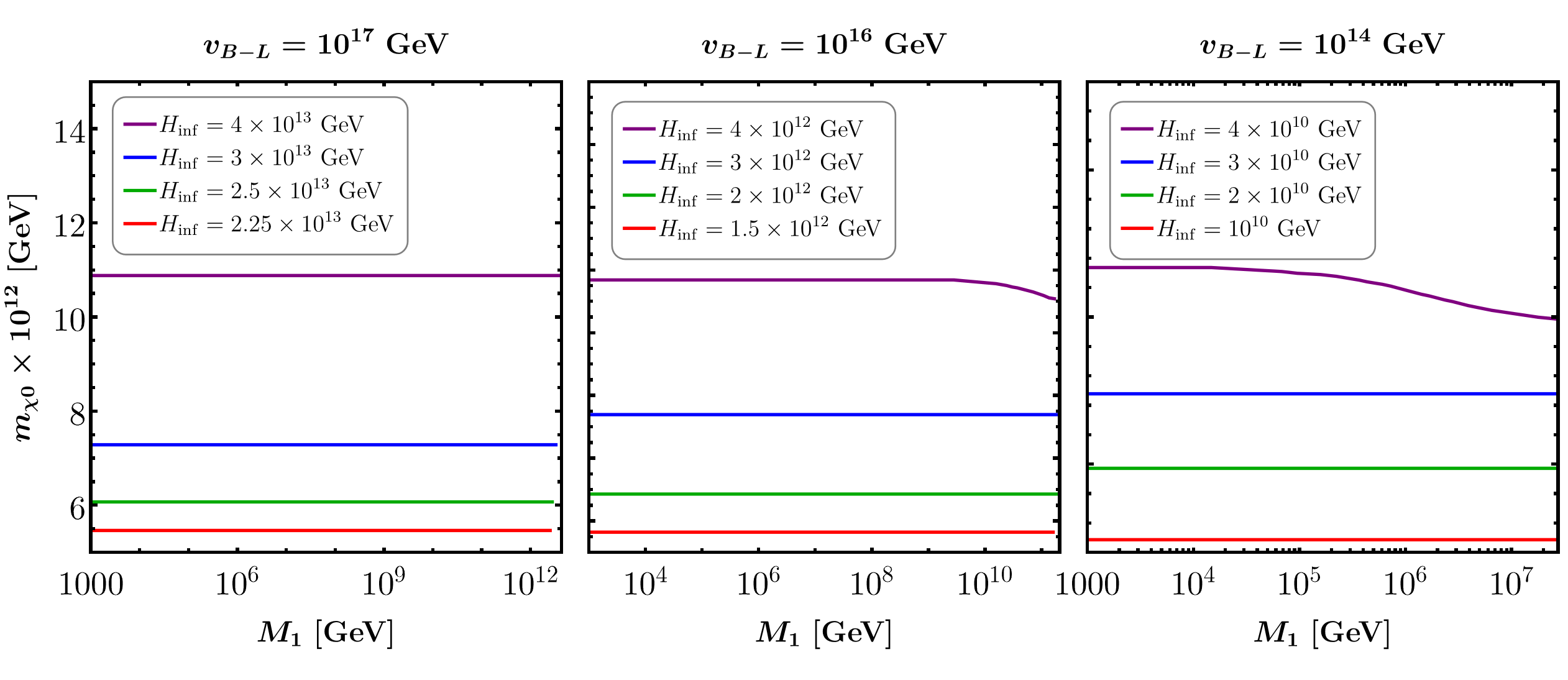} 
    \caption{$m_{\chi 0}$ as a function of the lightest RHN mass $M_1$ for three choices of the $B-L$ breaking scale $v_{B-L} = 10^{17}$ GeV (left), $10^{16}$ GeV (middle) and $10^{14}$ GeV (right). In each case we show results for four choices of $H_{\rm inf}$, shown in the inset.  
    }
    \label{fig:mchivsM}
\end{figure}

\begin{figure}[!ht]
    \centering        \includegraphics[width=0.99\textwidth]{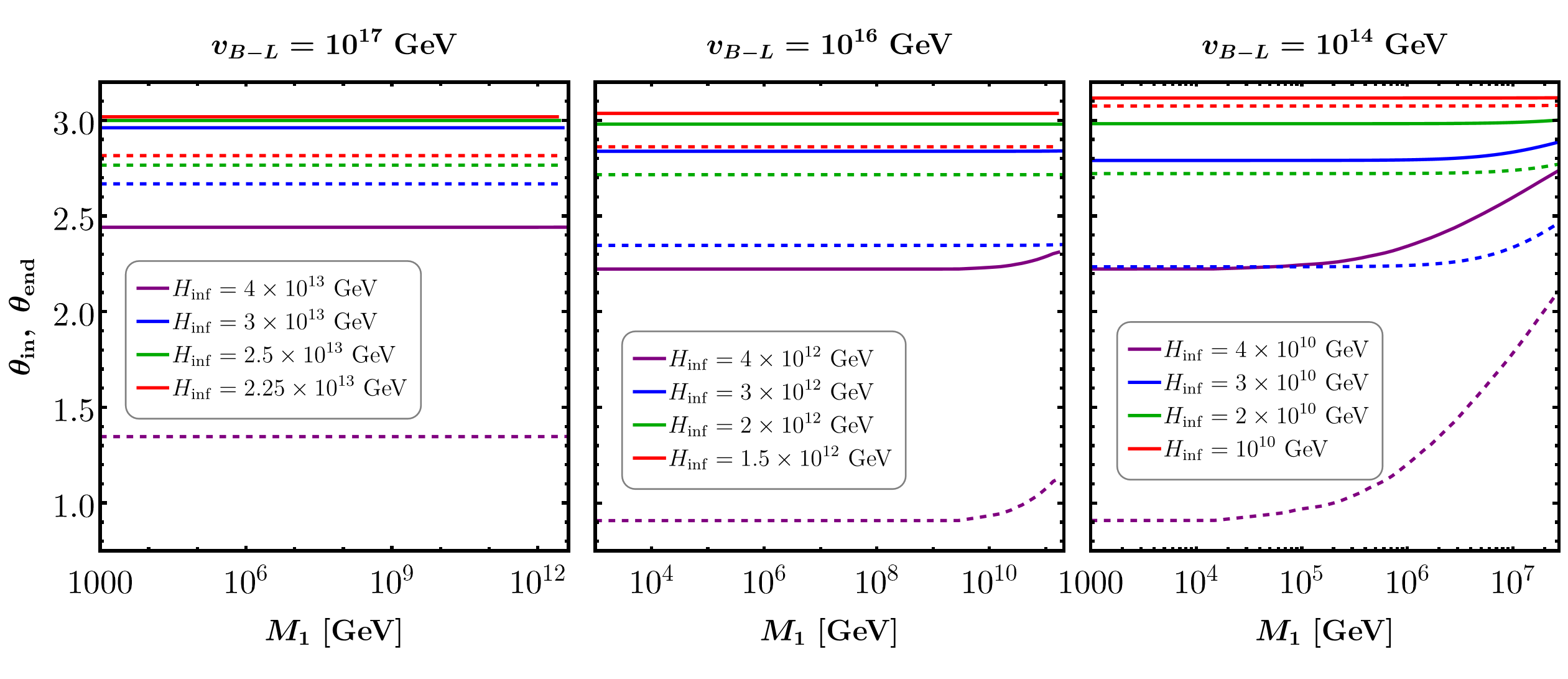} 
    \caption{Initial field value $\theta_{\rm in}$ $50$ e-folds before the end of inflation (solid lines), and field value at the end of inflation $\theta_{\rm end}$ (dashed lines) as a function of the lightest RHN mass $M_1$ for three choices of the $B-L$ breaking scale $v_{B-L} = 10^{17}$ GeV (left), $10^{16}$ GeV (middle) and $10^{14}$ GeV (right). }
    \label{fig:thetavsM}
\end{figure}

In figs.~\ref{fig:mchivsM} and \ref{fig:thetavsM}, respectively, we show the majoron mass, $m_{\chi 0}$, and the dimensionless field values $\theta_{\rm in}$ (50 e-folds before the end of inflation) and  $\theta_{\rm end}$ (at the end of inflation), corresponding to the parameter scans shown in fig.~
\ref{fig:fNLvsM}. Each point in the parameter space, specified by $(H_{\rm inf}, v_{B-L}), m_{\chi 0}, \theta_{\rm in}, M_1$, reproduces the observed power spectrum and spectral index at the CMB scales entirely from the majoron dynamics, and predicts observable non-Gaussianity. 

\newpage
\bibliography{reference}

\begin{thebibliography}{106}%
\makeatletter
\providecommand \@ifxundefined [1]{%
 \@ifx{#1\undefined}
}%
\providecommand \@ifnum [1]{%
 \ifnum #1\expandafter \@firstoftwo
 \else \expandafter \@secondoftwo
 \fi
}%
\providecommand \@ifx [1]{%
 \ifx #1\expandafter \@firstoftwo
 \else \expandafter \@secondoftwo
 \fi
}%
\providecommand \natexlab [1]{#1}%
\providecommand \enquote  [1]{``#1''}%
\providecommand \bibnamefont  [1]{#1}%
\providecommand \bibfnamefont [1]{#1}%
\providecommand \citenamefont [1]{#1}%
\providecommand \href@noop [0]{\@secondoftwo}%
\providecommand \href [0]{\begingroup \@sanitize@url \@href}%
\providecommand \@href[1]{\@@startlink{#1}\@@href}%
\providecommand \@@href[1]{\endgroup#1\@@endlink}%
\providecommand \@sanitize@url [0]{\catcode `\\12\catcode `\$12\catcode
  `\&12\catcode `\#12\catcode `\^12\catcode `\_12\catcode `\%12\relax}%
\providecommand \@@startlink[1]{}%
\providecommand \@@endlink[0]{}%
\providecommand \url  [0]{\begingroup\@sanitize@url \@url }%
\providecommand \@url [1]{\endgroup\@href {#1}{\urlprefix }}%
\providecommand \urlprefix  [0]{URL }%
\providecommand \Eprint [0]{\href }%
\providecommand \doibase [0]{http://dx.doi.org/}%
\providecommand \selectlanguage [0]{\@gobble}%
\providecommand \bibinfo  [0]{\@secondoftwo}%
\providecommand \bibfield  [0]{\@secondoftwo}%
\providecommand \translation [1]{[#1]}%
\providecommand \BibitemOpen [0]{}%
\providecommand \bibitemStop [0]{}%
\providecommand \bibitemNoStop [0]{.\EOS\space}%
\providecommand \EOS [0]{\spacefactor3000\relax}%
\providecommand \BibitemShut  [1]{\csname bibitem#1\endcsname}%
\let\auto@bib@innerbib\@empty
%</preamble>
\bibitem [{\citenamefont {Fukuda}\ \emph {et~al.}(1998)\citenamefont {Fukuda}
  \emph {et~al.}}]{Super-Kamiokande:1998kpq}%
  \BibitemOpen
  \bibfield  {author} {\bibinfo {author} {\bibfnamefont {Y.}~\bibnamefont
  {Fukuda}} \emph {et~al.} (\bibinfo {collaboration} {Super-Kamiokande}),\
  }\bibfield  {title} {\emph {\enquote {\bibinfo {title} {{Evidence for
  oscillation of atmospheric neutrinos}},}\ }}\href {\doibase
  10.1103/PhysRevLett.81.1562} {\bibfield  {journal} {\bibinfo  {journal}
  {Phys. Rev. Lett.}\ }\textbf {\bibinfo {volume} {81}},\ \bibinfo {pages}
  {1562} (\bibinfo {year} {1998})},\ \Eprint
  {http://arxiv.org/abs/hep-ex/9807003} {arXiv:hep-ex/9807003} \BibitemShut
  {NoStop}%
\bibitem [{\citenamefont {Ahmad}\ \emph {et~al.}(2001)\citenamefont {Ahmad}
  \emph {et~al.}}]{SNO:2001kpb}%
  \BibitemOpen
  \bibfield  {author} {\bibinfo {author} {\bibfnamefont {Q.~R.}\ \bibnamefont
  {Ahmad}} \emph {et~al.} (\bibinfo {collaboration} {SNO}),\ }\bibfield
  {title} {\emph {\enquote {\bibinfo {title} {{Measurement of the rate of
  $\nu_e+d \to p+p+e^-$ interactions produced by $^8$B solar neutrinos at the
  Sudbury Neutrino Observatory}},}\ }}\href {\doibase
  10.1103/PhysRevLett.87.071301} {\bibfield  {journal} {\bibinfo  {journal}
  {Phys. Rev. Lett.}\ }\textbf {\bibinfo {volume} {87}},\ \bibinfo {pages}
  {071301} (\bibinfo {year} {2001})},\ \Eprint
  {http://arxiv.org/abs/nucl-ex/0106015} {arXiv:nucl-ex/0106015} \BibitemShut
  {NoStop}%
\bibitem [{\citenamefont {Ahmad}\ \emph {et~al.}(2002)\citenamefont {Ahmad}
  \emph {et~al.}}]{SNO:2002tuh}%
  \BibitemOpen
  \bibfield  {author} {\bibinfo {author} {\bibfnamefont {Q.~R.}\ \bibnamefont
  {Ahmad}} \emph {et~al.} (\bibinfo {collaboration} {SNO}),\ }\bibfield
  {title} {\emph {\enquote {\bibinfo {title} {{Direct evidence for neutrino
  flavor transformation from neutral current interactions in the Sudbury
  Neutrino Observatory}},}\ }}\href {\doibase 10.1103/PhysRevLett.89.011301}
  {\bibfield  {journal} {\bibinfo  {journal} {Phys. Rev. Lett.}\ }\textbf
  {\bibinfo {volume} {89}},\ \bibinfo {pages} {011301} (\bibinfo {year}
  {2002})},\ \Eprint {http://arxiv.org/abs/nucl-ex/0204008}
  {arXiv:nucl-ex/0204008} \BibitemShut {NoStop}%
\bibitem [{\citenamefont {Esteban}\ \emph {et~al.}(2020)\citenamefont
  {Esteban}, \citenamefont {Gonzalez-Garcia}, \citenamefont {Maltoni},
  \citenamefont {Schwetz},\ and\ \citenamefont {Zhou}}]{Esteban:2020cvm}%
  \BibitemOpen
  \bibfield  {author} {\bibinfo {author} {\bibfnamefont {I.}~\bibnamefont
  {Esteban}}, \bibinfo {author} {\bibfnamefont {M.~C.}\ \bibnamefont
  {Gonzalez-Garcia}}, \bibinfo {author} {\bibfnamefont {M.}~\bibnamefont
  {Maltoni}}, \bibinfo {author} {\bibfnamefont {T.}~\bibnamefont {Schwetz}}, \
  and\ \bibinfo {author} {\bibfnamefont {A.}~\bibnamefont {Zhou}},\ }\bibfield
  {title} {\emph {\enquote {\bibinfo {title} {{The fate of hints: updated
  global analysis of three-flavor neutrino oscillations}},}\ }}\href {\doibase
  10.1007/JHEP09(2020)178} {\bibfield  {journal} {\bibinfo  {journal} {JHEP}\
  }\textbf {\bibinfo {volume} {09}},\ \bibinfo {pages} {178} (\bibinfo {year}
  {2020})},\ \Eprint {http://arxiv.org/abs/2007.14792} {arXiv:2007.14792
  [hep-ph]} \BibitemShut {NoStop}%
\bibitem [{NuF()}]{NuFIT}%
  \BibitemOpen
  \href {http://www.nu-fit.org/} {\enquote {\bibinfo {title} {{NuFIT 5.2:
  Three-neutrino fit based on data available in November 2022}},}\
  }\BibitemShut {NoStop}%
\bibitem [{\citenamefont {Fields}\ \emph {et~al.}(2020)\citenamefont {Fields},
  \citenamefont {Olive}, \citenamefont {Yeh},\ and\ \citenamefont
  {Young}}]{Fields:2019pfx}%
  \BibitemOpen
  \bibfield  {author} {\bibinfo {author} {\bibfnamefont {B.~D.}\ \bibnamefont
  {Fields}}, \bibinfo {author} {\bibfnamefont {K.~A.}\ \bibnamefont {Olive}},
  \bibinfo {author} {\bibfnamefont {T.-H.}\ \bibnamefont {Yeh}}, \ and\
  \bibinfo {author} {\bibfnamefont {C.}~\bibnamefont {Young}},\ }\bibfield
  {title} {\emph {\enquote {\bibinfo {title} {{Big-Bang Nucleosynthesis after
  Planck}},}\ }}\href {\doibase 10.1088/1475-7516/2020/03/010} {\bibfield
  {journal} {\bibinfo  {journal} {JCAP}\ }\textbf {\bibinfo {volume} {03}},\
  \bibinfo {pages} {010} (\bibinfo {year} {2020})},\ \bibinfo {note} {[Erratum:
  JCAP 11, E02 (2020)]},\ \Eprint {http://arxiv.org/abs/1912.01132}
  {arXiv:1912.01132 [astro-ph.CO]} \BibitemShut {NoStop}%
\bibitem [{\citenamefont {Aghanim}\ \emph {et~al.}(2020)\citenamefont {Aghanim}
  \emph {et~al.}}]{Planck:2018vyg}%
  \BibitemOpen
  \bibfield  {author} {\bibinfo {author} {\bibfnamefont {N.}~\bibnamefont
  {Aghanim}} \emph {et~al.} (\bibinfo {collaboration} {Planck}),\ }\bibfield
  {title} {\emph {\enquote {\bibinfo {title} {{Planck 2018 results. VI.
  Cosmological parameters}},}\ }}\href {\doibase 10.1051/0004-6361/201833910}
  {\bibfield  {journal} {\bibinfo  {journal} {Astron. Astrophys.}\ }\textbf
  {\bibinfo {volume} {641}},\ \bibinfo {pages} {A6} (\bibinfo {year} {2020})},\
  \bibinfo {note} {[Erratum: Astron.Astrophys. 652, C4 (2021)]},\ \Eprint
  {http://arxiv.org/abs/1807.06209} {arXiv:1807.06209 [astro-ph.CO]}
  \BibitemShut {NoStop}%
\bibitem [{\citenamefont {Minkowski}(1977)}]{Minkowski:1977sc}%
  \BibitemOpen
  \bibfield  {author} {\bibinfo {author} {\bibfnamefont {P.}~\bibnamefont
  {Minkowski}},\ }\bibfield  {title} {\emph {\enquote {\bibinfo {title} {{$\mu
  \to e\gamma$ at a Rate of One Out of $10^{9}$ Muon Decays?}}}\ }}\href
  {\doibase 10.1016/0370-2693(77)90435-X} {\bibfield  {journal} {\bibinfo
  {journal} {Phys. Lett. B}\ }\textbf {\bibinfo {volume} {67}},\ \bibinfo
  {pages} {421} (\bibinfo {year} {1977})}\BibitemShut {NoStop}%
\bibitem [{\citenamefont {Yanagida}(1979)}]{Yanagida:1979as}%
  \BibitemOpen
  \bibfield  {author} {\bibinfo {author} {\bibfnamefont {T.}~\bibnamefont
  {Yanagida}},\ }\bibfield  {title} {\emph {\enquote {\bibinfo {title}
  {{Horizontal gauge symmetry and masses of neutrinos}},}\ }}\href@noop {}
  {\bibfield  {journal} {\bibinfo  {journal} {Conf. Proc. C}\ }\textbf
  {\bibinfo {volume} {7902131}},\ \bibinfo {pages} {95} (\bibinfo {year}
  {1979})}\BibitemShut {NoStop}%
\bibitem [{\citenamefont {Gell-Mann}\ \emph {et~al.}()\citenamefont
  {Gell-Mann}, \citenamefont {Ramond},\ and\ \citenamefont
  {Slansky}}]{GellMann1979}%
  \BibitemOpen
  \bibfield  {author} {\bibinfo {author} {\bibfnamefont {M.}~\bibnamefont
  {Gell-Mann}}, \bibinfo {author} {\bibfnamefont {P.}~\bibnamefont {Ramond}}, \
  and\ \bibinfo {author} {\bibfnamefont {R.}~\bibnamefont {Slansky}},\
  }\href@noop {} {}\bibinfo {note} {{in} \emph{Sanibel talk}, {retroprinted as
  \href{https://arxiv.org/abs/hep-ph/9809459}{arXiv:hep-ph/9809459}, and in}
  \emph{Supergravity}, {North-Holland, Amsterdam (1979), PRINT-80-0576,
  retroprinted as
  \href{https://arxiv.org/abs/1306.4669}{arXiv:1306.4669[hep-th]}}}\BibitemShut
  {NoStop}%
\bibitem [{\citenamefont {Glashow}(1980)}]{Glashow:1979nm}%
  \BibitemOpen
  \bibfield  {author} {\bibinfo {author} {\bibfnamefont {S.~L.}\ \bibnamefont
  {Glashow}},\ }\bibfield  {title} {\emph {\enquote {\bibinfo {title} {{The
  Future of Elementary Particle Physics}},}\ }}\href {\doibase
  10.1007/978-1-4684-7197-7_15} {\bibfield  {journal} {\bibinfo  {journal}
  {NATO Sci. Ser. B}\ }\textbf {\bibinfo {volume} {61}},\ \bibinfo {pages}
  {687} (\bibinfo {year} {1980})}\BibitemShut {NoStop}%
\bibitem [{\citenamefont {Mohapatra}\ and\ \citenamefont
  {Senjanovic}(1980)}]{Mohapatra:1979ia}%
  \BibitemOpen
  \bibfield  {author} {\bibinfo {author} {\bibfnamefont {R.~N.}\ \bibnamefont
  {Mohapatra}}\ and\ \bibinfo {author} {\bibfnamefont {G.}~\bibnamefont
  {Senjanovic}},\ }\bibfield  {title} {\emph {\enquote {\bibinfo {title}
  {{Neutrino Mass and Spontaneous Parity Nonconservation}},}\ }}\href {\doibase
  10.1103/PhysRevLett.44.912} {\bibfield  {journal} {\bibinfo  {journal} {Phys.
  Rev. Lett.}\ }\textbf {\bibinfo {volume} {44}},\ \bibinfo {pages} {912}
  (\bibinfo {year} {1980})}\BibitemShut {NoStop}%
\bibitem [{\citenamefont {Fukugita}\ and\ \citenamefont
  {Yanagida}(1986)}]{Fukugita:1986hr}%
  \BibitemOpen
  \bibfield  {author} {\bibinfo {author} {\bibfnamefont {M.}~\bibnamefont
  {Fukugita}}\ and\ \bibinfo {author} {\bibfnamefont {T.}~\bibnamefont
  {Yanagida}},\ }\bibfield  {title} {\emph {\enquote {\bibinfo {title}
  {{Baryogenesis Without Grand Unification}},}\ }}\href {\doibase
  10.1016/0370-2693(86)91126-3} {\bibfield  {journal} {\bibinfo  {journal}
  {Phys. Lett. B}\ }\textbf {\bibinfo {volume} {174}},\ \bibinfo {pages} {45}
  (\bibinfo {year} {1986})}\BibitemShut {NoStop}%
\bibitem [{\citenamefont {Nardi}\ \emph {et~al.}(2006)\citenamefont {Nardi},
  \citenamefont {Nir}, \citenamefont {Roulet},\ and\ \citenamefont
  {Racker}}]{Nardi:2006fx}%
  \BibitemOpen
  \bibfield  {author} {\bibinfo {author} {\bibfnamefont {E.}~\bibnamefont
  {Nardi}}, \bibinfo {author} {\bibfnamefont {Y.}~\bibnamefont {Nir}}, \bibinfo
  {author} {\bibfnamefont {E.}~\bibnamefont {Roulet}}, \ and\ \bibinfo {author}
  {\bibfnamefont {J.}~\bibnamefont {Racker}},\ }\bibfield  {title} {\emph
  {\enquote {\bibinfo {title} {{The Importance of flavor in leptogenesis}},}\
  }}\href {\doibase 10.1088/1126-6708/2006/01/164} {\bibfield  {journal}
  {\bibinfo  {journal} {JHEP}\ }\textbf {\bibinfo {volume} {01}},\ \bibinfo
  {pages} {164} (\bibinfo {year} {2006})},\ \Eprint
  {http://arxiv.org/abs/hep-ph/0601084} {arXiv:hep-ph/0601084} \BibitemShut
  {NoStop}%
\bibitem [{\citenamefont {Abada}\ \emph
  {et~al.}(2006{\natexlab{a}})\citenamefont {Abada}, \citenamefont {Davidson},
  \citenamefont {Josse-Michaux}, \citenamefont {Losada},\ and\ \citenamefont
  {Riotto}}]{Abada:2006fw}%
  \BibitemOpen
  \bibfield  {author} {\bibinfo {author} {\bibfnamefont {A.}~\bibnamefont
  {Abada}}, \bibinfo {author} {\bibfnamefont {S.}~\bibnamefont {Davidson}},
  \bibinfo {author} {\bibfnamefont {F.-X.}\ \bibnamefont {Josse-Michaux}},
  \bibinfo {author} {\bibfnamefont {M.}~\bibnamefont {Losada}}, \ and\ \bibinfo
  {author} {\bibfnamefont {A.}~\bibnamefont {Riotto}},\ }\bibfield  {title}
  {\emph {\enquote {\bibinfo {title} {{Flavor issues in leptogenesis}},}\
  }}\href {\doibase 10.1088/1475-7516/2006/04/004} {\bibfield  {journal}
  {\bibinfo  {journal} {JCAP}\ }\textbf {\bibinfo {volume} {04}},\ \bibinfo
  {pages} {004} (\bibinfo {year} {2006}{\natexlab{a}})},\ \Eprint
  {http://arxiv.org/abs/hep-ph/0601083} {arXiv:hep-ph/0601083} \BibitemShut
  {NoStop}%
\bibitem [{\citenamefont {Abada}\ \emph
  {et~al.}(2006{\natexlab{b}})\citenamefont {Abada}, \citenamefont {Davidson},
  \citenamefont {Ibarra}, \citenamefont {Josse-Michaux}, \citenamefont
  {Losada},\ and\ \citenamefont {Riotto}}]{Abada:2006ea}%
  \BibitemOpen
  \bibfield  {author} {\bibinfo {author} {\bibfnamefont {A.}~\bibnamefont
  {Abada}}, \bibinfo {author} {\bibfnamefont {S.}~\bibnamefont {Davidson}},
  \bibinfo {author} {\bibfnamefont {A.}~\bibnamefont {Ibarra}}, \bibinfo
  {author} {\bibfnamefont {F.~X.}\ \bibnamefont {Josse-Michaux}}, \bibinfo
  {author} {\bibfnamefont {M.}~\bibnamefont {Losada}}, \ and\ \bibinfo {author}
  {\bibfnamefont {A.}~\bibnamefont {Riotto}},\ }\bibfield  {title} {\emph
  {\enquote {\bibinfo {title} {{Flavour Matters in Leptogenesis}},}\ }}\href
  {\doibase 10.1088/1126-6708/2006/09/010} {\bibfield  {journal} {\bibinfo
  {journal} {JHEP}\ }\textbf {\bibinfo {volume} {09}},\ \bibinfo {pages} {010}
  (\bibinfo {year} {2006}{\natexlab{b}})},\ \Eprint
  {http://arxiv.org/abs/hep-ph/0605281} {arXiv:hep-ph/0605281} \BibitemShut
  {NoStop}%
\bibitem [{\citenamefont {Davidson}\ and\ \citenamefont
  {Ibarra}(2002)}]{Davidson:2002qv}%
  \BibitemOpen
  \bibfield  {author} {\bibinfo {author} {\bibfnamefont {S.}~\bibnamefont
  {Davidson}}\ and\ \bibinfo {author} {\bibfnamefont {A.}~\bibnamefont
  {Ibarra}},\ }\bibfield  {title} {\emph {\enquote {\bibinfo {title} {{A Lower
  bound on the right-handed neutrino mass from leptogenesis}},}\ }}\href
  {\doibase 10.1016/S0370-2693(02)01735-5} {\bibfield  {journal} {\bibinfo
  {journal} {Phys. Lett. B}\ }\textbf {\bibinfo {volume} {535}},\ \bibinfo
  {pages} {25} (\bibinfo {year} {2002})},\ \Eprint
  {http://arxiv.org/abs/hep-ph/0202239} {arXiv:hep-ph/0202239} \BibitemShut
  {NoStop}%
\bibitem [{\citenamefont {Giudice}\ \emph {et~al.}(2004)\citenamefont
  {Giudice}, \citenamefont {Notari}, \citenamefont {Raidal}, \citenamefont
  {Riotto},\ and\ \citenamefont {Strumia}}]{Giudice:2003jh}%
  \BibitemOpen
  \bibfield  {author} {\bibinfo {author} {\bibfnamefont {G.~F.}\ \bibnamefont
  {Giudice}}, \bibinfo {author} {\bibfnamefont {A.}~\bibnamefont {Notari}},
  \bibinfo {author} {\bibfnamefont {M.}~\bibnamefont {Raidal}}, \bibinfo
  {author} {\bibfnamefont {A.}~\bibnamefont {Riotto}}, \ and\ \bibinfo {author}
  {\bibfnamefont {A.}~\bibnamefont {Strumia}},\ }\bibfield  {title} {\emph
  {\enquote {\bibinfo {title} {{Towards a complete theory of thermal
  leptogenesis in the SM and MSSM}},}\ }}\href {\doibase
  10.1016/j.nuclphysb.2004.02.019} {\bibfield  {journal} {\bibinfo  {journal}
  {Nucl. Phys. B}\ }\textbf {\bibinfo {volume} {685}},\ \bibinfo {pages} {89}
  (\bibinfo {year} {2004})},\ \Eprint {http://arxiv.org/abs/hep-ph/0310123}
  {arXiv:hep-ph/0310123} \BibitemShut {NoStop}%
\bibitem [{\citenamefont {Moffat}\ \emph {et~al.}(2018)\citenamefont {Moffat},
  \citenamefont {Pascoli}, \citenamefont {Petcov}, \citenamefont {Schulz},\
  and\ \citenamefont {Turner}}]{Moffat:2018wke}%
  \BibitemOpen
  \bibfield  {author} {\bibinfo {author} {\bibfnamefont {K.}~\bibnamefont
  {Moffat}}, \bibinfo {author} {\bibfnamefont {S.}~\bibnamefont {Pascoli}},
  \bibinfo {author} {\bibfnamefont {S.~T.}\ \bibnamefont {Petcov}}, \bibinfo
  {author} {\bibfnamefont {H.}~\bibnamefont {Schulz}}, \ and\ \bibinfo {author}
  {\bibfnamefont {J.}~\bibnamefont {Turner}},\ }\bibfield  {title} {\emph
  {\enquote {\bibinfo {title} {{Three-flavored nonresonant leptogenesis at
  intermediate scales}},}\ }}\href {\doibase 10.1103/PhysRevD.98.015036}
  {\bibfield  {journal} {\bibinfo  {journal} {Phys. Rev. D}\ }\textbf {\bibinfo
  {volume} {98}},\ \bibinfo {pages} {015036} (\bibinfo {year} {2018})},\
  \Eprint {http://arxiv.org/abs/1804.05066} {arXiv:1804.05066 [hep-ph]}
  \BibitemShut {NoStop}%
\bibitem [{\citenamefont {Covi}\ \emph {et~al.}(1996)\citenamefont {Covi},
  \citenamefont {Roulet},\ and\ \citenamefont {Vissani}}]{Covi:1996wh}%
  \BibitemOpen
  \bibfield  {author} {\bibinfo {author} {\bibfnamefont {L.}~\bibnamefont
  {Covi}}, \bibinfo {author} {\bibfnamefont {E.}~\bibnamefont {Roulet}}, \ and\
  \bibinfo {author} {\bibfnamefont {F.}~\bibnamefont {Vissani}},\ }\bibfield
  {title} {\emph {\enquote {\bibinfo {title} {{CP violating decays in
  leptogenesis scenarios}},}\ }}\href {\doibase 10.1016/0370-2693(96)00817-9}
  {\bibfield  {journal} {\bibinfo  {journal} {Phys. Lett. B}\ }\textbf
  {\bibinfo {volume} {384}},\ \bibinfo {pages} {169} (\bibinfo {year}
  {1996})},\ \Eprint {http://arxiv.org/abs/hep-ph/9605319}
  {arXiv:hep-ph/9605319} \BibitemShut {NoStop}%
\bibitem [{\citenamefont {Pilaftsis}(1997)}]{Pilaftsis:1997jf}%
  \BibitemOpen
  \bibfield  {author} {\bibinfo {author} {\bibfnamefont {A.}~\bibnamefont
  {Pilaftsis}},\ }\bibfield  {title} {\emph {\enquote {\bibinfo {title} {{CP
  violation and baryogenesis due to heavy Majorana neutrinos}},}\ }}\href
  {\doibase 10.1103/PhysRevD.56.5431} {\bibfield  {journal} {\bibinfo
  {journal} {Phys. Rev. D}\ }\textbf {\bibinfo {volume} {56}},\ \bibinfo
  {pages} {5431} (\bibinfo {year} {1997})},\ \Eprint
  {http://arxiv.org/abs/hep-ph/9707235} {arXiv:hep-ph/9707235} \BibitemShut
  {NoStop}%
\bibitem [{\citenamefont {Pilaftsis}\ and\ \citenamefont
  {Underwood}(2004)}]{Pilaftsis:2003gt}%
  \BibitemOpen
  \bibfield  {author} {\bibinfo {author} {\bibfnamefont {A.}~\bibnamefont
  {Pilaftsis}}\ and\ \bibinfo {author} {\bibfnamefont {T.~E.~J.}\ \bibnamefont
  {Underwood}},\ }\bibfield  {title} {\emph {\enquote {\bibinfo {title}
  {{Resonant leptogenesis}},}\ }}\href {\doibase
  10.1016/j.nuclphysb.2004.05.029} {\bibfield  {journal} {\bibinfo  {journal}
  {Nucl. Phys. B}\ }\textbf {\bibinfo {volume} {692}},\ \bibinfo {pages} {303}
  (\bibinfo {year} {2004})},\ \Eprint {http://arxiv.org/abs/hep-ph/0309342}
  {arXiv:hep-ph/0309342} \BibitemShut {NoStop}%
\bibitem [{\citenamefont {D'Onofrio}\ \emph {et~al.}(2014)\citenamefont
  {D'Onofrio}, \citenamefont {Rummukainen},\ and\ \citenamefont
  {Tranberg}}]{DOnofrio:2014rug}%
  \BibitemOpen
  \bibfield  {author} {\bibinfo {author} {\bibfnamefont {M.}~\bibnamefont
  {D'Onofrio}}, \bibinfo {author} {\bibfnamefont {K.}~\bibnamefont
  {Rummukainen}}, \ and\ \bibinfo {author} {\bibfnamefont {A.}~\bibnamefont
  {Tranberg}},\ }\bibfield  {title} {\emph {\enquote {\bibinfo {title}
  {{Sphaleron Rate in the Minimal Standard Model}},}\ }}\href {\doibase
  10.1103/PhysRevLett.113.141602} {\bibfield  {journal} {\bibinfo  {journal}
  {Phys. Rev. Lett.}\ }\textbf {\bibinfo {volume} {113}},\ \bibinfo {pages}
  {141602} (\bibinfo {year} {2014})},\ \Eprint {http://arxiv.org/abs/1404.3565}
  {arXiv:1404.3565 [hep-ph]} \BibitemShut {NoStop}%
\bibitem [{\citenamefont {Akhmedov}\ \emph {et~al.}(1998)\citenamefont
  {Akhmedov}, \citenamefont {Rubakov},\ and\ \citenamefont
  {Smirnov}}]{Akhmedov:1998qx}%
  \BibitemOpen
  \bibfield  {author} {\bibinfo {author} {\bibfnamefont {E.~K.}\ \bibnamefont
  {Akhmedov}}, \bibinfo {author} {\bibfnamefont {V.~A.}\ \bibnamefont
  {Rubakov}}, \ and\ \bibinfo {author} {\bibfnamefont {A.~Y.}\ \bibnamefont
  {Smirnov}},\ }\bibfield  {title} {\emph {\enquote {\bibinfo {title}
  {{Baryogenesis via neutrino oscillations}},}\ }}\href {\doibase
  10.1103/PhysRevLett.81.1359} {\bibfield  {journal} {\bibinfo  {journal}
  {Phys. Rev. Lett.}\ }\textbf {\bibinfo {volume} {81}},\ \bibinfo {pages}
  {1359} (\bibinfo {year} {1998})},\ \Eprint
  {http://arxiv.org/abs/hep-ph/9803255} {arXiv:hep-ph/9803255} \BibitemShut
  {NoStop}%
\bibitem [{\citenamefont {Klari\'c}\ \emph {et~al.}(2021)\citenamefont
  {Klari\'c}, \citenamefont {Shaposhnikov},\ and\ \citenamefont
  {Timiryasov}}]{Klaric:2020phc}%
  \BibitemOpen
  \bibfield  {author} {\bibinfo {author} {\bibfnamefont {J.}~\bibnamefont
  {Klari\'c}}, \bibinfo {author} {\bibfnamefont {M.}~\bibnamefont
  {Shaposhnikov}}, \ and\ \bibinfo {author} {\bibfnamefont {I.}~\bibnamefont
  {Timiryasov}},\ }\bibfield  {title} {\emph {\enquote {\bibinfo {title}
  {{Uniting Low-Scale Leptogenesis Mechanisms}},}\ }}\href {\doibase
  10.1103/PhysRevLett.127.111802} {\bibfield  {journal} {\bibinfo  {journal}
  {Phys. Rev. Lett.}\ }\textbf {\bibinfo {volume} {127}},\ \bibinfo {pages}
  {111802} (\bibinfo {year} {2021})},\ \Eprint
  {http://arxiv.org/abs/2008.13771} {arXiv:2008.13771 [hep-ph]} \BibitemShut
  {NoStop}%
\bibitem [{\citenamefont {Drewes}\ \emph {et~al.}(2022)\citenamefont {Drewes},
  \citenamefont {Georis},\ and\ \citenamefont {Klari\'c}}]{Drewes:2021nqr}%
  \BibitemOpen
  \bibfield  {author} {\bibinfo {author} {\bibfnamefont {M.}~\bibnamefont
  {Drewes}}, \bibinfo {author} {\bibfnamefont {Y.}~\bibnamefont {Georis}}, \
  and\ \bibinfo {author} {\bibfnamefont {J.}~\bibnamefont {Klari\'c}},\
  }\bibfield  {title} {\emph {\enquote {\bibinfo {title} {{Mapping the Viable
  Parameter Space for Testable Leptogenesis}},}\ }}\href {\doibase
  10.1103/PhysRevLett.128.051801} {\bibfield  {journal} {\bibinfo  {journal}
  {Phys. Rev. Lett.}\ }\textbf {\bibinfo {volume} {128}},\ \bibinfo {pages}
  {051801} (\bibinfo {year} {2022})},\ \Eprint
  {http://arxiv.org/abs/2106.16226} {arXiv:2106.16226 [hep-ph]} \BibitemShut
  {NoStop}%
\bibitem [{\citenamefont {Abdullahi}\ \emph {et~al.}(2023)\citenamefont
  {Abdullahi} \emph {et~al.}}]{Abdullahi:2022jlv}%
  \BibitemOpen
  \bibfield  {author} {\bibinfo {author} {\bibfnamefont {A.~M.}\ \bibnamefont
  {Abdullahi}} \emph {et~al.},\ }\bibfield  {title} {\emph {\enquote {\bibinfo
  {title} {{The present and future status of heavy neutral leptons}},}\ }}\href
  {\doibase 10.1088/1361-6471/ac98f9} {\bibfield  {journal} {\bibinfo
  {journal} {J. Phys. G}\ }\textbf {\bibinfo {volume} {50}},\ \bibinfo {pages}
  {020501} (\bibinfo {year} {2023})},\ \Eprint
  {http://arxiv.org/abs/2203.08039} {arXiv:2203.08039 [hep-ph]} \BibitemShut
  {NoStop}%
\bibitem [{\citenamefont {Beacham}\ \emph {et~al.}(2020)\citenamefont {Beacham}
  \emph {et~al.}}]{Beacham:2019nyx}%
  \BibitemOpen
  \bibfield  {author} {\bibinfo {author} {\bibfnamefont {J.}~\bibnamefont
  {Beacham}} \emph {et~al.},\ }\bibfield  {title} {\emph {\enquote {\bibinfo
  {title} {{Physics Beyond Colliders at CERN: Beyond the Standard Model Working
  Group Report}},}\ }}\href {\doibase 10.1088/1361-6471/ab4cd2} {\bibfield
  {journal} {\bibinfo  {journal} {J. Phys. G}\ }\textbf {\bibinfo {volume}
  {47}},\ \bibinfo {pages} {010501} (\bibinfo {year} {2020})},\ \Eprint
  {http://arxiv.org/abs/1901.09966} {arXiv:1901.09966 [hep-ex]} \BibitemShut
  {NoStop}%
\bibitem [{\citenamefont {Cirigliano}\ \emph {et~al.}(2022)\citenamefont
  {Cirigliano} \emph {et~al.}}]{Cirigliano:2022oqy}%
  \BibitemOpen
  \bibfield  {author} {\bibinfo {author} {\bibfnamefont {V.}~\bibnamefont
  {Cirigliano}} \emph {et~al.},\ }\bibfield  {title} {\emph {\enquote {\bibinfo
  {title} {{Neutrinoless Double-Beta Decay: A Roadmap for Matching Theory to
  Experiment}},}\ }}\href@noop {} {\  (\bibinfo {year} {2022})},\ \Eprint
  {http://arxiv.org/abs/2203.12169} {arXiv:2203.12169 [hep-ph]} \BibitemShut
  {NoStop}%
\bibitem [{\citenamefont {Endoh}\ \emph {et~al.}(2002)\citenamefont {Endoh},
  \citenamefont {Kaneko}, \citenamefont {Kang}, \citenamefont {Morozumi},\ and\
  \citenamefont {Tanimoto}}]{Endoh:2002wm}%
  \BibitemOpen
  \bibfield  {author} {\bibinfo {author} {\bibfnamefont {T.}~\bibnamefont
  {Endoh}}, \bibinfo {author} {\bibfnamefont {S.}~\bibnamefont {Kaneko}},
  \bibinfo {author} {\bibfnamefont {S.~K.}\ \bibnamefont {Kang}}, \bibinfo
  {author} {\bibfnamefont {T.}~\bibnamefont {Morozumi}}, \ and\ \bibinfo
  {author} {\bibfnamefont {M.}~\bibnamefont {Tanimoto}},\ }\bibfield  {title}
  {\emph {\enquote {\bibinfo {title} {{CP violation in neutrino oscillation and
  leptogenesis}},}\ }}\href {\doibase 10.1103/PhysRevLett.89.231601} {\bibfield
   {journal} {\bibinfo  {journal} {Phys. Rev. Lett.}\ }\textbf {\bibinfo
  {volume} {89}},\ \bibinfo {pages} {231601} (\bibinfo {year} {2002})},\
  \Eprint {http://arxiv.org/abs/hep-ph/0209020} {arXiv:hep-ph/0209020}
  \BibitemShut {NoStop}%
\bibitem [{\citenamefont {Di~Bari}\ and\ \citenamefont
  {Riotto}(2009)}]{DiBari:2008mp}%
  \BibitemOpen
  \bibfield  {author} {\bibinfo {author} {\bibfnamefont {P.}~\bibnamefont
  {Di~Bari}}\ and\ \bibinfo {author} {\bibfnamefont {A.}~\bibnamefont
  {Riotto}},\ }\bibfield  {title} {\emph {\enquote {\bibinfo {title}
  {{Successful type I Leptogenesis with SO(10)-inspired mass relations}},}\
  }}\href {\doibase 10.1016/j.physletb.2008.12.054} {\bibfield  {journal}
  {\bibinfo  {journal} {Phys. Lett. B}\ }\textbf {\bibinfo {volume} {671}},\
  \bibinfo {pages} {462} (\bibinfo {year} {2009})},\ \Eprint
  {http://arxiv.org/abs/0809.2285} {arXiv:0809.2285 [hep-ph]} \BibitemShut
  {NoStop}%
\bibitem [{\citenamefont {Bertuzzo}\ \emph {et~al.}(2011)\citenamefont
  {Bertuzzo}, \citenamefont {Di~Bari},\ and\ \citenamefont
  {Marzola}}]{Bertuzzo:2010et}%
  \BibitemOpen
  \bibfield  {author} {\bibinfo {author} {\bibfnamefont {E.}~\bibnamefont
  {Bertuzzo}}, \bibinfo {author} {\bibfnamefont {P.}~\bibnamefont {Di~Bari}}, \
  and\ \bibinfo {author} {\bibfnamefont {L.}~\bibnamefont {Marzola}},\
  }\bibfield  {title} {\emph {\enquote {\bibinfo {title} {{The problem of the
  initial conditions in flavoured leptogenesis and the tauon $N_2$-dominated
  scenario}},}\ }}\href {\doibase 10.1016/j.nuclphysb.2011.03.027} {\bibfield
  {journal} {\bibinfo  {journal} {Nucl. Phys. B}\ }\textbf {\bibinfo {volume}
  {849}},\ \bibinfo {pages} {521} (\bibinfo {year} {2011})},\ \Eprint
  {http://arxiv.org/abs/1007.1641} {arXiv:1007.1641 [hep-ph]} \BibitemShut
  {NoStop}%
\bibitem [{\citenamefont {Buccella}\ \emph {et~al.}(2012)\citenamefont
  {Buccella}, \citenamefont {Falcone}, \citenamefont {Fong}, \citenamefont
  {Nardi},\ and\ \citenamefont {Ricciardi}}]{Buccella:2012kc}%
  \BibitemOpen
  \bibfield  {author} {\bibinfo {author} {\bibfnamefont {F.}~\bibnamefont
  {Buccella}}, \bibinfo {author} {\bibfnamefont {D.}~\bibnamefont {Falcone}},
  \bibinfo {author} {\bibfnamefont {C.~S.}\ \bibnamefont {Fong}}, \bibinfo
  {author} {\bibfnamefont {E.}~\bibnamefont {Nardi}}, \ and\ \bibinfo {author}
  {\bibfnamefont {G.}~\bibnamefont {Ricciardi}},\ }\bibfield  {title} {\emph
  {\enquote {\bibinfo {title} {{Squeezing out predictions with leptogenesis
  from SO(10)}},}\ }}\href {\doibase 10.1103/PhysRevD.86.035012} {\bibfield
  {journal} {\bibinfo  {journal} {Phys. Rev. D}\ }\textbf {\bibinfo {volume}
  {86}},\ \bibinfo {pages} {035012} (\bibinfo {year} {2012})},\ \Eprint
  {http://arxiv.org/abs/1203.0829} {arXiv:1203.0829 [hep-ph]} \BibitemShut
  {NoStop}%
\bibitem [{\citenamefont {Altarelli}\ and\ \citenamefont
  {Meloni}(2013)}]{Altarelli:2013aqa}%
  \BibitemOpen
  \bibfield  {author} {\bibinfo {author} {\bibfnamefont {G.}~\bibnamefont
  {Altarelli}}\ and\ \bibinfo {author} {\bibfnamefont {D.}~\bibnamefont
  {Meloni}},\ }\bibfield  {title} {\emph {\enquote {\bibinfo {title} {{A non
  supersymmetric SO(10) grand unified model for all the physics below
  $M_{GUT}$}},}\ }}\href {\doibase 10.1007/JHEP08(2013)021} {\bibfield
  {journal} {\bibinfo  {journal} {JHEP}\ }\textbf {\bibinfo {volume} {08}},\
  \bibinfo {pages} {021} (\bibinfo {year} {2013})},\ \Eprint
  {http://arxiv.org/abs/1305.1001} {arXiv:1305.1001 [hep-ph]} \BibitemShut
  {NoStop}%
\bibitem [{\citenamefont {Fong}\ \emph {et~al.}(2015)\citenamefont {Fong},
  \citenamefont {Meloni}, \citenamefont {Meroni},\ and\ \citenamefont
  {Nardi}}]{Fong:2014gea}%
  \BibitemOpen
  \bibfield  {author} {\bibinfo {author} {\bibfnamefont {C.~S.}\ \bibnamefont
  {Fong}}, \bibinfo {author} {\bibfnamefont {D.}~\bibnamefont {Meloni}},
  \bibinfo {author} {\bibfnamefont {A.}~\bibnamefont {Meroni}}, \ and\ \bibinfo
  {author} {\bibfnamefont {E.}~\bibnamefont {Nardi}},\ }\bibfield  {title}
  {\emph {\enquote {\bibinfo {title} {{Leptogenesis in SO(10)}},}\ }}\href
  {\doibase 10.1007/JHEP01(2015)111} {\bibfield  {journal} {\bibinfo  {journal}
  {JHEP}\ }\textbf {\bibinfo {volume} {01}},\ \bibinfo {pages} {111} (\bibinfo
  {year} {2015})},\ \Eprint {http://arxiv.org/abs/1412.4776} {arXiv:1412.4776
  [hep-ph]} \BibitemShut {NoStop}%
\bibitem [{\citenamefont {Mummidi}\ and\ \citenamefont
  {Patel}(2021)}]{Mummidi:2021anm}%
  \BibitemOpen
  \bibfield  {author} {\bibinfo {author} {\bibfnamefont {V.~S.}\ \bibnamefont
  {Mummidi}}\ and\ \bibinfo {author} {\bibfnamefont {K.~M.}\ \bibnamefont
  {Patel}},\ }\bibfield  {title} {\emph {\enquote {\bibinfo {title}
  {{Leptogenesis and fermion mass fit in a renormalizable SO(10) model}},}\
  }}\href {\doibase 10.1007/JHEP12(2021)042} {\bibfield  {journal} {\bibinfo
  {journal} {JHEP}\ }\textbf {\bibinfo {volume} {12}},\ \bibinfo {pages} {042}
  (\bibinfo {year} {2021})},\ \Eprint {http://arxiv.org/abs/2109.04050}
  {arXiv:2109.04050 [hep-ph]} \BibitemShut {NoStop}%
\bibitem [{\citenamefont {Patel}(2023)}]{Patel:2022xxu}%
  \BibitemOpen
  \bibfield  {author} {\bibinfo {author} {\bibfnamefont {K.~M.}\ \bibnamefont
  {Patel}},\ }\bibfield  {title} {\emph {\enquote {\bibinfo {title} {{Minimal
  spontaneous CP-violating GUT and predictions for leptonic CP phases}},}\
  }}\href {\doibase 10.1103/PhysRevD.107.075041} {\bibfield  {journal}
  {\bibinfo  {journal} {Phys. Rev. D}\ }\textbf {\bibinfo {volume} {107}},\
  \bibinfo {pages} {075041} (\bibinfo {year} {2023})},\ \Eprint
  {http://arxiv.org/abs/2212.04095} {arXiv:2212.04095 [hep-ph]} \BibitemShut
  {NoStop}%
\bibitem [{\citenamefont {Ipek}\ \emph {et~al.}(2018)\citenamefont {Ipek},
  \citenamefont {Plascencia},\ and\ \citenamefont {Turner}}]{Ipek:2018sai}%
  \BibitemOpen
  \bibfield  {author} {\bibinfo {author} {\bibfnamefont {S.}~\bibnamefont
  {Ipek}}, \bibinfo {author} {\bibfnamefont {A.~D.}\ \bibnamefont
  {Plascencia}}, \ and\ \bibinfo {author} {\bibfnamefont {J.}~\bibnamefont
  {Turner}},\ }\bibfield  {title} {\emph {\enquote {\bibinfo {title}
  {{Assessing Perturbativity and Vacuum Stability in High-Scale
  Leptogenesis}},}\ }}\href {\doibase 10.1007/JHEP12(2018)111} {\bibfield
  {journal} {\bibinfo  {journal} {JHEP}\ }\textbf {\bibinfo {volume} {12}},\
  \bibinfo {pages} {111} (\bibinfo {year} {2018})},\ \Eprint
  {http://arxiv.org/abs/1806.00460} {arXiv:1806.00460 [hep-ph]} \BibitemShut
  {NoStop}%
\bibitem [{\citenamefont {Croon}\ \emph {et~al.}(2019)\citenamefont {Croon},
  \citenamefont {Fernandez}, \citenamefont {McKeen},\ and\ \citenamefont
  {White}}]{Croon:2019dfw}%
  \BibitemOpen
  \bibfield  {author} {\bibinfo {author} {\bibfnamefont {D.}~\bibnamefont
  {Croon}}, \bibinfo {author} {\bibfnamefont {N.}~\bibnamefont {Fernandez}},
  \bibinfo {author} {\bibfnamefont {D.}~\bibnamefont {McKeen}}, \ and\ \bibinfo
  {author} {\bibfnamefont {G.}~\bibnamefont {White}},\ }\bibfield  {title}
  {\emph {\enquote {\bibinfo {title} {{Stability, reheating and
  leptogenesis}},}\ }}\href {\doibase 10.1007/JHEP06(2019)098} {\bibfield
  {journal} {\bibinfo  {journal} {JHEP}\ }\textbf {\bibinfo {volume} {06}},\
  \bibinfo {pages} {098} (\bibinfo {year} {2019})},\ \Eprint
  {http://arxiv.org/abs/1903.08658} {arXiv:1903.08658 [hep-ph]} \BibitemShut
  {NoStop}%
\bibitem [{\citenamefont {Ghoshal}\ \emph
  {et~al.}(2022{\natexlab{a}})\citenamefont {Ghoshal}, \citenamefont {Nanda},\
  and\ \citenamefont {Saha}}]{Ghoshal:2022fud}%
  \BibitemOpen
  \bibfield  {author} {\bibinfo {author} {\bibfnamefont {A.}~\bibnamefont
  {Ghoshal}}, \bibinfo {author} {\bibfnamefont {D.}~\bibnamefont {Nanda}}, \
  and\ \bibinfo {author} {\bibfnamefont {A.~K.}\ \bibnamefont {Saha}},\
  }\bibfield  {title} {\emph {\enquote {\bibinfo {title} {{CMB footprints of
  high scale non-thermal leptogenesis}},}\ }}\href@noop {} {\  (\bibinfo {year}
  {2022}{\natexlab{a}})},\ \Eprint {http://arxiv.org/abs/2210.14176}
  {arXiv:2210.14176 [hep-ph]} \BibitemShut {NoStop}%
\bibitem [{\citenamefont {Dror}\ \emph {et~al.}(2020)\citenamefont {Dror},
  \citenamefont {Hiramatsu}, \citenamefont {Kohri}, \citenamefont {Murayama},\
  and\ \citenamefont {White}}]{Dror:2019syi}%
  \BibitemOpen
  \bibfield  {author} {\bibinfo {author} {\bibfnamefont {J.~A.}\ \bibnamefont
  {Dror}}, \bibinfo {author} {\bibfnamefont {T.}~\bibnamefont {Hiramatsu}},
  \bibinfo {author} {\bibfnamefont {K.}~\bibnamefont {Kohri}}, \bibinfo
  {author} {\bibfnamefont {H.}~\bibnamefont {Murayama}}, \ and\ \bibinfo
  {author} {\bibfnamefont {G.}~\bibnamefont {White}},\ }\bibfield  {title}
  {\emph {\enquote {\bibinfo {title} {{Testing the Seesaw Mechanism and
  Leptogenesis with Gravitational Waves}},}\ }}\href {\doibase
  10.1103/PhysRevLett.124.041804} {\bibfield  {journal} {\bibinfo  {journal}
  {Phys. Rev. Lett.}\ }\textbf {\bibinfo {volume} {124}},\ \bibinfo {pages}
  {041804} (\bibinfo {year} {2020})},\ \Eprint
  {http://arxiv.org/abs/1908.03227} {arXiv:1908.03227 [hep-ph]} \BibitemShut
  {NoStop}%
\bibitem [{\citenamefont {Saad}(2023)}]{Saad:2022mzu}%
  \BibitemOpen
  \bibfield  {author} {\bibinfo {author} {\bibfnamefont {S.}~\bibnamefont
  {Saad}},\ }\bibfield  {title} {\emph {\enquote {\bibinfo {title} {{Probing
  minimal grand unification through gravitational waves, proton decay, and
  fermion masses}},}\ }}\href {\doibase 10.1007/JHEP04(2023)058} {\bibfield
  {journal} {\bibinfo  {journal} {JHEP}\ }\textbf {\bibinfo {volume} {04}},\
  \bibinfo {pages} {058} (\bibinfo {year} {2023})},\ \Eprint
  {http://arxiv.org/abs/2212.05291} {arXiv:2212.05291 [hep-ph]} \BibitemShut
  {NoStop}%
\bibitem [{\citenamefont {Di~Bari}\ \emph {et~al.}(2023)\citenamefont
  {Di~Bari}, \citenamefont {King},\ and\ \citenamefont
  {Rahat}}]{DiBari:2023mwu}%
  \BibitemOpen
  \bibfield  {author} {\bibinfo {author} {\bibfnamefont {P.}~\bibnamefont
  {Di~Bari}}, \bibinfo {author} {\bibfnamefont {S.~F.}\ \bibnamefont {King}}, \
  and\ \bibinfo {author} {\bibfnamefont {M.~H.}\ \bibnamefont {Rahat}},\
  }\bibfield  {title} {\emph {\enquote {\bibinfo {title} {{Gravitational waves
  from phase transitions and cosmic strings in neutrino mass models with
  multiple Majorons}},}\ }}\href@noop {} {\  (\bibinfo {year} {2023})},\
  \Eprint {http://arxiv.org/abs/2306.04680} {arXiv:2306.04680 [hep-ph]}
  \BibitemShut {NoStop}%
\bibitem [{\citenamefont {Fu}\ \emph {et~al.}(2023)\citenamefont {Fu},
  \citenamefont {Ghoshal},\ and\ \citenamefont {King}}]{Fu:2023nrn}%
  \BibitemOpen
  \bibfield  {author} {\bibinfo {author} {\bibfnamefont {B.}~\bibnamefont
  {Fu}}, \bibinfo {author} {\bibfnamefont {A.}~\bibnamefont {Ghoshal}}, \ and\
  \bibinfo {author} {\bibfnamefont {S.}~\bibnamefont {King}},\ }\bibfield
  {title} {\emph {\enquote {\bibinfo {title} {{Cosmic string gravitational
  waves from global $U(1)_{B-L}$ symmetry breaking as a probe of the type I
  seesaw scale}},}\ }}\href@noop {} {\  (\bibinfo {year} {2023})},\ \Eprint
  {http://arxiv.org/abs/2306.07334} {arXiv:2306.07334 [hep-ph]} \BibitemShut
  {NoStop}%
\bibitem [{\citenamefont {Barman}\ \emph
  {et~al.}(2022{\natexlab{a}})\citenamefont {Barman}, \citenamefont {Borah},
  \citenamefont {Dasgupta},\ and\ \citenamefont {Ghoshal}}]{Barman:2022yos}%
  \BibitemOpen
  \bibfield  {author} {\bibinfo {author} {\bibfnamefont {B.}~\bibnamefont
  {Barman}}, \bibinfo {author} {\bibfnamefont {D.}~\bibnamefont {Borah}},
  \bibinfo {author} {\bibfnamefont {A.}~\bibnamefont {Dasgupta}}, \ and\
  \bibinfo {author} {\bibfnamefont {A.}~\bibnamefont {Ghoshal}},\ }\bibfield
  {title} {\emph {\enquote {\bibinfo {title} {{Probing high scale Dirac
  leptogenesis via gravitational waves from domain walls}},}\ }}\href {\doibase
  10.1103/PhysRevD.106.015007} {\bibfield  {journal} {\bibinfo  {journal}
  {Phys. Rev. D}\ }\textbf {\bibinfo {volume} {106}},\ \bibinfo {pages}
  {015007} (\bibinfo {year} {2022}{\natexlab{a}})},\ \Eprint
  {http://arxiv.org/abs/2205.03422} {arXiv:2205.03422 [hep-ph]} \BibitemShut
  {NoStop}%
\bibitem [{\citenamefont {King}\ \emph {et~al.}(2023)\citenamefont {King},
  \citenamefont {Marfatia},\ and\ \citenamefont {Rahat}}]{King:2023cgv}%
  \BibitemOpen
  \bibfield  {author} {\bibinfo {author} {\bibfnamefont {S.~F.}\ \bibnamefont
  {King}}, \bibinfo {author} {\bibfnamefont {D.}~\bibnamefont {Marfatia}}, \
  and\ \bibinfo {author} {\bibfnamefont {M.~H.}\ \bibnamefont {Rahat}},\
  }\bibfield  {title} {\emph {\enquote {\bibinfo {title} {{Towards
  distinguishing Dirac from Majorana neutrino mass with gravitational
  waves}},}\ }}\href@noop {} {\  (\bibinfo {year} {2023})},\ \Eprint
  {http://arxiv.org/abs/2306.05389} {arXiv:2306.05389 [hep-ph]} \BibitemShut
  {NoStop}%
\bibitem [{\citenamefont {Dasgupta}\ \emph {et~al.}(2022)\citenamefont
  {Dasgupta}, \citenamefont {Dev}, \citenamefont {Ghoshal},\ and\ \citenamefont
  {Mazumdar}}]{Dasgupta:2022isg}%
  \BibitemOpen
  \bibfield  {author} {\bibinfo {author} {\bibfnamefont {A.}~\bibnamefont
  {Dasgupta}}, \bibinfo {author} {\bibfnamefont {P.~S.~B.}\ \bibnamefont
  {Dev}}, \bibinfo {author} {\bibfnamefont {A.}~\bibnamefont {Ghoshal}}, \ and\
  \bibinfo {author} {\bibfnamefont {A.}~\bibnamefont {Mazumdar}},\ }\bibfield
  {title} {\emph {\enquote {\bibinfo {title} {{Gravitational wave pathway to
  testable leptogenesis}},}\ }}\href {\doibase 10.1103/PhysRevD.106.075027}
  {\bibfield  {journal} {\bibinfo  {journal} {Phys. Rev. D}\ }\textbf {\bibinfo
  {volume} {106}},\ \bibinfo {pages} {075027} (\bibinfo {year} {2022})},\
  \Eprint {http://arxiv.org/abs/2206.07032} {arXiv:2206.07032 [hep-ph]}
  \BibitemShut {NoStop}%
\bibitem [{\citenamefont {Borah}\ \emph {et~al.}(2022)\citenamefont {Borah},
  \citenamefont {Dasgupta},\ and\ \citenamefont {Saha}}]{Borah:2022cdx}%
  \BibitemOpen
  \bibfield  {author} {\bibinfo {author} {\bibfnamefont {D.}~\bibnamefont
  {Borah}}, \bibinfo {author} {\bibfnamefont {A.}~\bibnamefont {Dasgupta}}, \
  and\ \bibinfo {author} {\bibfnamefont {I.}~\bibnamefont {Saha}},\ }\bibfield
  {title} {\emph {\enquote {\bibinfo {title} {{Leptogenesis and dark matter
  through relativistic bubble walls with observable gravitational waves}},}\
  }}\href {\doibase 10.1007/JHEP11(2022)136} {\bibfield  {journal} {\bibinfo
  {journal} {JHEP}\ }\textbf {\bibinfo {volume} {11}},\ \bibinfo {pages} {136}
  (\bibinfo {year} {2022})},\ \Eprint {http://arxiv.org/abs/2207.14226}
  {arXiv:2207.14226 [hep-ph]} \BibitemShut {NoStop}%
\bibitem [{\citenamefont {Dunsky}\ \emph {et~al.}(2022)\citenamefont {Dunsky},
  \citenamefont {Ghoshal}, \citenamefont {Murayama}, \citenamefont
  {Sakakihara},\ and\ \citenamefont {White}}]{Dunsky:2021tih}%
  \BibitemOpen
  \bibfield  {author} {\bibinfo {author} {\bibfnamefont {D.~I.}\ \bibnamefont
  {Dunsky}}, \bibinfo {author} {\bibfnamefont {A.}~\bibnamefont {Ghoshal}},
  \bibinfo {author} {\bibfnamefont {H.}~\bibnamefont {Murayama}}, \bibinfo
  {author} {\bibfnamefont {Y.}~\bibnamefont {Sakakihara}}, \ and\ \bibinfo
  {author} {\bibfnamefont {G.}~\bibnamefont {White}},\ }\bibfield  {title}
  {\emph {\enquote {\bibinfo {title} {{GUTs, hybrid topological defects, and
  gravitational waves}},}\ }}\href {\doibase 10.1103/PhysRevD.106.075030}
  {\bibfield  {journal} {\bibinfo  {journal} {Phys. Rev. D}\ }\textbf {\bibinfo
  {volume} {106}},\ \bibinfo {pages} {075030} (\bibinfo {year} {2022})},\
  \Eprint {http://arxiv.org/abs/2111.08750} {arXiv:2111.08750 [hep-ph]}
  \BibitemShut {NoStop}%
\bibitem [{\citenamefont {Ghoshal}\ \emph
  {et~al.}(2022{\natexlab{b}})\citenamefont {Ghoshal}, \citenamefont
  {Samanta},\ and\ \citenamefont {White}}]{Ghoshal:2022kqp}%
  \BibitemOpen
  \bibfield  {author} {\bibinfo {author} {\bibfnamefont {A.}~\bibnamefont
  {Ghoshal}}, \bibinfo {author} {\bibfnamefont {R.}~\bibnamefont {Samanta}}, \
  and\ \bibinfo {author} {\bibfnamefont {G.}~\bibnamefont {White}},\ }\bibfield
   {title} {\emph {\enquote {\bibinfo {title} {{Bremsstrahlung High-frequency
  Gravitational Wave Signatures of High-scale Non-thermal Leptogenesis}},}\
  }}\href@noop {} {\  (\bibinfo {year} {2022}{\natexlab{b}})},\ \Eprint
  {http://arxiv.org/abs/2211.10433} {arXiv:2211.10433 [hep-ph]} \BibitemShut
  {NoStop}%
\bibitem [{\citenamefont {Berbig}\ and\ \citenamefont
  {Ghoshal}(2023)}]{Berbig:2023yyy}%
  \BibitemOpen
  \bibfield  {author} {\bibinfo {author} {\bibfnamefont {M.}~\bibnamefont
  {Berbig}}\ and\ \bibinfo {author} {\bibfnamefont {A.}~\bibnamefont
  {Ghoshal}},\ }\bibfield  {title} {\emph {\enquote {\bibinfo {title} {{Impact
  of high-scale Seesaw and Leptogenesis on inflationary tensor perturbations as
  detectable gravitational waves}},}\ }}\href {\doibase
  10.1007/JHEP05(2023)172} {\bibfield  {journal} {\bibinfo  {journal} {JHEP}\
  }\textbf {\bibinfo {volume} {05}},\ \bibinfo {pages} {172} (\bibinfo {year}
  {2023})},\ \Eprint {http://arxiv.org/abs/2301.05672} {arXiv:2301.05672
  [hep-ph]} \BibitemShut {NoStop}%
\bibitem [{\citenamefont {Perez-Gonzalez}\ and\ \citenamefont
  {Turner}(2021)}]{Perez-Gonzalez:2020vnz}%
  \BibitemOpen
  \bibfield  {author} {\bibinfo {author} {\bibfnamefont {Y.~F.}\ \bibnamefont
  {Perez-Gonzalez}}\ and\ \bibinfo {author} {\bibfnamefont {J.}~\bibnamefont
  {Turner}},\ }\bibfield  {title} {\emph {\enquote {\bibinfo {title}
  {{Assessing the tension between a black hole dominated early universe and
  leptogenesis}},}\ }}\href {\doibase 10.1103/PhysRevD.104.103021} {\bibfield
  {journal} {\bibinfo  {journal} {Phys. Rev. D}\ }\textbf {\bibinfo {volume}
  {104}},\ \bibinfo {pages} {103021} (\bibinfo {year} {2021})},\ \Eprint
  {http://arxiv.org/abs/2010.03565} {arXiv:2010.03565 [hep-ph]} \BibitemShut
  {NoStop}%
\bibitem [{\citenamefont {Datta}\ \emph {et~al.}(2021)\citenamefont {Datta},
  \citenamefont {Ghosal},\ and\ \citenamefont {Samanta}}]{Datta:2020bht}%
  \BibitemOpen
  \bibfield  {author} {\bibinfo {author} {\bibfnamefont {S.}~\bibnamefont
  {Datta}}, \bibinfo {author} {\bibfnamefont {A.}~\bibnamefont {Ghosal}}, \
  and\ \bibinfo {author} {\bibfnamefont {R.}~\bibnamefont {Samanta}},\
  }\bibfield  {title} {\emph {\enquote {\bibinfo {title} {{Baryogenesis from
  ultralight primordial black holes and strong gravitational waves from cosmic
  strings}},}\ }}\href {\doibase 10.1088/1475-7516/2021/08/021} {\bibfield
  {journal} {\bibinfo  {journal} {JCAP}\ }\textbf {\bibinfo {volume} {08}},\
  \bibinfo {pages} {021} (\bibinfo {year} {2021})},\ \Eprint
  {http://arxiv.org/abs/2012.14981} {arXiv:2012.14981 [hep-ph]} \BibitemShut
  {NoStop}%
\bibitem [{\citenamefont {Jyoti~Das}\ \emph {et~al.}(2021)\citenamefont
  {Jyoti~Das}, \citenamefont {Mahanta},\ and\ \citenamefont
  {Borah}}]{JyotiDas:2021shi}%
  \BibitemOpen
  \bibfield  {author} {\bibinfo {author} {\bibfnamefont {S.}~\bibnamefont
  {Jyoti~Das}}, \bibinfo {author} {\bibfnamefont {D.}~\bibnamefont {Mahanta}},
  \ and\ \bibinfo {author} {\bibfnamefont {D.}~\bibnamefont {Borah}},\
  }\bibfield  {title} {\emph {\enquote {\bibinfo {title} {{Low scale
  leptogenesis and dark matter in the presence of primordial black holes}},}\
  }}\href {\doibase 10.1088/1475-7516/2021/11/019} {\bibfield  {journal}
  {\bibinfo  {journal} {JCAP}\ }\textbf {\bibinfo {volume} {11}},\ \bibinfo
  {pages} {019} (\bibinfo {year} {2021})},\ \Eprint
  {http://arxiv.org/abs/2104.14496} {arXiv:2104.14496 [hep-ph]} \BibitemShut
  {NoStop}%
\bibitem [{\citenamefont {Barman}\ \emph
  {et~al.}(2022{\natexlab{b}})\citenamefont {Barman}, \citenamefont {Borah},
  \citenamefont {Das},\ and\ \citenamefont {Roshan}}]{Barman:2021ost}%
  \BibitemOpen
  \bibfield  {author} {\bibinfo {author} {\bibfnamefont {B.}~\bibnamefont
  {Barman}}, \bibinfo {author} {\bibfnamefont {D.}~\bibnamefont {Borah}},
  \bibinfo {author} {\bibfnamefont {S.~J.}\ \bibnamefont {Das}}, \ and\
  \bibinfo {author} {\bibfnamefont {R.}~\bibnamefont {Roshan}},\ }\bibfield
  {title} {\emph {\enquote {\bibinfo {title} {{Non-thermal origin of asymmetric
  dark matter from inflaton and primordial black holes}},}\ }}\href {\doibase
  10.1088/1475-7516/2022/03/031} {\bibfield  {journal} {\bibinfo  {journal}
  {JCAP}\ }\textbf {\bibinfo {volume} {03}},\ \bibinfo {pages} {031} (\bibinfo
  {year} {2022}{\natexlab{b}})},\ \Eprint {http://arxiv.org/abs/2111.08034}
  {arXiv:2111.08034 [hep-ph]} \BibitemShut {NoStop}%
\bibitem [{\citenamefont {Bernal}\ \emph {et~al.}(2022)\citenamefont {Bernal},
  \citenamefont {Fong}, \citenamefont {Perez-Gonzalez},\ and\ \citenamefont
  {Turner}}]{Bernal:2022pue}%
  \BibitemOpen
  \bibfield  {author} {\bibinfo {author} {\bibfnamefont {N.}~\bibnamefont
  {Bernal}}, \bibinfo {author} {\bibfnamefont {C.~S.}\ \bibnamefont {Fong}},
  \bibinfo {author} {\bibfnamefont {Y.~F.}\ \bibnamefont {Perez-Gonzalez}}, \
  and\ \bibinfo {author} {\bibfnamefont {J.}~\bibnamefont {Turner}},\
  }\bibfield  {title} {\emph {\enquote {\bibinfo {title} {{Rescuing high-scale
  leptogenesis using primordial black holes}},}\ }}\href {\doibase
  10.1103/PhysRevD.106.035019} {\bibfield  {journal} {\bibinfo  {journal}
  {Phys. Rev. D}\ }\textbf {\bibinfo {volume} {106}},\ \bibinfo {pages}
  {035019} (\bibinfo {year} {2022})},\ \Eprint
  {http://arxiv.org/abs/2203.08823} {arXiv:2203.08823 [hep-ph]} \BibitemShut
  {NoStop}%
\bibitem [{\citenamefont {Bhaumik}\ \emph {et~al.}(2022)\citenamefont
  {Bhaumik}, \citenamefont {Ghoshal},\ and\ \citenamefont
  {Lewicki}}]{Bhaumik:2022pil}%
  \BibitemOpen
  \bibfield  {author} {\bibinfo {author} {\bibfnamefont {N.}~\bibnamefont
  {Bhaumik}}, \bibinfo {author} {\bibfnamefont {A.}~\bibnamefont {Ghoshal}}, \
  and\ \bibinfo {author} {\bibfnamefont {M.}~\bibnamefont {Lewicki}},\
  }\bibfield  {title} {\emph {\enquote {\bibinfo {title} {{Doubly peaked
  induced stochastic gravitational wave background: testing baryogenesis from
  primordial black holes}},}\ }}\href {\doibase 10.1007/JHEP07(2022)130}
  {\bibfield  {journal} {\bibinfo  {journal} {JHEP}\ }\textbf {\bibinfo
  {volume} {07}},\ \bibinfo {pages} {130} (\bibinfo {year} {2022})},\ \Eprint
  {http://arxiv.org/abs/2205.06260} {arXiv:2205.06260 [astro-ph.CO]}
  \BibitemShut {NoStop}%
\bibitem [{\citenamefont {Cui}\ and\ \citenamefont
  {Xianyu}(2022)}]{Cui:2021iie}%
  \BibitemOpen
  \bibfield  {author} {\bibinfo {author} {\bibfnamefont {Y.}~\bibnamefont
  {Cui}}\ and\ \bibinfo {author} {\bibfnamefont {Z.-Z.}\ \bibnamefont
  {Xianyu}},\ }\bibfield  {title} {\emph {\enquote {\bibinfo {title} {{Probing
  Leptogenesis with the Cosmological Collider}},}\ }}\href {\doibase
  10.1103/PhysRevLett.129.111301} {\bibfield  {journal} {\bibinfo  {journal}
  {Phys. Rev. Lett.}\ }\textbf {\bibinfo {volume} {129}},\ \bibinfo {pages}
  {111301} (\bibinfo {year} {2022})},\ \Eprint
  {http://arxiv.org/abs/2112.10793} {arXiv:2112.10793 [hep-ph]} \BibitemShut
  {NoStop}%
\bibitem [{\citenamefont {Chikashige}\ \emph {et~al.}(1981)\citenamefont
  {Chikashige}, \citenamefont {Mohapatra},\ and\ \citenamefont
  {Peccei}}]{Chikashige:1980ui}%
  \BibitemOpen
  \bibfield  {author} {\bibinfo {author} {\bibfnamefont {Y.}~\bibnamefont
  {Chikashige}}, \bibinfo {author} {\bibfnamefont {R.~N.}\ \bibnamefont
  {Mohapatra}}, \ and\ \bibinfo {author} {\bibfnamefont {R.~D.}\ \bibnamefont
  {Peccei}},\ }\bibfield  {title} {\emph {\enquote {\bibinfo {title} {{Are
  There Real Goldstone Bosons Associated with Broken Lepton Number?}}}\ }}\href
  {\doibase 10.1016/0370-2693(81)90011-3} {\bibfield  {journal} {\bibinfo
  {journal} {Phys. Lett. B}\ }\textbf {\bibinfo {volume} {98}},\ \bibinfo
  {pages} {265} (\bibinfo {year} {1981})}\BibitemShut {NoStop}%
\bibitem [{\citenamefont {Schechter}\ and\ \citenamefont
  {Valle}(1982)}]{Schechter:1981cv}%
  \BibitemOpen
  \bibfield  {author} {\bibinfo {author} {\bibfnamefont {J.}~\bibnamefont
  {Schechter}}\ and\ \bibinfo {author} {\bibfnamefont {J.~W.~F.}\ \bibnamefont
  {Valle}},\ }\bibfield  {title} {\emph {\enquote {\bibinfo {title} {{Neutrino
  Decay and Spontaneous Violation of Lepton Number}},}\ }}\href {\doibase
  10.1103/PhysRevD.25.774} {\bibfield  {journal} {\bibinfo  {journal} {Phys.
  Rev. D}\ }\textbf {\bibinfo {volume} {25}},\ \bibinfo {pages} {774} (\bibinfo
  {year} {1982})}\BibitemShut {NoStop}%
\bibitem [{\citenamefont {Linde}\ and\ \citenamefont
  {Mukhanov}(1997)}]{Linde:1996gt}%
  \BibitemOpen
  \bibfield  {author} {\bibinfo {author} {\bibfnamefont {A.~D.}\ \bibnamefont
  {Linde}}\ and\ \bibinfo {author} {\bibfnamefont {V.~F.}\ \bibnamefont
  {Mukhanov}},\ }\bibfield  {title} {\emph {\enquote {\bibinfo {title}
  {{Nongaussian isocurvature perturbations from inflation}},}\ }}\href
  {\doibase 10.1103/PhysRevD.56.R535} {\bibfield  {journal} {\bibinfo
  {journal} {Phys. Rev. D}\ }\textbf {\bibinfo {volume} {56}},\ \bibinfo
  {pages} {R535} (\bibinfo {year} {1997})},\ \Eprint
  {http://arxiv.org/abs/astro-ph/9610219} {arXiv:astro-ph/9610219} \BibitemShut
  {NoStop}%
\bibitem [{\citenamefont {Enqvist}\ and\ \citenamefont
  {Sloth}(2002)}]{Enqvist:2001zp}%
  \BibitemOpen
  \bibfield  {author} {\bibinfo {author} {\bibfnamefont {K.}~\bibnamefont
  {Enqvist}}\ and\ \bibinfo {author} {\bibfnamefont {M.~S.}\ \bibnamefont
  {Sloth}},\ }\bibfield  {title} {\emph {\enquote {\bibinfo {title} {{Adiabatic
  CMB perturbations in pre - big bang string cosmology}},}\ }}\href {\doibase
  10.1016/S0550-3213(02)00043-3} {\bibfield  {journal} {\bibinfo  {journal}
  {Nucl. Phys. B}\ }\textbf {\bibinfo {volume} {626}},\ \bibinfo {pages} {395}
  (\bibinfo {year} {2002})},\ \Eprint {http://arxiv.org/abs/hep-ph/0109214}
  {arXiv:hep-ph/0109214} \BibitemShut {NoStop}%
\bibitem [{\citenamefont {Lyth}\ and\ \citenamefont
  {Wands}(2002)}]{Lyth:2001nq}%
  \BibitemOpen
  \bibfield  {author} {\bibinfo {author} {\bibfnamefont {D.~H.}\ \bibnamefont
  {Lyth}}\ and\ \bibinfo {author} {\bibfnamefont {D.}~\bibnamefont {Wands}},\
  }\bibfield  {title} {\emph {\enquote {\bibinfo {title} {{Generating the
  curvature perturbation without an inflaton}},}\ }}\href {\doibase
  10.1016/S0370-2693(01)01366-1} {\bibfield  {journal} {\bibinfo  {journal}
  {Phys. Lett. B}\ }\textbf {\bibinfo {volume} {524}},\ \bibinfo {pages} {5}
  (\bibinfo {year} {2002})},\ \Eprint {http://arxiv.org/abs/hep-ph/0110002}
  {arXiv:hep-ph/0110002} \BibitemShut {NoStop}%
\bibitem [{\citenamefont {Moroi}\ and\ \citenamefont
  {Takahashi}(2001)}]{Moroi:2001ct}%
  \BibitemOpen
  \bibfield  {author} {\bibinfo {author} {\bibfnamefont {T.}~\bibnamefont
  {Moroi}}\ and\ \bibinfo {author} {\bibfnamefont {T.}~\bibnamefont
  {Takahashi}},\ }\bibfield  {title} {\emph {\enquote {\bibinfo {title}
  {{Effects of cosmological moduli fields on cosmic microwave background}},}\
  }}\href {\doibase 10.1016/S0370-2693(01)01295-3} {\bibfield  {journal}
  {\bibinfo  {journal} {Phys. Lett. B}\ }\textbf {\bibinfo {volume} {522}},\
  \bibinfo {pages} {215} (\bibinfo {year} {2001})},\ \bibinfo {note} {[Erratum:
  Phys.Lett.B 539, 303--303 (2002)]},\ \Eprint
  {http://arxiv.org/abs/hep-ph/0110096} {arXiv:hep-ph/0110096} \BibitemShut
  {NoStop}%
\bibitem [{\citenamefont {Kobayashi}(2020)}]{Kobayashi:2020xhm}%
  \BibitemOpen
  \bibfield  {author} {\bibinfo {author} {\bibfnamefont {T.}~\bibnamefont
  {Kobayashi}},\ }\bibfield  {title} {\emph {\enquote {\bibinfo {title}
  {{Axionlike Origin of the Primordial Density Perturbation}},}\ }}\href
  {\doibase 10.1103/PhysRevLett.125.011302} {\bibfield  {journal} {\bibinfo
  {journal} {Phys. Rev. Lett.}\ }\textbf {\bibinfo {volume} {125}},\ \bibinfo
  {pages} {011302} (\bibinfo {year} {2020})},\ \Eprint
  {http://arxiv.org/abs/2005.01741} {arXiv:2005.01741 [astro-ph.CO]}
  \BibitemShut {NoStop}%
\bibitem [{\citenamefont {Sasaki}\ \emph {et~al.}(2006)\citenamefont {Sasaki},
  \citenamefont {Valiviita},\ and\ \citenamefont {Wands}}]{Sasaki:2006kq}%
  \BibitemOpen
  \bibfield  {author} {\bibinfo {author} {\bibfnamefont {M.}~\bibnamefont
  {Sasaki}}, \bibinfo {author} {\bibfnamefont {J.}~\bibnamefont {Valiviita}}, \
  and\ \bibinfo {author} {\bibfnamefont {D.}~\bibnamefont {Wands}},\ }\bibfield
   {title} {\emph {\enquote {\bibinfo {title} {{Non-Gaussianity of the
  primordial perturbation in the curvaton model}},}\ }}\href {\doibase
  10.1103/PhysRevD.74.103003} {\bibfield  {journal} {\bibinfo  {journal} {Phys.
  Rev. D}\ }\textbf {\bibinfo {volume} {74}},\ \bibinfo {pages} {103003}
  (\bibinfo {year} {2006})},\ \Eprint {http://arxiv.org/abs/astro-ph/0607627}
  {arXiv:astro-ph/0607627} \BibitemShut {NoStop}%
\bibitem [{\citenamefont {Enqvist}\ and\ \citenamefont
  {Takahashi}(2008)}]{Enqvist:2008gk}%
  \BibitemOpen
  \bibfield  {author} {\bibinfo {author} {\bibfnamefont {K.}~\bibnamefont
  {Enqvist}}\ and\ \bibinfo {author} {\bibfnamefont {T.}~\bibnamefont
  {Takahashi}},\ }\bibfield  {title} {\emph {\enquote {\bibinfo {title}
  {{Signatures of Non-Gaussianity in the Curvaton Model}},}\ }}\href {\doibase
  10.1088/1475-7516/2008/09/012} {\bibfield  {journal} {\bibinfo  {journal}
  {JCAP}\ }\textbf {\bibinfo {volume} {09}},\ \bibinfo {pages} {012} (\bibinfo
  {year} {2008})},\ \Eprint {http://arxiv.org/abs/0807.3069} {arXiv:0807.3069
  [astro-ph]} \BibitemShut {NoStop}%
\bibitem [{\citenamefont {Chingangbam}\ and\ \citenamefont
  {Huang}(2009)}]{Chingangbam:2009xi}%
  \BibitemOpen
  \bibfield  {author} {\bibinfo {author} {\bibfnamefont {P.}~\bibnamefont
  {Chingangbam}}\ and\ \bibinfo {author} {\bibfnamefont {Q.-G.}\ \bibnamefont
  {Huang}},\ }\bibfield  {title} {\emph {\enquote {\bibinfo {title} {{The
  Curvature Perturbation in the Axion-type Curvaton Model}},}\ }}\href
  {\doibase 10.1088/1475-7516/2009/04/031} {\bibfield  {journal} {\bibinfo
  {journal} {JCAP}\ }\textbf {\bibinfo {volume} {04}},\ \bibinfo {pages} {031}
  (\bibinfo {year} {2009})},\ \Eprint {http://arxiv.org/abs/0902.2619}
  {arXiv:0902.2619 [astro-ph.CO]} \BibitemShut {NoStop}%
\bibitem [{\citenamefont {Enqvist}\ \emph {et~al.}(2009)\citenamefont
  {Enqvist}, \citenamefont {Nurmi}, \citenamefont {Rigopoulos}, \citenamefont
  {Taanila},\ and\ \citenamefont {Takahashi}}]{Enqvist:2009zf}%
  \BibitemOpen
  \bibfield  {author} {\bibinfo {author} {\bibfnamefont {K.}~\bibnamefont
  {Enqvist}}, \bibinfo {author} {\bibfnamefont {S.}~\bibnamefont {Nurmi}},
  \bibinfo {author} {\bibfnamefont {G.}~\bibnamefont {Rigopoulos}}, \bibinfo
  {author} {\bibfnamefont {O.}~\bibnamefont {Taanila}}, \ and\ \bibinfo
  {author} {\bibfnamefont {T.}~\bibnamefont {Takahashi}},\ }\bibfield  {title}
  {\emph {\enquote {\bibinfo {title} {{The Subdominant Curvaton}},}\ }}\href
  {\doibase 10.1088/1475-7516/2009/11/003} {\bibfield  {journal} {\bibinfo
  {journal} {JCAP}\ }\textbf {\bibinfo {volume} {11}},\ \bibinfo {pages} {003}
  (\bibinfo {year} {2009})},\ \Eprint {http://arxiv.org/abs/0906.3126}
  {arXiv:0906.3126 [astro-ph.CO]} \BibitemShut {NoStop}%
\bibitem [{\citenamefont {Enqvist}\ \emph {et~al.}(2010)\citenamefont
  {Enqvist}, \citenamefont {Nurmi}, \citenamefont {Taanila},\ and\
  \citenamefont {Takahashi}}]{Enqvist:2009ww}%
  \BibitemOpen
  \bibfield  {author} {\bibinfo {author} {\bibfnamefont {K.}~\bibnamefont
  {Enqvist}}, \bibinfo {author} {\bibfnamefont {S.}~\bibnamefont {Nurmi}},
  \bibinfo {author} {\bibfnamefont {O.}~\bibnamefont {Taanila}}, \ and\
  \bibinfo {author} {\bibfnamefont {T.}~\bibnamefont {Takahashi}},\ }\bibfield
  {title} {\emph {\enquote {\bibinfo {title} {{Non-Gaussian Fingerprints of
  Self-Interacting Curvaton}},}\ }}\href {\doibase
  10.1088/1475-7516/2010/04/009} {\bibfield  {journal} {\bibinfo  {journal}
  {JCAP}\ }\textbf {\bibinfo {volume} {04}},\ \bibinfo {pages} {009} (\bibinfo
  {year} {2010})},\ \Eprint {http://arxiv.org/abs/0912.4657} {arXiv:0912.4657
  [astro-ph.CO]} \BibitemShut {NoStop}%
\bibitem [{\citenamefont {Mazumdar}\ and\ \citenamefont
  {Rocher}(2011)}]{Mazumdar:2010sa}%
  \BibitemOpen
  \bibfield  {author} {\bibinfo {author} {\bibfnamefont {A.}~\bibnamefont
  {Mazumdar}}\ and\ \bibinfo {author} {\bibfnamefont {J.}~\bibnamefont
  {Rocher}},\ }\bibfield  {title} {\emph {\enquote {\bibinfo {title} {{Particle
  physics models of inflation and curvaton scenarios}},}\ }}\href {\doibase
  10.1016/j.physrep.2010.08.001} {\bibfield  {journal} {\bibinfo  {journal}
  {Phys. Rept.}\ }\textbf {\bibinfo {volume} {497}},\ \bibinfo {pages} {85}
  (\bibinfo {year} {2011})},\ \Eprint {http://arxiv.org/abs/1001.0993}
  {arXiv:1001.0993 [hep-ph]} \BibitemShut {NoStop}%
\bibitem [{\citenamefont {Kawasaki}\ \emph {et~al.}(2011)\citenamefont
  {Kawasaki}, \citenamefont {Kobayashi},\ and\ \citenamefont
  {Takahashi}}]{Kawasaki:2011pd}%
  \BibitemOpen
  \bibfield  {author} {\bibinfo {author} {\bibfnamefont {M.}~\bibnamefont
  {Kawasaki}}, \bibinfo {author} {\bibfnamefont {T.}~\bibnamefont {Kobayashi}},
  \ and\ \bibinfo {author} {\bibfnamefont {F.}~\bibnamefont {Takahashi}},\
  }\bibfield  {title} {\emph {\enquote {\bibinfo {title} {{Non-Gaussianity from
  Curvatons Revisited}},}\ }}\href {\doibase 10.1103/PhysRevD.84.123506}
  {\bibfield  {journal} {\bibinfo  {journal} {Phys. Rev. D}\ }\textbf {\bibinfo
  {volume} {84}},\ \bibinfo {pages} {123506} (\bibinfo {year} {2011})},\
  \Eprint {http://arxiv.org/abs/1107.6011} {arXiv:1107.6011 [astro-ph.CO]}
  \BibitemShut {NoStop}%
\bibitem [{\citenamefont {Kawasaki}\ \emph {et~al.}(2013)\citenamefont
  {Kawasaki}, \citenamefont {Kobayashi},\ and\ \citenamefont
  {Takahashi}}]{Kawasaki:2012gg}%
  \BibitemOpen
  \bibfield  {author} {\bibinfo {author} {\bibfnamefont {M.}~\bibnamefont
  {Kawasaki}}, \bibinfo {author} {\bibfnamefont {T.}~\bibnamefont {Kobayashi}},
  \ and\ \bibinfo {author} {\bibfnamefont {F.}~\bibnamefont {Takahashi}},\
  }\bibfield  {title} {\emph {\enquote {\bibinfo {title} {{Non-Gaussianity from
  Axionic Curvaton}},}\ }}\href {\doibase 10.1088/1475-7516/2013/03/016}
  {\bibfield  {journal} {\bibinfo  {journal} {JCAP}\ }\textbf {\bibinfo
  {volume} {03}},\ \bibinfo {pages} {016} (\bibinfo {year} {2013})},\ \Eprint
  {http://arxiv.org/abs/1210.6595} {arXiv:1210.6595 [astro-ph.CO]} \BibitemShut
  {NoStop}%
\bibitem [{\citenamefont {Byrnes}\ \emph {et~al.}(2014)\citenamefont {Byrnes},
  \citenamefont {Cort\^es},\ and\ \citenamefont {Liddle}}]{Byrnes:2014xua}%
  \BibitemOpen
  \bibfield  {author} {\bibinfo {author} {\bibfnamefont {C.~T.}\ \bibnamefont
  {Byrnes}}, \bibinfo {author} {\bibfnamefont {M.}~\bibnamefont {Cort\^es}}, \
  and\ \bibinfo {author} {\bibfnamefont {A.~R.}\ \bibnamefont {Liddle}},\
  }\bibfield  {title} {\emph {\enquote {\bibinfo {title} {{Comprehensive
  analysis of the simplest curvaton model}},}\ }}\href {\doibase
  10.1103/PhysRevD.90.023523} {\bibfield  {journal} {\bibinfo  {journal} {Phys.
  Rev. D}\ }\textbf {\bibinfo {volume} {90}},\ \bibinfo {pages} {023523}
  (\bibinfo {year} {2014})},\ \Eprint {http://arxiv.org/abs/1403.4591}
  {arXiv:1403.4591 [astro-ph.CO]} \BibitemShut {NoStop}%
\bibitem [{\citenamefont {Takahashi}\ \emph {et~al.}(2022)\citenamefont
  {Takahashi}, \citenamefont {Yamada},\ and\ \citenamefont
  {Yokoyama}}]{Takahashi:2022bqc}%
  \BibitemOpen
  \bibfield  {author} {\bibinfo {author} {\bibfnamefont {T.}~\bibnamefont
  {Takahashi}}, \bibinfo {author} {\bibfnamefont {T.}~\bibnamefont {Yamada}}, \
  and\ \bibinfo {author} {\bibfnamefont {S.}~\bibnamefont {Yokoyama}},\
  }\bibfield  {title} {\emph {\enquote {\bibinfo {title} {{Sneutrinos as two
  inflatons and curvaton and leptogenesis}},}\ }}\href {\doibase
  10.1088/1475-7516/2022/11/021} {\bibfield  {journal} {\bibinfo  {journal}
  {JCAP}\ }\textbf {\bibinfo {volume} {11}},\ \bibinfo {pages} {021} (\bibinfo
  {year} {2022})},\ \Eprint {http://arxiv.org/abs/2208.08296} {arXiv:2208.08296
  [hep-ph]} \BibitemShut {NoStop}%
\bibitem [{\citenamefont {Ghoshal}\ and\ \citenamefont
  {Naskar}(2023)}]{Ghoshal:2023lly}%
  \BibitemOpen
  \bibfield  {author} {\bibinfo {author} {\bibfnamefont {A.}~\bibnamefont
  {Ghoshal}}\ and\ \bibinfo {author} {\bibfnamefont {A.}~\bibnamefont
  {Naskar}},\ }\bibfield  {title} {\emph {\enquote {\bibinfo {title}
  {{Generalising Axion-like particle as the curvaton: sourcing primordial
  density perturbation and non-Gaussianities}},}\ }}\href@noop {} {\  (\bibinfo
  {year} {2023})},\ \Eprint {http://arxiv.org/abs/2302.00668} {arXiv:2302.00668
  [hep-ph]} \BibitemShut {NoStop}%
\bibitem [{\citenamefont {Guth}(1981)}]{Guth:1980zm}%
  \BibitemOpen
  \bibfield  {author} {\bibinfo {author} {\bibfnamefont {A.~H.}\ \bibnamefont
  {Guth}},\ }\bibfield  {title} {\emph {\enquote {\bibinfo {title} {{The
  Inflationary Universe: A Possible Solution to the Horizon and Flatness
  Problems}},}\ }}\href {\doibase 10.1103/PhysRevD.23.347} {\bibfield
  {journal} {\bibinfo  {journal} {Phys. Rev. D}\ }\textbf {\bibinfo {volume}
  {23}},\ \bibinfo {pages} {347} (\bibinfo {year} {1981})}\BibitemShut
  {NoStop}%
\bibitem [{\citenamefont {Starobinsky}(1980)}]{Starobinsky:1980te}%
  \BibitemOpen
  \bibfield  {author} {\bibinfo {author} {\bibfnamefont {A.~A.}\ \bibnamefont
  {Starobinsky}},\ }\bibfield  {title} {\emph {\enquote {\bibinfo {title} {{A
  New Type of Isotropic Cosmological Models Without Singularity}},}\ }}\href
  {\doibase 10.1016/0370-2693(80)90670-X} {\bibfield  {journal} {\bibinfo
  {journal} {Phys. Lett. B}\ }\textbf {\bibinfo {volume} {91}},\ \bibinfo
  {pages} {99} (\bibinfo {year} {1980})}\BibitemShut {NoStop}%
\bibitem [{\citenamefont {Linde}(1982)}]{Linde:1981mu}%
  \BibitemOpen
  \bibfield  {author} {\bibinfo {author} {\bibfnamefont {A.~D.}\ \bibnamefont
  {Linde}},\ }\bibfield  {title} {\emph {\enquote {\bibinfo {title} {{A New
  Inflationary Universe Scenario: A Possible Solution of the Horizon, Flatness,
  Homogeneity, Isotropy and Primordial Monopole Problems}},}\ }}\href {\doibase
  10.1016/0370-2693(82)91219-9} {\bibfield  {journal} {\bibinfo  {journal}
  {Phys. Lett. B}\ }\textbf {\bibinfo {volume} {108}},\ \bibinfo {pages} {389}
  (\bibinfo {year} {1982})}\BibitemShut {NoStop}%
\bibitem [{\citenamefont {Albrecht}\ and\ \citenamefont
  {Steinhardt}(1982)}]{Albrecht:1982wi}%
  \BibitemOpen
  \bibfield  {author} {\bibinfo {author} {\bibfnamefont {A.}~\bibnamefont
  {Albrecht}}\ and\ \bibinfo {author} {\bibfnamefont {P.~J.}\ \bibnamefont
  {Steinhardt}},\ }\bibfield  {title} {\emph {\enquote {\bibinfo {title}
  {{Cosmology for Grand Unified Theories with Radiatively Induced Symmetry
  Breaking}},}\ }}\href {\doibase 10.1103/PhysRevLett.48.1220} {\bibfield
  {journal} {\bibinfo  {journal} {Phys. Rev. Lett.}\ }\textbf {\bibinfo
  {volume} {48}},\ \bibinfo {pages} {1220} (\bibinfo {year}
  {1982})}\BibitemShut {NoStop}%
\bibitem [{\citenamefont {Linde}(1983)}]{Linde:1983gd}%
  \BibitemOpen
  \bibfield  {author} {\bibinfo {author} {\bibfnamefont {A.~D.}\ \bibnamefont
  {Linde}},\ }\bibfield  {title} {\emph {\enquote {\bibinfo {title} {{Chaotic
  Inflation}},}\ }}\href {\doibase 10.1016/0370-2693(83)90837-7} {\bibfield
  {journal} {\bibinfo  {journal} {Phys. Lett. B}\ }\textbf {\bibinfo {volume}
  {129}},\ \bibinfo {pages} {177} (\bibinfo {year} {1983})}\BibitemShut
  {NoStop}%
\bibitem [{\citenamefont {M\"unchmeyer}\ \emph {et~al.}(2019)\citenamefont
  {M\"unchmeyer}, \citenamefont {Madhavacheril}, \citenamefont {Ferraro},
  \citenamefont {Johnson},\ and\ \citenamefont {Smith}}]{Munchmeyer:2018eey}%
  \BibitemOpen
  \bibfield  {author} {\bibinfo {author} {\bibfnamefont {M.}~\bibnamefont
  {M\"unchmeyer}}, \bibinfo {author} {\bibfnamefont {M.~S.}\ \bibnamefont
  {Madhavacheril}}, \bibinfo {author} {\bibfnamefont {S.}~\bibnamefont
  {Ferraro}}, \bibinfo {author} {\bibfnamefont {M.~C.}\ \bibnamefont
  {Johnson}}, \ and\ \bibinfo {author} {\bibfnamefont {K.~M.}\ \bibnamefont
  {Smith}},\ }\bibfield  {title} {\emph {\enquote {\bibinfo {title}
  {{Constraining local non-Gaussianities with kinetic
  Sunyaev-Zel\textquoteright{}dovich tomography}},}\ }}\href {\doibase
  10.1103/PhysRevD.100.083508} {\bibfield  {journal} {\bibinfo  {journal}
  {Phys. Rev. D}\ }\textbf {\bibinfo {volume} {100}},\ \bibinfo {pages}
  {083508} (\bibinfo {year} {2019})},\ \Eprint
  {http://arxiv.org/abs/1810.13424} {arXiv:1810.13424 [astro-ph.CO]}
  \BibitemShut {NoStop}%
\bibitem [{\citenamefont {Mu\~noz}\ \emph {et~al.}(2015)\citenamefont
  {Mu\~noz}, \citenamefont {Ali-Ha\"\i{}moud},\ and\ \citenamefont
  {Kamionkowski}}]{Munoz:2015eqa}%
  \BibitemOpen
  \bibfield  {author} {\bibinfo {author} {\bibfnamefont {J.~B.}\ \bibnamefont
  {Mu\~noz}}, \bibinfo {author} {\bibfnamefont {Y.}~\bibnamefont
  {Ali-Ha\"\i{}moud}}, \ and\ \bibinfo {author} {\bibfnamefont
  {M.}~\bibnamefont {Kamionkowski}},\ }\bibfield  {title} {\emph {\enquote
  {\bibinfo {title} {{Primordial non-gaussianity from the bispectrum of 21-cm
  fluctuations in the dark ages}},}\ }}\href {\doibase
  10.1103/PhysRevD.92.083508} {\bibfield  {journal} {\bibinfo  {journal} {Phys.
  Rev. D}\ }\textbf {\bibinfo {volume} {92}},\ \bibinfo {pages} {083508}
  (\bibinfo {year} {2015})},\ \Eprint {http://arxiv.org/abs/1506.04152}
  {arXiv:1506.04152 [astro-ph.CO]} \BibitemShut {NoStop}%
\bibitem [{\citenamefont {Di~Luzio}\ \emph {et~al.}(2020)\citenamefont
  {Di~Luzio}, \citenamefont {Giannotti}, \citenamefont {Nardi},\ and\
  \citenamefont {Visinelli}}]{DiLuzio:2020wdo}%
  \BibitemOpen
  \bibfield  {author} {\bibinfo {author} {\bibfnamefont {L.}~\bibnamefont
  {Di~Luzio}}, \bibinfo {author} {\bibfnamefont {M.}~\bibnamefont {Giannotti}},
  \bibinfo {author} {\bibfnamefont {E.}~\bibnamefont {Nardi}}, \ and\ \bibinfo
  {author} {\bibfnamefont {L.}~\bibnamefont {Visinelli}},\ }\bibfield  {title}
  {\emph {\enquote {\bibinfo {title} {{The landscape of QCD axion models}},}\
  }}\href {\doibase 10.1016/j.physrep.2020.06.002} {\bibfield  {journal}
  {\bibinfo  {journal} {Phys. Rept.}\ }\textbf {\bibinfo {volume} {870}},\
  \bibinfo {pages} {1} (\bibinfo {year} {2020})},\ \Eprint
  {http://arxiv.org/abs/2003.01100} {arXiv:2003.01100 [hep-ph]} \BibitemShut
  {NoStop}%
\bibitem [{\citenamefont {Kitajima}\ \emph {et~al.}(2014)\citenamefont
  {Kitajima}, \citenamefont {Langlois}, \citenamefont {Takahashi},
  \citenamefont {Takesako},\ and\ \citenamefont {Yokoyama}}]{Kitajima:2014xna}%
  \BibitemOpen
  \bibfield  {author} {\bibinfo {author} {\bibfnamefont {N.}~\bibnamefont
  {Kitajima}}, \bibinfo {author} {\bibfnamefont {D.}~\bibnamefont {Langlois}},
  \bibinfo {author} {\bibfnamefont {T.}~\bibnamefont {Takahashi}}, \bibinfo
  {author} {\bibfnamefont {T.}~\bibnamefont {Takesako}}, \ and\ \bibinfo
  {author} {\bibfnamefont {S.}~\bibnamefont {Yokoyama}},\ }\bibfield  {title}
  {\emph {\enquote {\bibinfo {title} {{Thermal Effects and Sudden Decay
  Approximation in the Curvaton Scenario}},}\ }}\href {\doibase
  10.1088/1475-7516/2014/10/032} {\bibfield  {journal} {\bibinfo  {journal}
  {JCAP}\ }\textbf {\bibinfo {volume} {10}},\ \bibinfo {pages} {032} (\bibinfo
  {year} {2014})},\ \Eprint {http://arxiv.org/abs/1407.5148} {arXiv:1407.5148
  [astro-ph.CO]} \BibitemShut {NoStop}%
\bibitem [{\citenamefont {Starobinsky}(1982)}]{Starobinsky:1982ee}%
  \BibitemOpen
  \bibfield  {author} {\bibinfo {author} {\bibfnamefont {A.~A.}\ \bibnamefont
  {Starobinsky}},\ }\bibfield  {title} {\emph {\enquote {\bibinfo {title}
  {{Dynamics of Phase Transition in the New Inflationary Universe Scenario and
  Generation of Perturbations}},}\ }}\href {\doibase
  10.1016/0370-2693(82)90541-X} {\bibfield  {journal} {\bibinfo  {journal}
  {Phys. Lett. B}\ }\textbf {\bibinfo {volume} {117}},\ \bibinfo {pages} {175}
  (\bibinfo {year} {1982})}\BibitemShut {NoStop}%
\bibitem [{\citenamefont {Salopek}\ and\ \citenamefont
  {Bond}(1990)}]{Salopek:1990jq}%
  \BibitemOpen
  \bibfield  {author} {\bibinfo {author} {\bibfnamefont {D.~S.}\ \bibnamefont
  {Salopek}}\ and\ \bibinfo {author} {\bibfnamefont {J.~R.}\ \bibnamefont
  {Bond}},\ }\bibfield  {title} {\emph {\enquote {\bibinfo {title} {{Nonlinear
  evolution of long wavelength metric fluctuations in inflationary models}},}\
  }}\href {\doibase 10.1103/PhysRevD.42.3936} {\bibfield  {journal} {\bibinfo
  {journal} {Phys. Rev. D}\ }\textbf {\bibinfo {volume} {42}},\ \bibinfo
  {pages} {3936} (\bibinfo {year} {1990})}\BibitemShut {NoStop}%
\bibitem [{\citenamefont {Sasaki}\ and\ \citenamefont
  {Stewart}(1996)}]{Sasaki:1995aw}%
  \BibitemOpen
  \bibfield  {author} {\bibinfo {author} {\bibfnamefont {M.}~\bibnamefont
  {Sasaki}}\ and\ \bibinfo {author} {\bibfnamefont {E.~D.}\ \bibnamefont
  {Stewart}},\ }\bibfield  {title} {\emph {\enquote {\bibinfo {title} {{A
  General analytic formula for the spectral index of the density perturbations
  produced during inflation}},}\ }}\href {\doibase 10.1143/PTP.95.71}
  {\bibfield  {journal} {\bibinfo  {journal} {Prog. Theor. Phys.}\ }\textbf
  {\bibinfo {volume} {95}},\ \bibinfo {pages} {71} (\bibinfo {year} {1996})},\
  \Eprint {http://arxiv.org/abs/astro-ph/9507001} {arXiv:astro-ph/9507001}
  \BibitemShut {NoStop}%
\bibitem [{\citenamefont {Lyth}\ \emph {et~al.}(2005)\citenamefont {Lyth},
  \citenamefont {Malik},\ and\ \citenamefont {Sasaki}}]{Lyth:2004gb}%
  \BibitemOpen
  \bibfield  {author} {\bibinfo {author} {\bibfnamefont {D.~H.}\ \bibnamefont
  {Lyth}}, \bibinfo {author} {\bibfnamefont {K.~A.}\ \bibnamefont {Malik}}, \
  and\ \bibinfo {author} {\bibfnamefont {M.}~\bibnamefont {Sasaki}},\
  }\bibfield  {title} {\emph {\enquote {\bibinfo {title} {{A General proof of
  the conservation of the curvature perturbation}},}\ }}\href {\doibase
  10.1088/1475-7516/2005/05/004} {\bibfield  {journal} {\bibinfo  {journal}
  {JCAP}\ }\textbf {\bibinfo {volume} {05}},\ \bibinfo {pages} {004} (\bibinfo
  {year} {2005})},\ \Eprint {http://arxiv.org/abs/astro-ph/0411220}
  {arXiv:astro-ph/0411220} \BibitemShut {NoStop}%
\bibitem [{\citenamefont {Lyth}\ and\ \citenamefont
  {Rodriguez}(2005)}]{Lyth:2005fi}%
  \BibitemOpen
  \bibfield  {author} {\bibinfo {author} {\bibfnamefont {D.~H.}\ \bibnamefont
  {Lyth}}\ and\ \bibinfo {author} {\bibfnamefont {Y.}~\bibnamefont
  {Rodriguez}},\ }\bibfield  {title} {\emph {\enquote {\bibinfo {title} {{The
  Inflationary prediction for primordial non-Gaussianity}},}\ }}\href {\doibase
  10.1103/PhysRevLett.95.121302} {\bibfield  {journal} {\bibinfo  {journal}
  {Phys. Rev. Lett.}\ }\textbf {\bibinfo {volume} {95}},\ \bibinfo {pages}
  {121302} (\bibinfo {year} {2005})},\ \Eprint
  {http://arxiv.org/abs/astro-ph/0504045} {arXiv:astro-ph/0504045} \BibitemShut
  {NoStop}%
\bibitem [{\citenamefont {Sugiyama}\ \emph {et~al.}(2013)\citenamefont
  {Sugiyama}, \citenamefont {Komatsu},\ and\ \citenamefont
  {Futamase}}]{Sugiyama:2012tj}%
  \BibitemOpen
  \bibfield  {author} {\bibinfo {author} {\bibfnamefont {N.~S.}\ \bibnamefont
  {Sugiyama}}, \bibinfo {author} {\bibfnamefont {E.}~\bibnamefont {Komatsu}}, \
  and\ \bibinfo {author} {\bibfnamefont {T.}~\bibnamefont {Futamase}},\
  }\bibfield  {title} {\emph {\enquote {\bibinfo {title} {{$\delta$N
  formalism}},}\ }}\href {\doibase 10.1103/PhysRevD.87.023530} {\bibfield
  {journal} {\bibinfo  {journal} {Phys. Rev. D}\ }\textbf {\bibinfo {volume}
  {87}},\ \bibinfo {pages} {023530} (\bibinfo {year} {2013})},\ \Eprint
  {http://arxiv.org/abs/1208.1073} {arXiv:1208.1073 [gr-qc]} \BibitemShut
  {NoStop}%
\bibitem [{\citenamefont {Maldacena}(2003)}]{Maldacena:2002vr}%
  \BibitemOpen
  \bibfield  {author} {\bibinfo {author} {\bibfnamefont {J.~M.}\ \bibnamefont
  {Maldacena}},\ }\bibfield  {title} {\emph {\enquote {\bibinfo {title}
  {{Non-Gaussian features of primordial fluctuations in single field
  inflationary models}},}\ }}\href {\doibase 10.1088/1126-6708/2003/05/013}
  {\bibfield  {journal} {\bibinfo  {journal} {JHEP}\ }\textbf {\bibinfo
  {volume} {05}},\ \bibinfo {pages} {013} (\bibinfo {year} {2003})},\ \Eprint
  {http://arxiv.org/abs/astro-ph/0210603} {arXiv:astro-ph/0210603} \BibitemShut
  {NoStop}%
\bibitem [{\citenamefont {Lyth}(2005)}]{Lyth:2005qk}%
  \BibitemOpen
  \bibfield  {author} {\bibinfo {author} {\bibfnamefont {D.~H.}\ \bibnamefont
  {Lyth}},\ }\bibfield  {title} {\emph {\enquote {\bibinfo {title} {{Generating
  the curvature perturbation at the end of inflation}},}\ }}\href {\doibase
  10.1088/1475-7516/2005/11/006} {\bibfield  {journal} {\bibinfo  {journal}
  {JCAP}\ }\textbf {\bibinfo {volume} {11}},\ \bibinfo {pages} {006} (\bibinfo
  {year} {2005})},\ \Eprint {http://arxiv.org/abs/astro-ph/0510443}
  {arXiv:astro-ph/0510443} \BibitemShut {NoStop}%
\bibitem [{\citenamefont {Byrnes}\ and\ \citenamefont
  {Wands}(2006)}]{Byrnes:2006fr}%
  \BibitemOpen
  \bibfield  {author} {\bibinfo {author} {\bibfnamefont {C.~T.}\ \bibnamefont
  {Byrnes}}\ and\ \bibinfo {author} {\bibfnamefont {D.}~\bibnamefont {Wands}},\
  }\bibfield  {title} {\emph {\enquote {\bibinfo {title} {{Curvature and
  isocurvature perturbations from two-field inflation in a slow-roll
  expansion}},}\ }}\href {\doibase 10.1103/PhysRevD.74.043529} {\bibfield
  {journal} {\bibinfo  {journal} {Phys. Rev. D}\ }\textbf {\bibinfo {volume}
  {74}},\ \bibinfo {pages} {043529} (\bibinfo {year} {2006})},\ \Eprint
  {http://arxiv.org/abs/astro-ph/0605679} {arXiv:astro-ph/0605679} \BibitemShut
  {NoStop}%
\bibitem [{\citenamefont {Remmen}\ and\ \citenamefont
  {Carroll}(2014)}]{Remmen:2014mia}%
  \BibitemOpen
  \bibfield  {author} {\bibinfo {author} {\bibfnamefont {G.~N.}\ \bibnamefont
  {Remmen}}\ and\ \bibinfo {author} {\bibfnamefont {S.~M.}\ \bibnamefont
  {Carroll}},\ }\bibfield  {title} {\emph {\enquote {\bibinfo {title} {{How
  Many $e$-Folds Should We Expect from High-Scale Inflation?}}}\ }}\href
  {\doibase 10.1103/PhysRevD.90.063517} {\bibfield  {journal} {\bibinfo
  {journal} {Phys. Rev. D}\ }\textbf {\bibinfo {volume} {90}},\ \bibinfo
  {pages} {063517} (\bibinfo {year} {2014})},\ \Eprint
  {http://arxiv.org/abs/1405.5538} {arXiv:1405.5538 [hep-th]} \BibitemShut
  {NoStop}%
\bibitem [{\citenamefont {Akrami}\ \emph
  {et~al.}(2020{\natexlab{a}})\citenamefont {Akrami} \emph
  {et~al.}}]{Planck:2019kim}%
  \BibitemOpen
  \bibfield  {author} {\bibinfo {author} {\bibfnamefont {Y.}~\bibnamefont
  {Akrami}} \emph {et~al.} (\bibinfo {collaboration} {Planck}),\ }\bibfield
  {title} {\emph {\enquote {\bibinfo {title} {{Planck 2018 results. IX.
  Constraints on primordial non-Gaussianity}},}\ }}\href {\doibase
  10.1051/0004-6361/201935891} {\bibfield  {journal} {\bibinfo  {journal}
  {Astron. Astrophys.}\ }\textbf {\bibinfo {volume} {641}},\ \bibinfo {pages}
  {A9} (\bibinfo {year} {2020}{\natexlab{a}})},\ \Eprint
  {http://arxiv.org/abs/1905.05697} {arXiv:1905.05697 [astro-ph.CO]}
  \BibitemShut {NoStop}%
\bibitem [{\citenamefont {Akrami}\ \emph
  {et~al.}(2020{\natexlab{b}})\citenamefont {Akrami} \emph
  {et~al.}}]{Planck:2018jri}%
  \BibitemOpen
  \bibfield  {author} {\bibinfo {author} {\bibfnamefont {Y.}~\bibnamefont
  {Akrami}} \emph {et~al.} (\bibinfo {collaboration} {Planck}),\ }\bibfield
  {title} {\emph {\enquote {\bibinfo {title} {{Planck 2018 results. X.
  Constraints on inflation}},}\ }}\href {\doibase 10.1051/0004-6361/201833887}
  {\bibfield  {journal} {\bibinfo  {journal} {Astron. Astrophys.}\ }\textbf
  {\bibinfo {volume} {641}},\ \bibinfo {pages} {A10} (\bibinfo {year}
  {2020}{\natexlab{b}})},\ \Eprint {http://arxiv.org/abs/1807.06211}
  {arXiv:1807.06211 [astro-ph.CO]} \BibitemShut {NoStop}%
\bibitem [{\citenamefont {Kinney}\ \emph {et~al.}(2012)\citenamefont {Kinney},
  \citenamefont {Moradinezhad~Dizgah}, \citenamefont {Powell},\ and\
  \citenamefont {Riotto}}]{Kinney:2012ik}%
  \BibitemOpen
  \bibfield  {author} {\bibinfo {author} {\bibfnamefont {W.~H.}\ \bibnamefont
  {Kinney}}, \bibinfo {author} {\bibfnamefont {A.}~\bibnamefont
  {Moradinezhad~Dizgah}}, \bibinfo {author} {\bibfnamefont {B.~A.}\
  \bibnamefont {Powell}}, \ and\ \bibinfo {author} {\bibfnamefont
  {A.}~\bibnamefont {Riotto}},\ }\bibfield  {title} {\emph {\enquote {\bibinfo
  {title} {{Inflaton or Curvaton? Constraints on Bimodal Primordial Spectra
  from Mixed Perturbations}},}\ }}\href {\doibase 10.1103/PhysRevD.86.023527}
  {\bibfield  {journal} {\bibinfo  {journal} {Phys. Rev. D}\ }\textbf {\bibinfo
  {volume} {86}},\ \bibinfo {pages} {023527} (\bibinfo {year} {2012})},\
  \Eprint {http://arxiv.org/abs/1203.0693} {arXiv:1203.0693 [astro-ph.CO]}
  \BibitemShut {NoStop}%
\bibitem [{\citenamefont {Fonseca}\ and\ \citenamefont
  {Wands}(2012)}]{Fonseca:2012cj}%
  \BibitemOpen
  \bibfield  {author} {\bibinfo {author} {\bibfnamefont {J.}~\bibnamefont
  {Fonseca}}\ and\ \bibinfo {author} {\bibfnamefont {D.}~\bibnamefont
  {Wands}},\ }\bibfield  {title} {\emph {\enquote {\bibinfo {title}
  {{Primordial non-Gaussianity from mixed inflaton-curvaton perturbations}},}\
  }}\href {\doibase 10.1088/1475-7516/2012/06/028} {\bibfield  {journal}
  {\bibinfo  {journal} {JCAP}\ }\textbf {\bibinfo {volume} {06}},\ \bibinfo
  {pages} {028} (\bibinfo {year} {2012})},\ \Eprint
  {http://arxiv.org/abs/1204.3443} {arXiv:1204.3443 [astro-ph.CO]} \BibitemShut
  {NoStop}%
\bibitem [{\citenamefont {Enqvist}\ and\ \citenamefont
  {Takahashi}(2013)}]{Enqvist:2013paa}%
  \BibitemOpen
  \bibfield  {author} {\bibinfo {author} {\bibfnamefont {K.}~\bibnamefont
  {Enqvist}}\ and\ \bibinfo {author} {\bibfnamefont {T.}~\bibnamefont
  {Takahashi}},\ }\bibfield  {title} {\emph {\enquote {\bibinfo {title} {{Mixed
  Inflaton and Spectator Field Models after Planck}},}\ }}\href {\doibase
  10.1088/1475-7516/2013/10/034} {\bibfield  {journal} {\bibinfo  {journal}
  {JCAP}\ }\textbf {\bibinfo {volume} {10}},\ \bibinfo {pages} {034} (\bibinfo
  {year} {2013})},\ \Eprint {http://arxiv.org/abs/1306.5958} {arXiv:1306.5958
  [astro-ph.CO]} \BibitemShut {NoStop}%
\bibitem [{\citenamefont {Ellis}\ \emph {et~al.}(2014)\citenamefont {Ellis},
  \citenamefont {Fairbairn},\ and\ \citenamefont {Sueiro}}]{Ellis:2013iea}%
  \BibitemOpen
  \bibfield  {author} {\bibinfo {author} {\bibfnamefont {J.}~\bibnamefont
  {Ellis}}, \bibinfo {author} {\bibfnamefont {M.}~\bibnamefont {Fairbairn}}, \
  and\ \bibinfo {author} {\bibfnamefont {M.}~\bibnamefont {Sueiro}},\
  }\bibfield  {title} {\emph {\enquote {\bibinfo {title} {{Rescuing Quadratic
  Inflation}},}\ }}\href {\doibase 10.1088/1475-7516/2014/02/044} {\bibfield
  {journal} {\bibinfo  {journal} {JCAP}\ }\textbf {\bibinfo {volume} {02}},\
  \bibinfo {pages} {044} (\bibinfo {year} {2014})},\ \Eprint
  {http://arxiv.org/abs/1312.1353} {arXiv:1312.1353 [astro-ph.CO]} \BibitemShut
  {NoStop}%
\bibitem [{\citenamefont {Lodman}\ \emph {et~al.}(2023)\citenamefont {Lodman},
  \citenamefont {Lu},\ and\ \citenamefont {Randall}}]{Lodman:2023yrc}%
  \BibitemOpen
  \bibfield  {author} {\bibinfo {author} {\bibfnamefont {J.}~\bibnamefont
  {Lodman}}, \bibinfo {author} {\bibfnamefont {Q.}~\bibnamefont {Lu}}, \ and\
  \bibinfo {author} {\bibfnamefont {L.}~\bibnamefont {Randall}},\ }\bibfield
  {title} {\emph {\enquote {\bibinfo {title} {{Savior Curvatons and Large
  non-Gaussianity}},}\ }}\href@noop {} {\  (\bibinfo {year} {2023})},\ \Eprint
  {http://arxiv.org/abs/2306.13128} {arXiv:2306.13128 [astro-ph.CO]}
  \BibitemShut {NoStop}%
\bibitem [{\citenamefont {Acquaviva}\ \emph {et~al.}(2003)\citenamefont
  {Acquaviva}, \citenamefont {Bartolo}, \citenamefont {Matarrese},\ and\
  \citenamefont {Riotto}}]{Acquaviva:2002ud}%
  \BibitemOpen
  \bibfield  {author} {\bibinfo {author} {\bibfnamefont {V.}~\bibnamefont
  {Acquaviva}}, \bibinfo {author} {\bibfnamefont {N.}~\bibnamefont {Bartolo}},
  \bibinfo {author} {\bibfnamefont {S.}~\bibnamefont {Matarrese}}, \ and\
  \bibinfo {author} {\bibfnamefont {A.}~\bibnamefont {Riotto}},\ }\bibfield
  {title} {\emph {\enquote {\bibinfo {title} {{Second order cosmological
  perturbations from inflation}},}\ }}\href {\doibase
  10.1016/S0550-3213(03)00550-9} {\bibfield  {journal} {\bibinfo  {journal}
  {Nucl. Phys. B}\ }\textbf {\bibinfo {volume} {667}},\ \bibinfo {pages} {119}
  (\bibinfo {year} {2003})},\ \Eprint {http://arxiv.org/abs/astro-ph/0209156}
  {arXiv:astro-ph/0209156} \BibitemShut {NoStop}%
\bibitem [{\citenamefont {Cabass}\ \emph {et~al.}(2017)\citenamefont {Cabass},
  \citenamefont {Pajer},\ and\ \citenamefont {Schmidt}}]{Cabass:2016cgp}%
  \BibitemOpen
  \bibfield  {author} {\bibinfo {author} {\bibfnamefont {G.}~\bibnamefont
  {Cabass}}, \bibinfo {author} {\bibfnamefont {E.}~\bibnamefont {Pajer}}, \
  and\ \bibinfo {author} {\bibfnamefont {F.}~\bibnamefont {Schmidt}},\
  }\bibfield  {title} {\emph {\enquote {\bibinfo {title} {{How Gaussian can our
  Universe be?}}}\ }}\href {\doibase 10.1088/1475-7516/2017/01/003} {\bibfield
  {journal} {\bibinfo  {journal} {JCAP}\ }\textbf {\bibinfo {volume} {01}},\
  \bibinfo {pages} {003} (\bibinfo {year} {2017})},\ \Eprint
  {http://arxiv.org/abs/1612.00033} {arXiv:1612.00033 [hep-th]} \BibitemShut
  {NoStop}%
\bibitem [{\citenamefont {Ade}\ \emph {et~al.}(2021)\citenamefont {Ade} \emph
  {et~al.}}]{BICEP:2021xfz}%
  \BibitemOpen
  \bibfield  {author} {\bibinfo {author} {\bibfnamefont {P.~A.~R.}\
  \bibnamefont {Ade}} \emph {et~al.} (\bibinfo {collaboration} {BICEP, Keck}),\
  }\bibfield  {title} {\emph {\enquote {\bibinfo {title} {{Improved Constraints
  on Primordial Gravitational Waves using Planck, WMAP, and BICEP/Keck
  Observations through the 2018 Observing Season}},}\ }}\href {\doibase
  10.1103/PhysRevLett.127.151301} {\bibfield  {journal} {\bibinfo  {journal}
  {Phys. Rev. Lett.}\ }\textbf {\bibinfo {volume} {127}},\ \bibinfo {pages}
  {151301} (\bibinfo {year} {2021})},\ \Eprint
  {http://arxiv.org/abs/2110.00483} {arXiv:2110.00483 [astro-ph.CO]}
  \BibitemShut {NoStop}%
\bibitem [{\citenamefont {Tristram}\ \emph {et~al.}(2022)\citenamefont
  {Tristram} \emph {et~al.}}]{Tristram:2021tvh}%
  \BibitemOpen
  \bibfield  {author} {\bibinfo {author} {\bibfnamefont {M.}~\bibnamefont
  {Tristram}} \emph {et~al.},\ }\bibfield  {title} {\emph {\enquote {\bibinfo
  {title} {{Improved limits on the tensor-to-scalar ratio using BICEP and
  Planck data}},}\ }}\href {\doibase 10.1103/PhysRevD.105.083524} {\bibfield
  {journal} {\bibinfo  {journal} {Phys. Rev. D}\ }\textbf {\bibinfo {volume}
  {105}},\ \bibinfo {pages} {083524} (\bibinfo {year} {2022})},\ \Eprint
  {http://arxiv.org/abs/2112.07961} {arXiv:2112.07961 [astro-ph.CO]}
  \BibitemShut {NoStop}%
\end{thebibliography}%
\bibliographystyle{apsrev4-1mod}
\end{document}